\documentclass[prb,10pt,twocolumn,superscriptaddress,longbibliography]{revtex4-1}
\usepackage[dvipdfmx]{graphicx}
\usepackage{dcolumn}
\usepackage{bm}
\usepackage{color}
\usepackage{ulem}
\usepackage{natbib}
\usepackage{amsmath,amssymb}
\usepackage{hyperref}

\begin{document}
\newcommand{\bR}{\mbox{\boldmath $R$}}
\newcommand{\tr}[1]{\textcolor{black}{#1}}
\newcommand{\txr}[1]{\textcolor{black}{#1}}
\newcommand{\txrs}[1]{\textcolor{red}{\sout{#1}}}
\newcommand{\txm}[1]{\textcolor{magenta}{#1}}
\newcommand{\trs}[1]{\textcolor{black}{\sout{#1}}}
\newcommand{\tb}[1]{\textcolor{black}{#1}}
\newcommand{\tbs}[1]{\textcolor{black}{\sout{#1}}}
\newcommand{\Ha}{\mathcal{H}}
\newcommand{\mh}{\mathsf{h}}
\newcommand{\mA}{\mathsf{A}}
\newcommand{\mB}{\mathsf{B}}
\newcommand{\mC}{\mathsf{C}}
\newcommand{\mS}{\mathsf{S}}
\newcommand{\mU}{\mathsf{U}}
\newcommand{\mX}{\mathsf{X}}
\newcommand{\sP}{\mathcal{P}}
\newcommand{\sL}{\mathcal{L}}
\newcommand{\sO}{\mathcal{O}}
\newcommand{\la}{\langle}
\newcommand{\ra}{\rangle}
\newcommand{\ga}{\alpha}
\newcommand{\gb}{\beta}
\newcommand{\gc}{\gamma}
\newcommand{\gs}{\sigma}
\newcommand{\vk}{{\bm{k}}}
\newcommand{\vq}{{\bm{q}}}
\newcommand{\vR}{{\bm{R}}}
\newcommand{\vQ}{{\bm{Q}}}
\newcommand{\vga}{{\bm{\alpha}}}
\newcommand{\vgc}{{\bm{\gamma}}}
\newcommand{\Ns}{N_{\text{s}}}
\newcommand{\avrg}[1]{\left\langle #1 \right\rangle}
\newcommand{\eqsa}[1]{\begin{eqnarray} #1 \end{eqnarray}}
\newcommand{\eqwd}[1]{\begin{widetext}\begin{eqnarray} #1 \end{eqnarray}\end{widetext}}
\newcommand{\hatd}[2]{\hat{ #1 }^{\dagger}_{ #2 }}
\newcommand{\hatn}[2]{\hat{ #1 }^{\ }_{ #2 }}
\newcommand{\wdtd}[2]{\widetilde{ #1 }^{\dagger}_{ #2 }}
\newcommand{\wdtn}[2]{\widetilde{ #1 }^{\ }_{ #2 }}
\newcommand{\cond}[1]{\overline{ #1 }_{0}}
\newcommand{\conp}[2]{\overline{ #1 }_{0#2}}
\newcommand{\nn}{\nonumber\\}
\newcommand{\cdt}{$\cdot$}
\newcommand{\bra}[1]{\langle#1|}
\newcommand{\ket}[1]{|#1\rangle}
\newcommand{\braket}[2]{\langle #1 | #2 \rangle}
\newcommand{\bvec}[1]{\mbox{\boldmath$#1$}}
\newcommand{\blue}[1]{{#1}}
\newcommand{\bl}[1]{{#1}}
\newcommand{\red}[1]{\textcolor{black}{#1}}
\newcommand{\rr}[1]{{#1}}
\newcommand{\bu}[1]{\textcolor{black}{#1}}
\newcommand{\cyan}[1]{\textcolor{black}{#1}}
\newcommand{\fj}[1]{{#1}}
\newcommand{\green}[1]{{#1}}
\newcommand{\gr}[1]{\textcolor{green}{#1}}
\newcommand{\tm}[1]{\textcolor{magenta}{#1}}
\newcommand{\tgr}[1]{\textcolor{green1}{#1}}
\newcommand{\tgrs}[1]{\textcolor{green1}{\sout{#1}}}
\newcommand{\tmg}[1]{\textcolor{black}{#1}}
\newcommand{\tmgs}[1]{\textcolor{magenta}{\sout{#1}}}
\newcommand{\pUp}{\ensuremath{
  \mathchoice{\vcenter{\hbox{\includegraphics[height=2.0ex]{fig001.eps}}}}
    {\vcenter{\hbox{\includegraphics[height=2.0ex]{fig001.eps}}}}
    {\vcenter{\hbox{\includegraphics[height=1.5ex]{fig001.eps}}}}
    {\vcenter{\hbox{\includegraphics[height=1.0ex]{fig001.eps}}}}
}}
\newcommand{\pDown}{\ensuremath{
  \mathchoice{\vcenter{\hbox{\includegraphics[height=2.0ex]{fig002.eps}}}}
    {\vcenter{\hbox{\includegraphics[height=2.0ex]{fig002.eps}}}}
    {\vcenter{\hbox{\includegraphics[height=1.5ex]{fig002.eps}}}}
    {\vcenter{\hbox{\includegraphics[height=1.0ex]{fig002.eps}}}}
}}
\newcommand{\pSpinp}{\ensuremath{
  \mathchoice{\vcenter{\hbox{\includegraphics[height=2.0ex]{fig003.eps}}}}
    {\vcenter{\hbox{\includegraphics[height=2.0ex]{fig003.eps}}}}
    {\vcenter{\hbox{\includegraphics[height=1.5ex]{fig003.eps}}}}
    {\vcenter{\hbox{\includegraphics[height=1.0ex]{fig003.eps}}}}
}}

\definecolor{green}{rgb}{0,0.5,0.1}
\definecolor{green1}{rgb}{0,1.0,0.0}
\definecolor{blue}{rgb}{0,0,0.8}
\definecolor{cyan}{rgb}{0,0.8,0.9}
\definecolor{grey}{rgb}{0.3,0.3,0.3}
\definecolor{orange}{rgb}{1,0.5,0.25}
\preprint{APS/123-QED}

\title{
Numerical Algorithm for Exact Finite Temperature Spectra and\\
Its Application to Frustrated Quantum Spin Systems 
}
\author{Youhei Yamaji}
\email{yamaji@ap.t.u-tokyo.ac.jp}
\affiliation{Department of Applied Physics, The University of Tokyo, Hongo, Bunkyo-ku, Tokyo, 113-8656, Japan}
\affiliation{JST PRESTO, Hongo, Bunkyo-ku, Tokyo, 113-8656, Japan}
\author{Takafumi Suzuki}
\affiliation{Graduate School of Engineering, University of Hyogo, Hyogo, 670-2280, Japan}
\author{Mitsuaki Kawamura}
\affiliation{The Institute for Solid State Physics, The University of Tokyo, Kashiwa-shi, Chiba, 277-8581, Japan}

\date{\today}

\begin{abstract}
A numerical algorithm to calculate exact finite-temperature spectra of many-body lattice Hamiltonians
is formulated by combining the typicality approach and the shifted Krylov subspace method.
The combined algorithm, which we name finite-temperature shifted Krylov subspace method for simulating spectra (FTK$\omega$),
efficiently constructs
typical pure states in microcanonical shells
and reproduces the canonical-ensemble probability distribution at finite temperatures
with the computational cost proportional to the Fock space dimension.
The present 
FTK$\omega$
enables us to exactly calculate finite-temperature spectra of
many-body systems
whose system sizes are twice larger than
those handled by the canonical ensemble average
and allows us to access the frequency domain directly without sequential real-time evolution
often used in previous studies.
By employing the reweighting method with the present algorithm,
we obtain significant reduction of the numerical costs for temperature sweeps.
Application to a representative frustrated quantum spin system,
namely the Kiteav-Heisenberg model on a honeycomb lattice,
demonstrates the capability of the FTK$\omega$.
The Kitaev-Heisenberg model shows quantum phase transitions from the quantum spin liquid phase
exactly obtained for the Kitaev model to magnetically ordered phases when
the finite amplitude of the Heisenberg exchange coupling is introduced.
We examine temperature dependence of dynamical spin structure factors of the Kitaev-Heisenberg model in proximity
to the quantum spin liquid.
It is clarified that the crossover from a spin-excitation continuum, which is a characteristic of the quantum spin liquid,
to a damped high-energy magnon mode occurs at temperatures
higher than the energy scale of the Heisenberg exchange couplings or
the spin gap that is a signature of the quantum spin liquid at zero temperature.
The crossover and the closeness to
the Kitaev's quantum spin liquid are quantitatively measured by
the width of the excitation continuum or the magnon spectrum.
The present results shed new light on analysis of neutron scattering and other spectroscopy measurements on spin-liquid candidates.
\end{abstract}
\pacs{
}
\maketitle

\section{Introduction}

An expectation value of an observable $\hat{O}$
in equilibrium at inverse temperature $\beta$
is given by the canonical ensemble average,
\eqsa{
\langle\hat{O}\rangle_{\beta}^{\rm ens}=
\sum_{\nu}\frac{e^{-\beta E_{\nu}}}{Z(\beta)}\bra{\nu}\hat{O}\ket{\nu}
}
for the many-body quantum system described by the Hamiltonian $\hat{H}$,
where $E_{\nu}$ and $\ket{\nu}$ are eigenvalues and orthonormalized eigenvectors of $\hat{H}$, respectively.
Here, $Z(\beta)$ is the partition function given by $Z(\beta)=\sum_{\nu}e^{-\beta E_{\nu}}$.
Although the formula is simple, the evaluation of it is not straightforward.
Even if modern supercomputers are employed, it remains difficult
due to exponential walls~\cite{RevModPhys.71.1253}.

The Fock space dimension $N_{\rm F}$ of the many-body quantum system
increases exponentially with the total number of particles in
or the size of the target system.
When the system is composed of mutually interacting $S$=$1/2$ quantum spins (or interacting qubits),
the Fock space dimension is given as $N_{\rm F}=2^{N}$,
where $N$ is the total number of the spins (or qubits).
Since the straightforward evaluation of the canonical ensemble average
requires every eigenvalue and eigenvector of $\hat{H}$,
the memory cost for it is
scaled by $N_{\rm F}^2$ and
the computational cost of it is scaled by $N_{\rm F}^3$.
For example, when the target system consists of $N=24$ $S$=$1/2$ quantum spins,
it requires storing $2^{48}\sim 3\times 10^{14}$ complex numbers ($\sim$ 4PB) on memory and,
at least,
$2^{72}\sim 4\times 10^{21}$ floating-point operations.
Thus, even for the largest-ever supercomputer, it has remained a formidable task.

Although the ensemble average is the canonical prescription of statistical mechanics,
it has been found and elucidated that the canonical or microcanonical ensemble is neither the only way
to calculate the equilibrium expectation value~\cite{Imada_Takahashi,skilling2013maximum,
PhysRevB.47.7929,
PhysRevB.49.5065,
PhysRevE.62.4365}
of the observable at finite temperatures
\footnote{The studies may recall
the maximum-entropy approach~\cite{mead1984maximum,skilling2013maximum}
and
the forced oscillator method~\cite{PhysRevB.31.4508,PhysRevB.36.8933}
as predecessors of them.}
nor to construct
statistical mechanics~\cite{PhysRevLett.80.1373,popescu2006entanglement,PhysRevLett.96.050403,
sugita2007basis,PhysRevLett.99.160404,PhysRevLett.108.240401,PhysRevLett.111.010401}.

Indeed, in the thermodynamic limit, it has been proven that a single pure state, which is called a typical pure state,
replaces the canonical ensemble~\cite{PhysRevE.62.4365,PhysRevLett.99.160404,PhysRevLett.108.240401}.
Even for finite-size and finite $N_{\rm F}$ systems,
the computational cost of calculating $\langle\hat{O}\rangle_{\beta}^{\rm ens}$ is reduced from $\mathcal{O}(N_{\rm F}^3)$ to $\mathcal{O}(N_{\rm F})$~\cite{PhysRevE.62.4365,PhysRevLett.99.160404,PhysRevLett.108.240401}.
A typical pure state is constructed by utilizing imaginary time evolution of a random vector~\cite{Imada_Takahashi,PhysRevB.49.5065,PhysRevLett.108.240401,PhysRevLett.111.010401}.
The pure state approach may recall the thermo field dynamics that replaces the canonical ensemble with a statistical pure state~\cite{Takahashi_Umezawa}.
However, we note that the pure state approach does not assume equal {\it a priori} probability while the thermo field dynamics is constructed
by keeping the equal {\it a priori} probability.

The $\mathcal{O}(N_{\rm F})$ method enables us to simulate the systems with twice larger number of particles or spins than the conventional ensemble average.
If every eigenstate is stored on memory, the memory cost is scaled by $N_{\rm F}^2$.
However, if a single pure-state wave function reproduces finite-temperature expectation values of observables,
the memory cost is scaled by $N_{\rm F}$.
Therefore, even though several pure states are required in practical simulation, due to the exponential dependence of $N_{\rm F}$ on $N$,
the typical pure state approach can handle twice larger system size.

In addition to thermodynamic quantities in equilibrium, excitation spectra
at finite temperatures are also accessible with computational costs of $\mathcal{O}(N_{\rm F})$.
The excitation spectra at finite temperatures
are 
calculated by
constructing a set of excited states by the Lanczos method~\cite{PhysRevB.49.5065},
by simulating real-time evolution of a typical pure state~\cite{PhysRevLett.90.047203,
PhysRevLett.102.110403,PhysRevLett.110.070404,
PhysRevLett.112.120601,PhysRevLett.112.130403,
doi:10.7566/JPSJ.83.094001,PhysRevB.90.155104,PhysRevB.92.205103},
or by constructing a microcanonical ensemble~\cite{PhysRevB.68.235106,PhysRevLett.92.067202,doi:10.7566/JPSJ.83.094001},
whose computational costs are of $\mathcal{O}(N_{\rm F})$.
The first method has been considered
that hundreds of initial random vectors are necessary to obtain accurate results.
In the second method,
the excitation spectra are achieved by the Fourier transformation of the real-time evolution with an appropriate perturbation.
The accuracy of the approach is guaranteed by typicality in the real-time evolution~\cite{PhysRevLett.102.110403,PhysRevLett.110.070404,
PhysRevLett.112.130403,doi:10.7566/JPSJ.83.094001}.
However, longer real-time simulation is required to obtain lower energy spectra in this approach.
The third method requires determining the temperature that corresponds to the obtained microcanonical shell independently.
A more efficient and self-contained method has been desirable.

A growing demand for
finite-temperature simulation of excitation spectra has originated from experimental researches on many-body quantum systems.
As a typical example, the electron spin resonance in strongly correlated electrons has stimulated
not only theoretical studies~\cite{doi:10.1143/JPSJ.9.888,doi:10.1143/PTP.27.529,PhysRevLett.26.1186,PhysRevB.65.134410},
but also, numerical studies~\cite{PhysRevLett.90.047203,PhysRevB.86.224412,PhysRevB.81.224421} on
the finite-temperature excitation spectra.

Raman scattering and inelastic neutron scattering
measurements on a class of quantum magnets, so-called Kitaev materials~\cite{Note_Kitaev_materials},
have brought renewed attention to 
the temperature dependence of the excitation spectra
~\cite{PhysRevLett.114.147201,banerjee2016proximate,banerjee2016neutron}.
Pioneered by Kitaev,
a family of exactly solvable two-dimensional quantum spin Hamiltonians
has been found~\cite{AnnalsofPhysics321.2,PhysRevB.79.024426},
which
does not exhibit any spotaneous symmetry breaking down to zero temperature
and, thus, shows spin liquid ground state~\cite{review_Balents}.
The models in the family are generically called the Kitaev models. 
Shortly after the findings,
it has been proposed~\cite{Jackeli,PhysRevLett.105.027204}
that the Kitaev model on the two dimensional honeycomb lattice
captures
low-energy spin degrees of freedom
in heavy transition metal oxides with honeycomb networks of transition metal ions,
which are typified by an iridium oxide $\alpha$-Na$_2$IrO$_3$~\cite{PhysRevLett.102.256403,Singh_Gegenwart}.
So far, the Kitaev materials including
$\alpha$-$A_2$IrO$_3$ ($A$=Na, Li), $\alpha$-RuCl$_3$, hyperhoneycomb iridate $\beta$-Li$_2$IrO$_3$~\cite{PhysRevLett.114.077202}, and
stripy-honeycomb iridate $\gamma$-Li$_2$IrO$_3$~\cite{modic2014new}
exhibit
spontaneous time-reversal symmetry breakings.
However,
these materials expected in proximity to the Kitaev's spin liquid stimulate
the experimental research on the excitation spectra at finite temperatures, which requires theoretical counterparts.
In addition to the theoretical studies on Raman spectra and dynamical spin structure factors of the simple Kitaev model
at zero temperature~\cite{PhysRevLett.112.207203,PhysRevLett.113.187201} and finite temperatures~\cite{nasu2016fermionic,PhysRevLett.117.157203},
theoretical and numerical
studies on
the Kitaev-like Hamiltonian on variety of tricoordinate networks~\cite{PhysRevB.89.235102} and
more realistic effective Hamiltonians~\cite{PhysRevLett.105.027204,PhysRevLett.110.097204,arXiv:1312.7437,PhysRevLett.112.077204,PhysRevLett.113.107201} are highly desirable.

In this paper, we propose an $\mathcal{O}(N_{\rm F})$ algorithm for simulating exact finite-temperature excitation spectra in frequency domain
by combining the typical pure state approach and the shifted Krylov subspace method~\cite{frommer2003bicgstab}.
The present algorithm rewrites the Lehmann representation of the Green's function by utilizing
randomly-taken linear combination of eigenstates in an equi-energy shell (a microcanonical shell), instead of eigenstates.
The combination of the typical pure states and the shifted Krylov subspace method leads to $\mathcal{O}(N_{\rm F})$ construction of
the linear combination of eigenstates in a single equi-energy shell.
The linear combination is a typical pure state that corresponds to a microcanonical shell~\footnote{The construction of the microcanonical shell may remind the readers of the microcanonical thermal pure quantum (TPQ) state proposed in Ref.\onlinecite{PhysRevLett.108.240401}. 
To avoid possible confusion, we note that the microcanonical TPQ state
does not correspond to a microcannonical shell.
As proven in Ref.\onlinecite{PhysRevLett.111.010401}, the microcanonical thermal pure quantum state reproduces
probability distribution of the canonical ensemble}.
We name the present algorithm finite-temperature shifted Krylov subspace method for
simulating spectra (FTK$\omega$).
The construction of the microcanonical shells is
compatible with the reweighting method~\cite{PhysRevB.44.5081} that reduces computational costs for tuning temperature.  
It also ensures parallelizability of the FTK$\omega$ algorithm and makes it suitable for massively parallel environments.

We show an application of the FTK$\omega$
to the simplest effective Hamiltonian
of the two-dimensional Kitaev-like systems on honeycomb lattices, namely,
the Kitaev-Heisenberg model~\cite{PhysRevLett.110.097204}.
When the Heisenberg exchange coupling, which breaks the integrability
of the Kitaev model, is introduced,
quantum phase transitions between the Kitaev's spin liquid~\cite{AnnalsofPhysics321.2} and magnetically ordered states are realized~\cite{PhysRevLett.105.027204,PhysRevLett.110.097204,PhysRevB.83.245104,PhysRevB.90.195102}. 
We focus on the proximity of the phase boundary between the Kitaev's spin liquid phase and a magnetically ordered phase,
where characteristics of the Kitaev's spin liquid, such as the thermal fractionalization~\cite{PhysRevB.92.115122}, are observed by heating the magnetically ordered ground states~\cite{PhysRevB.93.174425}. 
We examine temperature dependence of dynamical spin structure factors of the Kitaev-Heisenberg model
and
clarify that the crossover from a spin-excitation continuum, which is a characteristics of the quantum spin liquid,
to a damped high-energy magnon mode occurs at temperatures
higher than the energy scale of the Heisenberg exchange couplings or the spin gap that is a signature of the quantum spin liquid at zero temperature.
The crossover and the closeness to the quantum spin liquid are quantitatively measured by
a dimensionless ratio of
the width of the excitation continuum or the damped magnon spectrum
and the energy at which the spectral weight becomes maximum.
The present results shed new light on analysis of neutron scattering and other spectroscopy measurements on the spin-liquid candidates.

The rest of the paper is organized as follows.
We review the typical pure state approach and the shifted Krylov subspace method in section \ref{secII}.
In section \ref{secIII}, the $\mathcal{O}(N_{\rm F})$ algorithm is detailed.
The computational costs and parallelizabitiy are examined in section \ref{secIV}.
The application of the present algorithm to the Kitaev and Kitaev-Heisenberg models
is shown in section \ref{secV}.
Section \ref{secVI} is devoted to the summary and discussion.

\section{Preliminaries}
\label{secII}
Before going to the formulation of
the $\mathcal{O}(N_{\rm F})$ FTK$\omega$ algorithm,
we briefly review the two building blocks of the algorithm
to make this paper self-contained:
The typical pure state approach and the shifted Krylov subspace method
are explained in the following.
\subsection{Typical pure state approach}

First, we explain that a typical pure state indeed replaces the canonical ensemble at infinite temperature.
At $\beta=0$, a typical pure state is nothing but a random vector~\cite{Imada_Takahashi,PhysRevE.62.4365} as shown below.
A random vector is easily constructed by employing real space configurations $\{\ket{x}\}$
as,
\eqsa{
\ket{\phi_0} = \sum_{x} c_x \ket{x},
}
where $\{c_x\}$ is a set of random complex numbers that satisfies the normalization condition $\sum_x |c_x|^2=1$.
If we focus on a quantum lattice model consisting of $N$ $S$=$1/2$ spins, the real space configurations are simply
given as sets of binary bits, such as $\ket{x}=\ket{\sigma_0 \sigma_1 \cdots \sigma_{N-1}}$ ($\sigma_j=0,1$ or $\sigma_j=\uparrow,\downarrow$), which are easy to implement.
Uniform distribution on the unit sphere in $\mathbb{R}^{2N_{\rm F}}$ is often used as
probability distribution for the set of the random numbers $\{c_x\}$~\cite{ULLAH196465,PhysRevE.62.4365}.
Then, the average of the expectation value $\bra{\phi_0}\hat{O}\ket{\phi_0}$ over the uniform probability is trivially equal to 
the ensemble average $\langle\hat{O}\rangle_{\beta=0}^{\rm ens}$ as
\eqsa{
\mathbb{E}[\bra{\phi_0}\hat{O}\ket{\phi_0}]=
N_{\rm F}^{-1}\sum_{\nu} \bra{\nu}\hat{O}\ket{\nu}=
\langle\hat{O}\rangle_{\beta=0}^{\rm ens},
}
where
$\mathbb{E}[f(\{c_x\})]$ denotes the average of a function $f(\{c_x\})$ over the uniform probability,
$\mathbb{E}[|c_x|^2]=N_{\rm F}^{-1}$,
and the unitary transformation
$\ket{\nu}=\sum_x U_{x\nu}\ket{x}$ are used.

A non-trivial fact is exponentially small 
variance of the difference
$\delta\hat{O}=\bra{\phi_0}\hat{O}\ket{\phi_0}-\langle\hat{O}\rangle_{\beta=0}^{\rm ens}$,
which is given by
\eqsa{
\mathbb{E}[\delta\hat{O}^{\dagger}\delta\hat{O}]
=\frac{N_{\rm F}^{-1}{\rm Tr}[\hat{O}^{\dagger}\hat{O}]-
N_{\rm F}^{-2}|{\rm Tr}\ \hat{O}|^2}{N_{\rm F}+1},\label{variance_HDR}
}
where ${\rm Tr}\ \hat{O}=\sum_{\nu} \bra{\nu}\hat{O}\ket{\nu}$.
The above formula Eq.(\ref{variance_HDR}) is found by Hams and De Raedt~\cite{PhysRevE.62.4365}.
Later, Sugita~\cite{sugita2007basis} and Reimann~\cite{PhysRevLett.99.160404} obtained
essentially the same results independently.  
If we set $\hat{O}=\hat{H}$ and recall that width of energy distribution is scaled by $N$
as
\eqsa{
N_{\rm F}^{-1}{\rm Tr}[\hat{H}^2] - N_{\rm F}^{-2}{\rm Tr}[\hat{H}]^2 \propto N,
}
we obtain exponentially small variance of energy estimated by the typical state $\ket{\phi_0}$:
The variance $\mathbb{E}[(\delta\hat{H})^2]$ turns out to be proportional to $N/(N_{\rm F}+1)< N 2^{-N}$ for $N$ quantum spins
by utilizing Eq.(\ref{variance_HDR}). 


The partition function at a finite temperature is also obtained by the typical state $\ket{\phi_0}$~\cite{PhysRevE.62.4365}.
If we set $\hat{O}=e^{-\beta \hat{H}}$, we obtain $Z(\beta)=N_{\rm F}\mathbb{E}[\bra{\phi_0}e^{-\beta\hat{H}}\ket{\phi_0}]$. 
The variance of the typical-state estimate $N_{\rm F}\mathbb{E}[\bra{\phi_0}e^{-\beta\hat{H}}\ket{\phi_0}]$
is also given by Eq.(\ref{variance_HDR}):
The upper bound of the variance is estimated~\cite{Miyashita_DeRaedt} as
\eqsa{
\frac{\mathbb{E}\left[\left(N_{\rm F} \bra{\phi_0}e^{-\beta\hat{H}}\ket{\phi_0} - Z(\beta)\right)^2\right]}{Z(\beta)^2}
<
e^{-S(\beta^{\ast})},
}
where $S(\beta)$ is entropy at the inverse temperature $\beta$ and
$\beta^{\ast}$ is a constant that satisfies $\beta < \beta^{\ast} < 2\beta$.

Ensemble average of an observable $\hat{O}$ other than $e^{-\beta\hat{H}}$ is replaced by
the expectation value with the following
typical pure state~\cite{Imada_Takahashi,PhysRevE.62.4365,PhysRevLett.111.010401},
\eqsa{
\ket{\phi_{\beta}}=e^{-\beta\hat{H}/2}\ket{\phi_0},
\label{tau}
}
which is obtained through imaginary-time evolution initialized with $\ket{\phi_0}$.
The details of the imaginary-time evolution are given in Appendix \ref{appendix_imaginary}.
The ensemble average of $\hat{O}$
is given by
\eqsa{
\langle\hat{O}\rangle_{\beta}^{\rm ens}=
\frac{\displaystyle\mathbb{E}\left[\bra{\phi_{\beta}}\hat{O}\ket{\phi_{\beta}}\right]}
{\displaystyle\mathbb{E}\left[\braket{\phi_{\beta}}{\phi_{\beta}}\right]}.
}
Variance of the estimate by the typical pure state $\ket{\phi_{\beta}}$,
\eqsa{
\sigma_{O}^2=\mathbb{E}
\left[
\left(
\frac{\left\langle \phi_{\beta} \right|\hat{O}\left|\phi_{\beta}\right\rangle}{\langle \phi_{\beta}|\phi_{\beta}\rangle}
-
\langle \hat{O}\rangle^{\rm ens}_{\beta}
\right)^2
\right],
}
is also bounded as
\eqsa{
\sigma_{O}^2
\leq
\frac{\langle (\hat{O}-\langle \hat{O} \rangle_{2\beta}^{\rm ens})^2 \rangle_{2\beta}^{\rm ens}
+(\langle \hat{O} \rangle_{2\beta}^{\rm ens}
-\langle \hat{O} \rangle_{\beta}^{\rm ens})^2}
{\exp [2\beta \{F(2\beta)-F(\beta)\}]},
}
which is derived by Sugiura and Shimizu~\cite{PhysRevLett.111.010401}.
Here, $F(\beta)$ is free energy
given by $F(\beta)=-k_{\rm B}T \ln Z(\beta)$.

For later use, we briefly explain the evaluations of entropy and heat capacity by the typical pure states.
The strict definition for entropy and heat capacity is as follows:
Entropy $S(\beta)$ is given by using the identity $F(\beta)=\langle \hat{H} \rangle_{\beta}^{\rm ens}-TS(\beta)$
as
\eqsa{
S(\beta)=
\frac{\mathbb{E}[\langle \phi_{\beta}|\hat{H} | \phi_{\beta} \rangle]}{T\mathbb{E}\left[\langle \phi_{\beta}| \phi_{\beta} \rangle\right]}
+k_{\rm B}\ln \left( N_{\rm F}\mathbb{E}\left[\langle \phi_{\beta}| \phi_{\beta} \rangle\right]\right),
\label{simplifiedS}
}
and heat capacity $C(\beta)$ is simply given by
\eqsa{
C(\beta)=
\frac{
\displaystyle
\frac{\mathbb{E}[\langle \phi_{\beta}|\hat{H}^2 | \phi_{\beta} \rangle]}{\mathbb{E}\left[\langle \phi_{\beta}| \phi_{\beta} \rangle\right]}
-\frac{\mathbb{E}[\langle \phi_{\beta}|\hat{H} | \phi_{\beta} \rangle]^2}{\mathbb{E}\left[\langle \phi_{\beta}| \phi_{\beta} \rangle\right]^2}
}{k_{\rm B}T^2}.
\label{simplifiedC}
}
However, for simplicity, we use simplified estimates of
entropy and heat capacity as 
\eqsa{
S(\beta)=\mathbb{E}\left[
\frac{\langle \phi_{\beta}| \hat{H}| \phi_{\beta} \rangle}{T\langle \phi_{\beta}| \phi_{\beta} \rangle}
+k_{\rm B}\ln \left( N_{\rm F}\langle \phi_{\beta}| \phi_{\beta} \rangle\right)
\right],
}
and
\eqsa{
C(\beta)=
\frac{
\displaystyle
\mathbb{E}\left[
\frac{\langle \phi_{\beta}| \hat{H}^2| \phi_{\beta} \rangle}{\langle \phi_{\beta}| \phi_{\beta} \rangle}
-\frac{\langle \phi_{\beta}| \hat{H}| \phi_{\beta} \rangle^2}{\langle \phi_{\beta}| \phi_{\beta} \rangle^2}
\right]
}{k_{\rm B}T^2}.
}
These simplified formulae do not cause deviation from the exact ones given by Eqs.(\ref{simplifiedS}) and (\ref{simplifiedC})
beyond standard deviation and standard errors at least in the application to
frustrated magnets.

\subsection{Shifted Krylov subspace method}
Excitation spectra are given by taking imaginary parts of Green's functions in the linear response theory.
We start with the following Green's function at zero temperature, 
\eqsa{
G^{AB} (\zeta)=\bra{0}\hat{A}^{\dagger}(\zeta-\hat{H})^{-1}\hat{B}\ket{0},
}
where $\ket{0}$ is the ground state.
Although the standard $\mathcal{O}(N_{\rm F})$ approach to calculate the excitation spectra of correlated electrons systems is the Lanczos method~\cite{PhysRevLett.59.2999},
here, we review an alternative approach below.

To evaluate the above formula, we solve a linear equation
by employing a conjugate gradient (CG) method,
instead of calculating the expectation value of the resolvent of $\hat{H}$ by the Lanczos method.
The CG methods find
a solution in a
subspace of the Fock space as follows.
First, by introducing the following three vectors,
\eqsa{
\ket{\lambda}
&=&
\hat{A}\ket{0},
\\
\ket{\rho}
&=&
\hat{B}\ket{0},
\\
\ket{\chi (\zeta)}
&=&(\zeta-\hat{H})^{-1}\ket{\rho},
}
we rewrite $G^{AB} (\zeta)$ as
\eqsa{
G^{AB} (\zeta)
= \braket{\lambda}{\chi(\zeta)},
}
where $\ket{\chi (\zeta)}$ is unkown.
To obtain the unknown vector $\ket{\chi(\zeta)}$, we solve the following linear equation,
\eqsa{
(\zeta-\hat{H})\ket{\chi (\zeta)} = \ket{\rho}.\label{linearSQomega}
}

When the matrix $\hat{H}$ is not able to be stored in the memory but a few wave functions can be stored,
the linear equation is solved iteratively, for example, by using the CG methods.
At $n$th iteration, the CG algorithm initialized with $\ket{\chi_0(\zeta)}=\ket{\rho}$
finds an approximate solution $\ket{\chi_n (\zeta)}$ within a $n$-dimensional Krylov subspace
$\mathcal{K}_n (\zeta-\hat{H},\ket{\rho})={\rm span}\{\ket{\rho},(\zeta-\hat{H})\ket{\rho},\dots,
(\zeta-\hat{H})^{n-1}\ket{\rho}\}$.

For complex matrices and vectors, a variation of the CG algorithm, the biconjugate gradient (BiCG), is employed.
At each steps, the BiCG algorithm searches
an approximate solution $\ket{\chi_n (\zeta)}$ by utilizing
a biorthogonal basis set.
The biorthogonal basis set consists of
the residual vectors, $\{\ket{\rho_0 (\zeta)},\ket{\rho_1 (\zeta)},\dots,\ket{\rho_{n-1} (\zeta)}\}$
and $\{\bra{\widetilde{\rho}_0 (\zeta)},\bra{\widetilde{\rho}_1 (\zeta)},\dots,\bra{\widetilde{\rho}_{n-1} (\zeta)}\}$ 
that satisfy $\braket{\widetilde{\rho}_k (\zeta)}{\rho_{k'} (\zeta)}\propto \delta_{kk'}$,
is
iteratively generated by
\eqsa{
\ket{\rho_k (\zeta)}=
(\zeta-\hat{H})\ket{\chi_k (\zeta)} - \ket{\rho},
}
and
\eqsa{
\bra{\widetilde{\rho}_k (\zeta)}=
\bra{\widetilde{\chi}_k (\zeta)}
(\zeta^{\ast}-\hat{H})
 - \bra{\widetilde{\rho}},
}
where $\bra{\widetilde{\rho}}$ is an arbitrary vector with a non-zero 2-norm and a finite internal product
$\braket{\widetilde{\rho}}{\rho}$. 
The approximate solution $\ket{\chi_n (\zeta)}$ ($\bra{\widetilde{\chi}_n (\zeta)}$)
is found in the basis set $\{\ket{\rho_0 (\zeta)},\ket{\rho_1 (\zeta)},\dots,\ket{\rho_{n-1} (\zeta)}\}$
($\{\bra{\widetilde{\rho}_0 (\zeta)},\bra{\widetilde{\rho}_1 (\zeta)},\dots,\bra{\widetilde{\rho}_{n-1} (\zeta)}\}$).

We note that one needs to solve Eq.(\ref{linearSQomega})
essentially once at a fixed complex number $\zeta=\hbar\omega+i\eta$
to obtain whole spectrum $-{\rm Im}G^{AB} (\hbar\omega+i\eta)$.
Due to the shift invariance of the Krylov subspace, namely,
$\mathcal{K}_n (\zeta-\hat{H},\ket{\phi})=\mathcal{K}_n (\zeta'-\hat{H},\ket{\phi})$ for any complex number $\zeta'\neq \zeta$,
the biorthogonal bases $\ket{\rho_k (\zeta)}$ and $\bra{\widetilde{\rho}_k (\zeta)}$ are proportional to
the other biorthogonal bases $\ket{\rho_k (\zeta')}$ and $\bra{\widetilde{\rho}_k (\zeta')}$, respectively~\cite{frommer2003bicgstab}.
Then, one can obtain $\ket{\chi_n (\zeta')}$ from $\ket{\chi_n (\zeta)}$ without the matrix-vector multiplication~\cite{frommer2003bicgstab}.
The Krylov subspace methods utilizing the shift invariance are called the shifted Krylov subspace methods.

In this study,
we employ
the shifted BiCG method~\cite{frommer2003bicgstab} 
implemented in a numerical library $K\omega$ for the shifted Krylov subspace method~\cite{Komega}.
The condition for truncating the shifted BiCG iteration and
the dimension of the Krylov subspace required for the convergence
are examined later.
Typical dimension of the Krylov subspace is of the order of ten thousand at most in the present application.

\section{Finite-temperature dynamical spin structure factor}
\label{secIII}
If every eigenvalue $\{E_{\nu} \}$ and eigenvector $\{\ket{\nu}\}$ of the Hamiltonian $\hat{H}$
are known, the Green's function at a finite temperature $\beta^{-1}$
is given as 
\eqsa{
\displaystyle 
\mathcal{G}^{AB}_{\beta}(\hbar\omega+i\eta)
=
\sum_{\nu,\mu}
\frac{e^{-\beta E_{\nu}}}{Z(\beta)}
\frac{\bra{\nu}
\hat{A}^{\dagger}
\ket{\mu}
\bra{\mu}
\hat{B}^{\ }
\ket{\nu}}
{\hbar\omega+i\eta+E_{\nu}-E_{\mu}},
}
where $Z$ is the partition function of the system defined as $\displaystyle Z(\beta)=\sum_{\nu}e^{-\beta E_{\nu}}$.
In the following sections, we will formulate an algorithm to estimate $\mathcal{G}^{AB}_{\beta}$
with computational costs of $\mathcal{O}(N_{\rm F})$, and give upper bounds of errors in the estimate. 
For later use, we rewrite the above expression of $\mathcal{G}^{AB}_{\beta}$ as
\eqsa{
\mathcal{G}^{AB}_{\beta}(\hbar\omega+i\eta)
=
\displaystyle 
\sum_{\nu}
\frac{e^{-\beta E_{\nu}}}{Z(\beta)}
\bra{\nu}
\hat{A}^{\dagger}
\frac{1}{\hbar\omega+i\eta+E_{\nu}-\hat{H}}
\hat{B}^{\ }
\ket{\nu}.
\label{chiAB}
\nonumber\\
}

\subsection{Intuitive overview}
Here, we reformulate Eq.(\ref{chiAB}) with a typical pure state $\ket{\phi_{\beta}}$
to avoid using the whole set of $E_{\nu}$ and $\ket{\nu}$.
First, we note that the normalized typical state is naively expected to behave as
\eqsa{
\ket{\psi_{\beta}}\equiv
\frac{\ket{\phi_{\beta}}}{\sqrt{\braket{\phi_{\beta}}{\phi_{\beta}}}}
\sim \sum_{\nu}e^{i\varphi_{\nu}}\frac{e^{-\frac{\beta}{2} E_{\nu}}}{\sqrt{Z(\beta)}}\ket{\nu},
}
where each $\varphi_{\nu}$ is a random variable distributed over the interval $[0,2\pi)$.
By introducing a projection operator,
\eqsa{
\hat{P}_{\nu} = \ket{\nu}\bra{\nu},
}
we rewrite the formula based on canonical ensemble, Eq.(\ref{chiAB}), as
\eqsa{
\mathcal{G}^{AB}_{\beta}(\zeta)
\sim
\sum_{\nu}
\bra{\psi_{\beta}}
\hat{P}_{\nu}
\hat{A}^{\dagger}
\frac{1}{\zeta+E_{\nu}-\hat{H}}
\hat{B}
\hat{P}_{\nu}
\ket{\psi_{\beta}}.
}
Thus far, there is no reduction of computational costs from the original formula Eq.(\ref{chiAB}),
since the exact projection operator $\hat{P}_{\nu}$ requires the whole set of $\ket{\nu}$. 

The important step is to find an economical and practical implementation of the projection operator $\hat{P}_{\nu}$. 
Although there is no $\mathcal{O}(N_{\rm F})$ implementation of the exact $\hat{P}_{\nu}$ in the literature as far as we know,
there is a filter operator~\cite{doi:10.1143/ptp/4.4.514,Sakurai2003119,ikegami2010filter,Shimizu201613}
that constructs equi-energy shells and is realizable with the computational cost of $\mathcal{O}(N_{\rm F})$
by employing the shifted Krylov method, as follows.

\subsection{Filter operator and equi-energy shells}
\begin{figure}[thb]
\centering
\includegraphics[width=8cm]{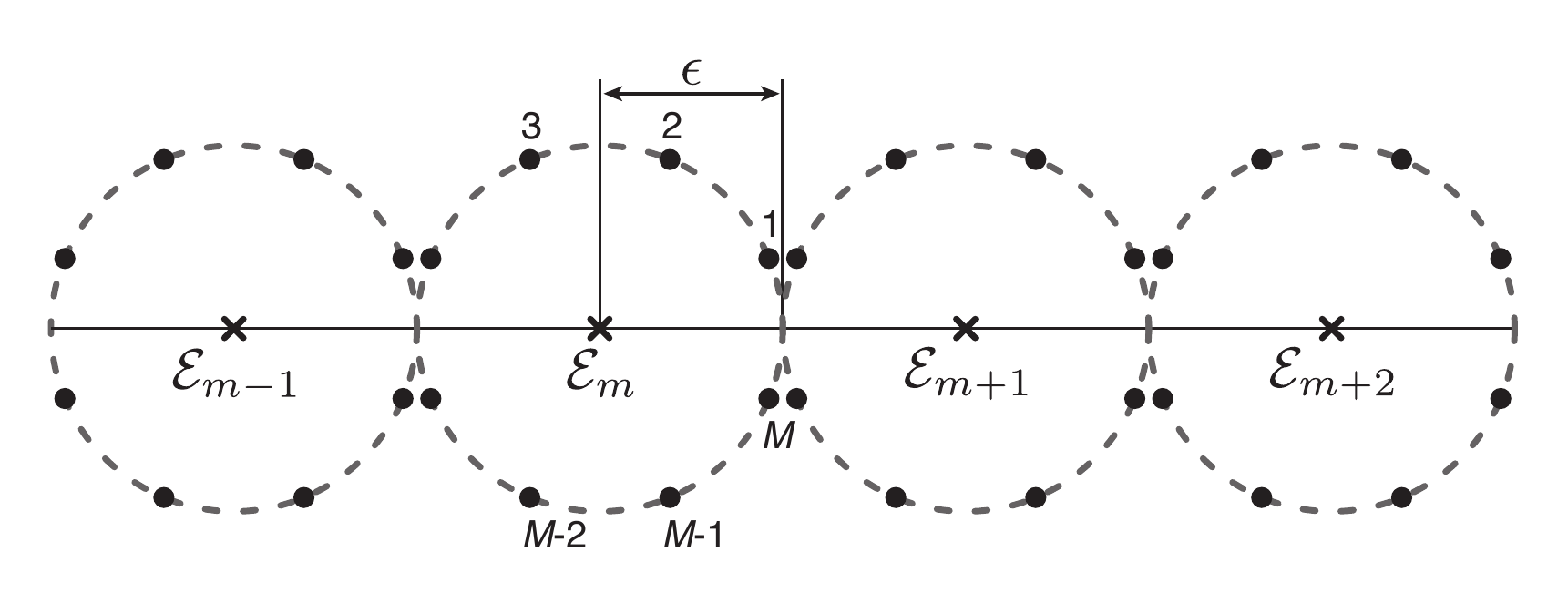}
\caption{(color online):
Discretized contours for the filter operators defined in Eq.(\ref{dfilterP}).}
\label{fig_contour}
\end{figure}

The filter operator~\cite{doi:10.1143/ptp/4.4.514} is defined by integrating the resolvent of $\hat{H}$ along a contour $C_{\gamma,\rho}$ defined
by $z=\rho e^{i\theta} + \gamma$ with $0\leq \theta < 2\pi $ as
\eqsa{
\hat{P}_{\gamma,\rho}=\frac{1}{2\pi i}\oint_{C_{\gamma,\rho}}\frac{dz}{z-\hat{H}}.
}
If the filter operator is applied to an arbitrary wave function $\ket{\phi}=\sum_{\nu} d_{\nu} \ket{\nu}$,
the operator filters the eigenvectors with the eigenvalues $E_{\nu} \not\in (\gamma-\rho,\gamma+\rho)$ as
\eqsa{
\hat{P}_{\gamma,\rho}\ket{\phi}=\sum_{E_{\nu}\in  (\gamma-\rho,\gamma+\rho)} d_{\nu} \ket{\nu}.
}
When a small $\rho$ limit is taken, the filter operator realizes a microcanonical ensemble.
The filter operator is practically implemented as a Reimann sum~\cite{Sakurai2003119,ikegami2010filter}:
The discretized filter operator is defined as
\eqsa{
\hat{P}_{\gamma,\rho,M}=
\frac{1}{M}
\sum_{j=1}^{M}
\frac{\rho e^{i\theta_j}}{\rho e^{i\theta_j}+\gamma -\hat{H}},
\label{dfilterP}
}
where $\theta_j = 2\pi (j-1/2)/M$.
The discretized contour is illustrated in Fig.~\ref{fig_contour}.
Multiplication of $\hat{P}_{\gamma,\rho,M}$ to
a wave function is simply realized by the shifted Krylov subspace method
while it is hardly achievable by the standard Lanczos algorithm.

By introducing an appropriate grid measured from the low-energy onset $E_{\rm b}$ in energy axis,
\eqsa{
\mathcal{E}_m=E_{\rm b}+(2m+1)\epsilon,
}
the set of the filter operators $\{\hat{P}_{\mathcal{E}_m,\epsilon,M}\}$
with the discretization parameters,
\eqsa{\mbox{\boldmath$\delta$}=(E_{\rm b},\epsilon,M),}
indeed replace the projection operators $\{\hat{P}_{\nu}\}$.
The filtered typical state given by
\eqsa{
\ket{\phi_{\beta,\mbox{\boldmath$\delta$}}^{m}}
=
\hat{P}_{\mathcal{E}_m,\epsilon,M}
\ket{\phi_{\beta}}\label{filtered}
}
is a random vector residing in an equi-energy shell $(\mathcal{E}_m-\epsilon,\mathcal{E}_m+\epsilon)$,
which corresponds to a microcanonical ensemble.
From the filtered typical pure states, we obtain a discretized formula for
probability distribution as
\eqsa{
\widetilde{\mathcal{P}}_{\beta,\mbox{\boldmath$\delta$}}(\mathcal{E}_m)
=
\frac{1}{2\epsilon}
\frac{
\braket{\phi_{\beta,\mbox{\boldmath$\delta$}}^{m}}{\phi_{\beta,\mbox{\boldmath$\delta$}}^{m}}}
{\braket{\phi_{\beta}}{\phi_{\beta}}}.
}

In the following application, we prepare the $L$ filter operators to cover an energy range
$[E_{\rm b},E_{\rm b}+2L\epsilon]$.
Here,
$E_{\rm b}$ and $L$ are chosen to keep the probability distribution $\widetilde{\mathcal{P}}_{\beta,\mbox{\boldmath$\delta$}}(E)$ smaller than $10^{-14}$
outside the energy range $[E_{\rm b},E_{\rm b}+2L\epsilon]$.

\subsection{Green's function}
A representation of the Green's function is thus achieved by employing the filtered typical pure states
$\{\ket{\psi_{\beta,\mbox{\boldmath$\delta$}}^{m}}\}$
as
\eqsa{
\widetilde{\mathcal{G}}_{\beta,\mbox{\boldmath$\delta$}}^{AB}(\zeta)
=
\frac{\displaystyle
\sum_{m = 0}^{L-1}
\bra{\phi_{\beta,\mbox{\boldmath$\delta$}}^{m}}
\hat{A}^{\dagger}
\frac{1}{\zeta+\mathcal{E}_m -\hat{H}}
\hat{B}
\ket{\phi_{\beta,\mbox{\boldmath$\delta$}}^{m}}}
{\braket{\phi_{\beta}}{\phi_{\beta}}}.\label{GAB_TPS}
}
After taking appropriate limits and average over the distribution of initial random vectors $\ket{\phi_0}$,
we indeed replace the canonical ensemble prescription by combining the typical pure states and
the shifted Krylov subspace method as
\eqsa{
\mathcal{G}^{AB}_{\beta}(\zeta)=
\lim_{\epsilon\rightarrow + 0} \lim_{M\rightarrow +\infty}
\frac{\displaystyle
\mathbb{E}\left[
\braket{\phi_{\beta}}{\phi_{\beta}}
\widetilde{\mathcal{G}}_{\beta,\mbox{\boldmath$\delta$}}^{AB}(\zeta)
\right]}
{\mathbb{E}[\braket{\phi_{\beta}}{\phi_{\beta}}]}.
\label{limGAB}
}
For simplicity, we use the normalized filtered typical pure state,
\eqsa{
\ket{\psi_{\beta,\mbox{\boldmath$\delta$}}^{m}}
=\frac{\ket{\phi_{\beta,\mbox{\boldmath$\delta$}}^{m}}}{\sqrt{\braket{\phi_{\beta}}{\phi_{\beta}}}},
}
instead of $\ket{\phi_{\beta,\mbox{\boldmath$\delta$}}^{m}}$ in Eq.(\ref{GAB_TPS})
and replace the denominator of the righthand side with unity.

The deviation between the pure-sate representation $\widetilde{\mathcal{G}}_{\beta,\mbox{\boldmath$\delta$}}^{AB}(\zeta)$
and the canonical-ensemble representation $\mathcal{G}^{AB}_{\beta}(\zeta)$ is bounded as follows.
The source of the deviation is twofold:
The discretization parameters $\mbox{\boldmath$\delta$}=(E_{\rm b},\epsilon,M)$ and the variance of the stochastic variables $\{c_n \}$. 
The former source can be examined by changing the set of the discretization parameters $\mbox{\boldmath$\delta$}$.
By following
Refs.\onlinecite{PhysRevE.62.4365} and \onlinecite{PhysRevLett.111.010401},
the upper bound of the variance between $\widetilde{\mathcal{G}}_{\beta,\mbox{\boldmath$\delta$}}^{AB}(\zeta)$
and $\mathcal{G}^{AB}_{\beta}(\zeta)$ due to the variance of $\{c_n \}$
is estimated
as
\eqsa{
&&\mathbb{E}[|\widetilde{\mathcal{G}}_{\beta,\mbox{\boldmath$\delta$}}^{AB}(\hbar\omega+i\eta)-\mathcal{G}^{AB}_{\beta}(\hbar\omega+i\eta)|^2]
\nonumber\\
&&
\lesssim 
e^{-2\beta [F(2\beta)-F(\beta)]}
\Biggl[
\frac{\pi}{\overline{\eta}}\frac{
\langle\hat{B}^{\dagger}\hat{A}
\hat{A}^{\dagger}\hat{B}\rangle_{2\beta}^{\rm ens}}{\eta + \hbar/\tau}
\Biggr.
\nonumber\\
&&
\Biggl.
+
|\mathcal{G}_{\beta}^{AB}(\hbar\omega+i\eta)
-
\mathcal{G}_{2\beta}^{AB}(\hbar\omega+i\eta)|^2
\Biggr],\label{upperboundofdeviation}
}
where constants $\overline{\eta}$ and $\tau$
satisfy
${\rm min}\{\eta,\epsilon\}<\overline{\eta}<\mathcal{O}(NJ_0)$
and $\tau>0$, respectively.
The details of the derivation is given in Appendix \ref{appendix_bound}.
As pointed out in Ref.\onlinecite{PhysRevLett.111.010401},
the prefactor $\exp [-2\beta \{F(2\beta)-F(\beta)\}]$ in the upper bound of the standard deviation
is estimated by entropy:
Due to convex nature of free energy, there is an inverse temperature
$\beta^{\ast}\in (\beta, 2\beta)$ that satisfies $S(\beta^{\ast})=2\beta [F(2\beta)-F(\beta)]$,
where $S$ is entropy.
Therefore, the prefactor exponentially decreases with increasing $N$, since entropy is extensive quantity.
The variance Eq.(\ref{upperboundofdeviation}) is a counterpart in frequency domain
of the variance in time domain~\cite{PhysRevLett.102.110403,PhysRevLett.110.070404,
PhysRevLett.112.130403,doi:10.7566/JPSJ.83.094001} although the relation between them is unknown as far as we know.

\subsection{Reweighting}
The present FTK$\omega$ algorithm based on the filtered typical states resembles histogram techniques~\cite{
FALCIONI1982331,MARINARI1984123}, which were introduced to exploit Monte Carlo simulation data.
Indeed, by reweighting the Boltzmann factors in the filtered typical states
$\left\{\ket{\phi_{\beta,\mbox{\boldmath$\delta$}}^{m}}\right\}$,
we can calculate a finite-temperature expectation value of any operator $\hat{O}$ at an inverse temperature $\beta'$ different
from $\beta$~\cite{PhysRevB.44.5081}.
The filtered typical states for $\beta'$ are given by
\eqsa{
\ket{\phi_{\beta',\mbox{\boldmath$\delta$}}^{m}}
=
e^{\frac{\beta-\beta'}{2}\mathcal{E}_m}\ket{\phi_{\beta,\mbox{\boldmath$\delta$}}^{m}},
}
where the factor $\exp \left[(\beta-\beta')\mathcal{E}_m /2 \right]$ is a c-number.
Since the c-number and any operators commute,
the expectation value of any operator $\hat{O}$ taken by the filtered state $\ket{\phi_{\beta',\mbox{\boldmath$\delta$}}^{m}}$
is given by
\eqsa{
\bra{\phi_{\beta',\mbox{\boldmath$\delta$}}^{m}}\hat{O}\ket{\phi_{\beta',\mbox{\boldmath$\delta$}}^{m}}
=
e^{(\beta-\beta')\mathcal{E}_m}
\bra{\phi_{\beta,\mbox{\boldmath$\delta$}}^{m}}\hat{O}\ket{\phi_{\beta,\mbox{\boldmath$\delta$}}^{m}}.
}
Therefore, if once we calculate $\bra{\phi_{\beta,\mbox{\boldmath$\delta$}}^{m}}\hat{O}\ket{\phi_{\beta,\mbox{\boldmath$\delta$}}^{m}}$
and $\braket{\phi_{\beta,\mbox{\boldmath$\delta$}}^{m}}{\phi_{\beta,\mbox{\boldmath$\delta$}}^{m}}$ for every $m \in [0,L)$,
we can estimate the expectation value of $\hat{O}$ at $\beta'\ (\neq \beta)$ by the following simple expression,
\eqsa{
\langle\hat{O}\rangle_{\beta',\mbox{\boldmath$\delta$}}
\equiv
\frac{\displaystyle \sum_{m=0}^{L-1} e^{(\beta-\beta')\mathcal{E}_m}\bra{\phi_{\beta,\mbox{\boldmath$\delta$}}^{m}}\hat{O}\ket{\phi_{\beta,\mbox{\boldmath$\delta$}}^{m}}}
{\displaystyle\sum_{\ell=0}^{L-1}e^{(\beta-\beta')\mathcal{E}_{\ell}}
\braket{\phi_{\beta,\mbox{\boldmath$\delta$}}^{\ell}}{\phi_{\beta,\mbox{\boldmath$\delta$}}^{\ell}}}.
\label{FRW1}
}
The probability distribution for $\beta'$ is also obtained as 
\eqsa{
\widetilde{\mathcal{P}}_{\beta',\mbox{\boldmath$\delta$}}(\mathcal{E}_m)
=
\frac{1}{2\epsilon}
\frac{\displaystyle e^{(\beta-\beta')\mathcal{E}_m}\braket{\phi_{\beta,\mbox{\boldmath$\delta$}}^{m}}{\phi_{\beta,\mbox{\boldmath$\delta$}}^{m}}}
{\displaystyle\sum_{\ell=0}^{L-1}e^{(\beta-\beta')\mathcal{E}_{\ell}}
\braket{\phi_{\beta,\mbox{\boldmath$\delta$}}^{\ell}}{\phi_{\beta,\mbox{\boldmath$\delta$}}^{\ell}}}.
\label{FRW2}
}

The reweighting method works and significantly reduces the computational cost for tuning temperature if
the probability distributions at
the original temperature $1/\beta$ and the target temperature $1/\beta'$ are overlapped each other~\cite{PhysRevB.44.5081}.
The ratio of the difference in internal energy and width of these distributions determines whether these distributions
are overlapped or not:
If the ratio,
\eqsa{
r=
\frac{
|\langle \hat{H}\rangle^{\rm ens}_{\beta}-\langle\hat{H}\rangle^{\rm ens}_{\beta'}|}{\sqrt{C(\beta)/(k_{\rm B}\beta^2)}+\sqrt{C(\beta')/(k_{\rm B}\beta'^2)}},
}
is small, the probability distributions at $\beta$ and $\beta'$ are overlapped each other.
Here, we approximate the probability distribution at $\beta$ by a Gaussian distribution $N(X)=e^{-(X-\langle\hat{H}\rangle^{\rm ens}_{\beta})^2/2\sigma^2}/\sqrt{2\pi\sigma^2}$
with the variance $\sigma^2=C(\beta)/(k_{\rm B}\beta^2)$. 
When the system size $N$ increases with fixed $\beta$, the ratio $r$ increases at the square root of $N$.
Therefore, if we enlarge the system size with the fixed difference $1/\beta-1/\beta'$, the overlap decreases.
As shown in Sec.\ref{secV}, the reweighting method works for the finite-size spin clusters up to $N=24$
if the difference $1/\beta-1/\beta'$ is appropriate.
In the present application, we choose the difference $1/\beta-1/\beta'$ that corresponds to $r \sim 0.6$
and confirm the accuracy of the reweighting by comparing the reweighted spectrum at $\beta'$ starting from $\beta$
with the spectrum directly calculated at $\beta'$.

\section{Costs and Parallelizability}
\label{secIV}
\subsection{Numerical cost of the present algorithm}
We examine numerical costs 
and parallelizability 
of the present method summarized in Eqs.(\ref{filtered}) and (\ref{GAB_TPS}).
The most time-consuming part is shared by
the present method and other related ones~\cite{PhysRevLett.90.047203,
PhysRevLett.112.120601,doi:10.7566/JPSJ.83.094001,PhysRevB.90.155104,PhysRevB.92.205103,PhysRevB.68.235106,PhysRevLett.92.067202}:
It is multiplication between the Hamiltonian matrix $\hat{H}$ and a wave function,
which is the most time-consuming operation of a single Lanczos step in the Lanczos method.
Thus, the numerical costs of these methods are measured by the number of
the Lanczos steps (or matrix-vector multiplications).

The numerical cost of the present method is determined by
the number of the Lanczos steps for the imaginary time evolution in Eq.(\ref{tau}),
the filter operation in Eq.(\ref{filtered}), the calculation of the equi-energy Green's function
$\bra{\psi_{\beta,\mbox{\boldmath$\delta$}}^{m}}
\hat{A}^{\dagger}(\zeta+\mathcal{E}_m-\hat{H})^{-1}\hat{B}\ket{\psi_{\beta,\mbox{\boldmath$\delta$}}^{m}}$
in Eq.(\ref{GAB_TPS}), and the number of the equi-energy shells, where these numbers are denoted by $N_{\rm L}^{(\tau)}$,
$N_{\rm L}^{(P)}$, $N_{\rm L}^{({\rm e}G)}$, and $N_E$ $(=L)$, respectively.
Then the numerical cost is scaled by
\eqsa{
N_{\rm L}^{(\tau)}+(\alpha_{\rm K}N_{\rm L}^{(P)}+N_{\rm L}^{({\rm e}G)})\times N_E,
}
where $\alpha_{\rm K}$ is a factor larger than $1$ due to the additional cost of the shifted Krylov subspace method.
Here, we note that the naive implementation of the imaginary time evolution (see Appendix \ref{appendix_imaginary}) is
accurate but less efficient than
the implementation by polynomial expansion of the imaginary-time-evolution operators~\cite{PhysRevLett.90.047203}
and the microcanonical TPQ algorithm~\cite{PhysRevLett.108.240401}.
However, $N_{\rm L}^{(\tau)}$ is negligible in the practical simulations at high and moderate temperatures.

\subsection{Numerical cost of related methods}
In comparison, we also estimate the numerical costs of
the closely related methods, namely,
the Boltzmann-weighted time-dependent method (BWTDM)~\cite{PhysRevLett.90.047203} 
and the microcanonical Lanczos method (MCLM)~\cite{PhysRevB.68.235106}.
The BWTDM consists of the imaginary-time and real-time evolution.
Therefore, the numerical cost of the BWTDM is scaled by $N_{\rm L}^{(\tau)}+N_{\rm L}^{(T)}$,
where $N_{\rm L}^{(T)}$ is the number of the Lanczos steps for the real-time evolution
and is proportional to the number of the time steps.
The computational cost of the other related methods may be the same~\cite{PhysRevLett.112.120601,doi:10.7566/JPSJ.83.094001,PhysRevB.90.155104}.
The MCLM, instead, consists of projection to obtain a pure state in a microcanonical shell and the calculation of Green's function by
employing the standard Lanczos method.
The projection is realized through obtaining an eigenstate with the lowest eigenvalue of $(\hat{H}-\lambda)^2$ by the Lanczos method,
where $\lambda$ is the target energy.
In the practical applications of the MCLM, a single microcanonical shell is used for finite-temperature simulations~\cite{PhysRevB.68.235106,PhysRevLett.92.067202} although the MCLM can be used to take a canonical ensemble average by constructing multiple microcanonical shells~\cite{prelovvsek2013ground}.
Thus, the numerical cost of MCLM is scaled by $N_{\rm L}^{\rm (MC)}+N_{\rm L}^{({\rm e}G)}$,
where $N_{\rm L}^{\rm (MC)}$ is the number of the Lanczos step for the projection by $(\hat{H}-\lambda)^2$.
If we set $N_{\rm L}^{(\tau)}=0$ and $N_E=1$, the present method essentially reproduces the same results obtained by the MCLM
although the projection and the present filter operation are quantitatively different. 

\subsection{Advantage of the present algorithm}
The present algorithm seems to be more computationally demanding than the previous related methods~\cite{PhysRevB.49.5065,PhysRevLett.90.047203,
PhysRevLett.112.120601,doi:10.7566/JPSJ.83.094001,PhysRevB.90.155104,PhysRevB.92.205103,PhysRevB.68.235106,PhysRevLett.92.067202,doi:10.7566/JPSJ.83.094001}.
However,
the present method has several advantages over the previous ones. 
The most striking difference between the present algorithm and the related previous approaches
is the computational cost for tuning temperature.
The previous approaches require to repeat the entire simulation to obtain the linear responses at different temperature.
None of the previous approaches is compatible with the reweighting method.
In contrast, the present algorithm interpolates potentially exact spectra at temperatures
between two adjacent temperature points with negligible numerical costs by employing the reweighting method.
The reweighting method works if the overlap of the probability distributions at these two temperatures is significant.

\subsection{Parallelizability}
The present algorithm is more parallelizable than
the real-time evolution of the typical pure state.
The difference in the parallelizability of these two approaches becomes evident, when higher resolution (or smaller broadening factor $\eta$) is required.
To obtain higher resolution in frequency, the present FTK$\omega$ needs more filter operators, or larger $N_{E}$ $(=L)$ and smaller $\epsilon$.
On the other hand, the methods based on real-time evolution need longer time steps, or larger $N_{\rm L}^{(T)}$.
While the construction of each filter operator is parallelizable, the real-time evolution is sequential and not parallelizable.

By taking a simple hybrid parallelization scheme, we examine the parallelizability.
Aside from parallel efficiency, the numerical costs with the $N_{\rm th}$ threads and $N_{\rm pr}$ processes
are scaled as follows if certain schemes of parallelization are chosen.
Here, we choose a simple scheme that
parallelizes a single Lanczos step by shared memory parallelization with $N_{\rm th}$ threads.
The numerical cost of the present method may be scaled as
\eqsa{
\frac{N_{\rm L}^{(\tau)}}{N_{\rm th}}+\left(\alpha_{\rm K}\frac{N_{\rm L}^{(P)}}{N_{\rm th}}+\frac{N_{\rm L}^{({\rm e}G)}}{N_{\rm th}}\right)\times \frac{N_E}{N_{\rm pr}}.
}
In addition to parallelization of an every single Lanczos step, the summation over $m$ in Eq.(\ref{GAB_TPS}) can be parallelized efficiently.
The only way to parallelize the BWTDM, in contrast, is the parallelization of the Lanczos step.
Therefore, the long-time sequential simulation of the BWTDM 
required to obtain the accurate low-energy spectra, 
is bottlenecked by parallelization efficiency of the single Lanczos step.

\begin{figure}[tbh]
\centering
\includegraphics[width=5.5cm]{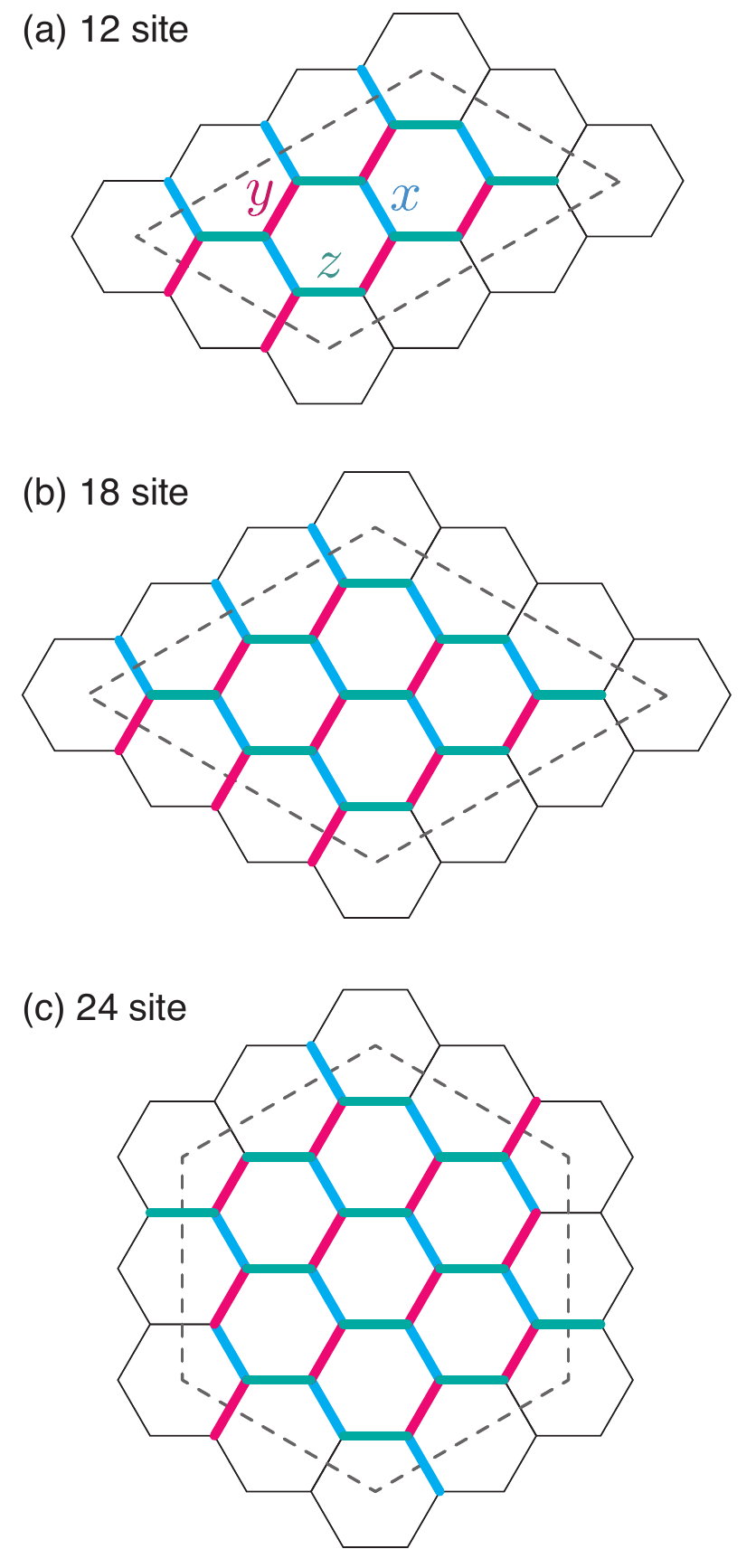}
\caption{(color online):
Finite-size honeycomb clusters with periodic boundary conditions.
Bonds along the three different directions are labeled as
the $x$, $y$, and $z$ bond, which are along -60$^{\circ}$, 60$^{\circ}$,
and $0^{\circ}$ (horizontal) directions, respectively.
The 12 site and 18 site clusters shown in (a) and (b), respectively, are used for comparison with
results by canonical ensemble.
Comparison with the thermodynamic limit of the Kitaev model and
application to the Kitaev-Heisenberg model are done for
the 24 site cluster depicted in (c).
}
\label{fig_honeycomb_18_24}
\end{figure}

\section{Numerical results}
\label{secV}
Here, we examine the accuracy of the present $\mathcal{O}(N_{\rm F})$ algorithm
with practical choices of the parameter $\mbox{\boldmath$\delta$}$
and show an application to the Kitaev-Heisenberg model.
We will calculate
dynamical spin structure factors (DSFs)
at finite temperature
on finite-size honeycomb clusters with periodic boundary conditions, which
are illustrated in Fig.~\ref{fig_honeycomb_18_24}.
By setting $\zeta=\hbar\omega+i\eta$ and
\eqsa{
\hat{A}=\hat{B}=\hat{S}_{+\mbox{\boldmath$Q$}}^{\alpha}\equiv
N^{-1/2}\sum_{\ell}e^{+i\mbox{\boldmath$Q$}\cdot\mbox{\boldmath$R$}_{\ell}}\hat{S}^{\alpha}_{\ell},
}
in Eq.(\ref{GAB_TPS}),
we obtain the DSFs
at a momentum $\mbox{\boldmath$Q$}$ and a frequency $\omega$ as
\eqsa{
\widetilde{S}_{\beta,\mbox{\boldmath$\delta$}}(\mbox{\boldmath$Q$},\omega)
&=&
-\frac{1}{\pi}
{\rm Im}
\sum_{\alpha=x,y,z}
\sum_{m=0}^{L-1}
\bra{\psi_{\beta,\mbox{\boldmath$\delta$}}^{m}}
\hat{S}^{\alpha}_{-\mbox{\boldmath$Q$}}
\nonumber\\
&&\times
\frac{1}{\hbar\omega+i\eta+\mathcal{E}_m -\hat{H}}
\hat{S}^{\alpha}_{+\mbox{\boldmath$Q$}}
\ket{\psi_{\beta,\mbox{\boldmath$\delta$}}^{m}},\label{SQomega_TPS}
}
where $\hat{S}_{\ell}^{\alpha}$ ($\alpha=x,y,z$) is an $S$=1/2 spin operator.
We note that $\mbox{\boldmath$R$}_{\ell}$ is the real space coordinate of the $\ell$ th spin, instead of the position
of the unit cell that contains the $\ell$ th spin.
 
\subsection{Target Hamiltonian}
\begin{figure}[b]
\centering
\includegraphics[width=8cm]{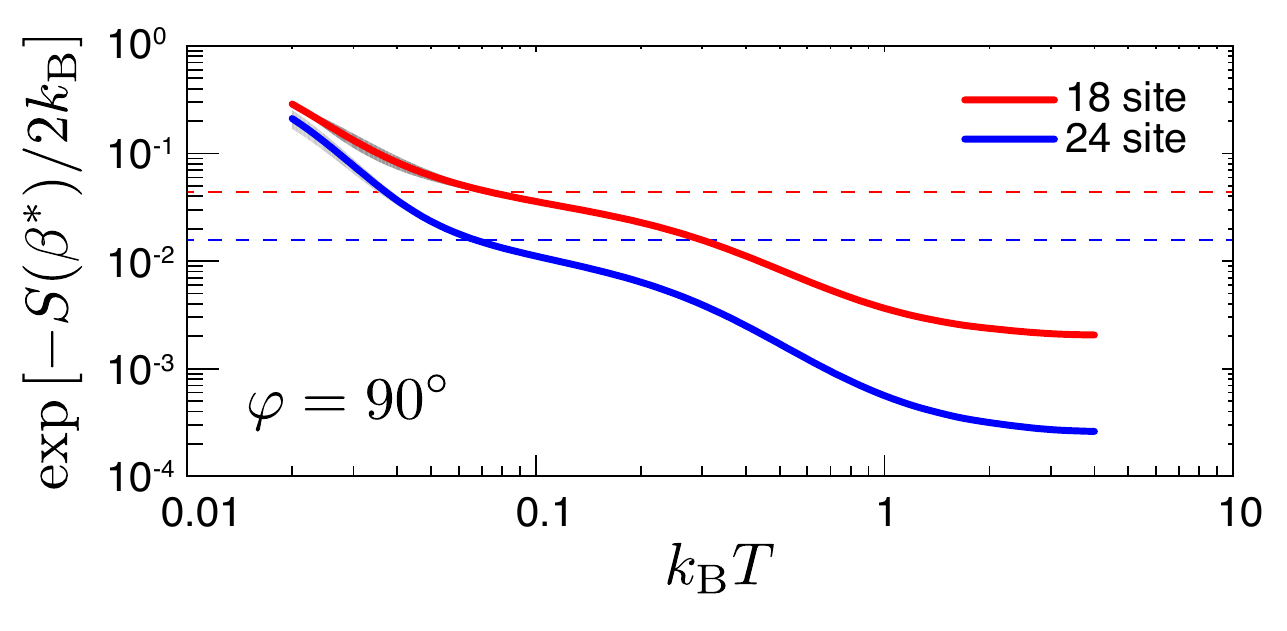}
\caption{(color online):
Temperature dependence of $\exp [-S(\beta^{\ast})/2k_{\rm B}]$ for $\varphi=90^{\circ}$.
The red (blue) solid curve represents $\exp [-S(\beta^{\ast})/2k_{\rm B}]$ of the 18 site (24 site) cluster with the periodic boundary condition
illustrated in Fig.~\ref{fig_honeycomb_18_24}.
The shaded gray belts illustrate the standard deviation estimated by four initial random vectors.
The vertical red (blue) dashed line indicates $(\exp [- (N\ln2) / 2])^{1/2}$ for $N=18$ ($N=24$).
}
\label{fig_expShalf}
\end{figure}

The Kitaev-Heisenberg model on a honeycomb lattice~\cite{PhysRevLett.105.027204} consists of $S$=1/2 spins
that mutually interact with 
two types of the nearest-neighbor exchange couplings:
The Kitaev coupling~\cite{AnnalsofPhysics321.2} and the Heisenberg exchange coupling.
The nearest-neighbor bonds on the honeycomb lattice have three different directions.
When the three bonds are labeled as $x$, $y$, and $z$, the Kitaev-Heisenberg Hamiltonian,
\eqsa{
\hat{H}=\sum_{\gamma=x,y,z}\sum_{\langle i,j \rangle\in \gamma}\hat{H}_{ij}^{(\gamma)},
}
is defined by the exchange coupling for the $\gamma$ ($=x,y,z$) bond,
\eqsa{
\hat{H}_{ij}^{(\gamma)} = 
J\vec{\hat{S}}_{i}\cdot\vec{\hat{S}}_{j}
+
K \hat{S}_{i}^{\gamma}\hat{S}_{j}^{\gamma},
}
where $K=J_0 \sin\varphi$ is the Kitaev coupling constant and
$J=(J_0 /2)\cos \varphi$
is the Heisenberg exchange coupling constant. 
Below, we set the energy unit as $J_0=1$.

The phase diagram of the Kitaev-Heisenberg model
has been numerically clarified~\cite{PhysRevLett.110.097204,PhysRevB.90.195102}. 
The exact diagonalization for a 24 site cluster shows
that the stripy, N\'eel, zigzag, and ferromagnetic ordered phases are the ground states
for $-76.1^{\circ} \lesssim \varphi \lesssim -33.8^{\circ}$,
$-33.8^{\circ} \lesssim \varphi \lesssim 87.7^{\circ}$,
$92.2^{\circ} \lesssim \varphi \lesssim 161.8^{\circ}$,
and
$161.8^{\circ} \lesssim \varphi \lesssim 251.8^{\circ}$, respectively.
The spin liquid phase is stablized for $87.7^{\circ} \lesssim \varphi \lesssim 92.2^{\circ}$
and $251.8^{\circ} \lesssim \varphi \lesssim 283.9^{\circ}=-76.1^{\circ}$~\cite{PhysRevLett.110.097204}.
These phase boundaries are consistent with the previous tensor-network study~\cite{PhysRevB.90.195102}. 

In the following section,
we demonstrate capability and efficiency
of the present algorithm
by calculating DSFs
around the phase boundary between the spin liquid phase and the zigzag ordered phase at $\varphi\sim 92.2^{\circ}$.

\subsection{Variance and statistical errors}
\begin{figure}[b]
\centering
\includegraphics[width=8.5cm]{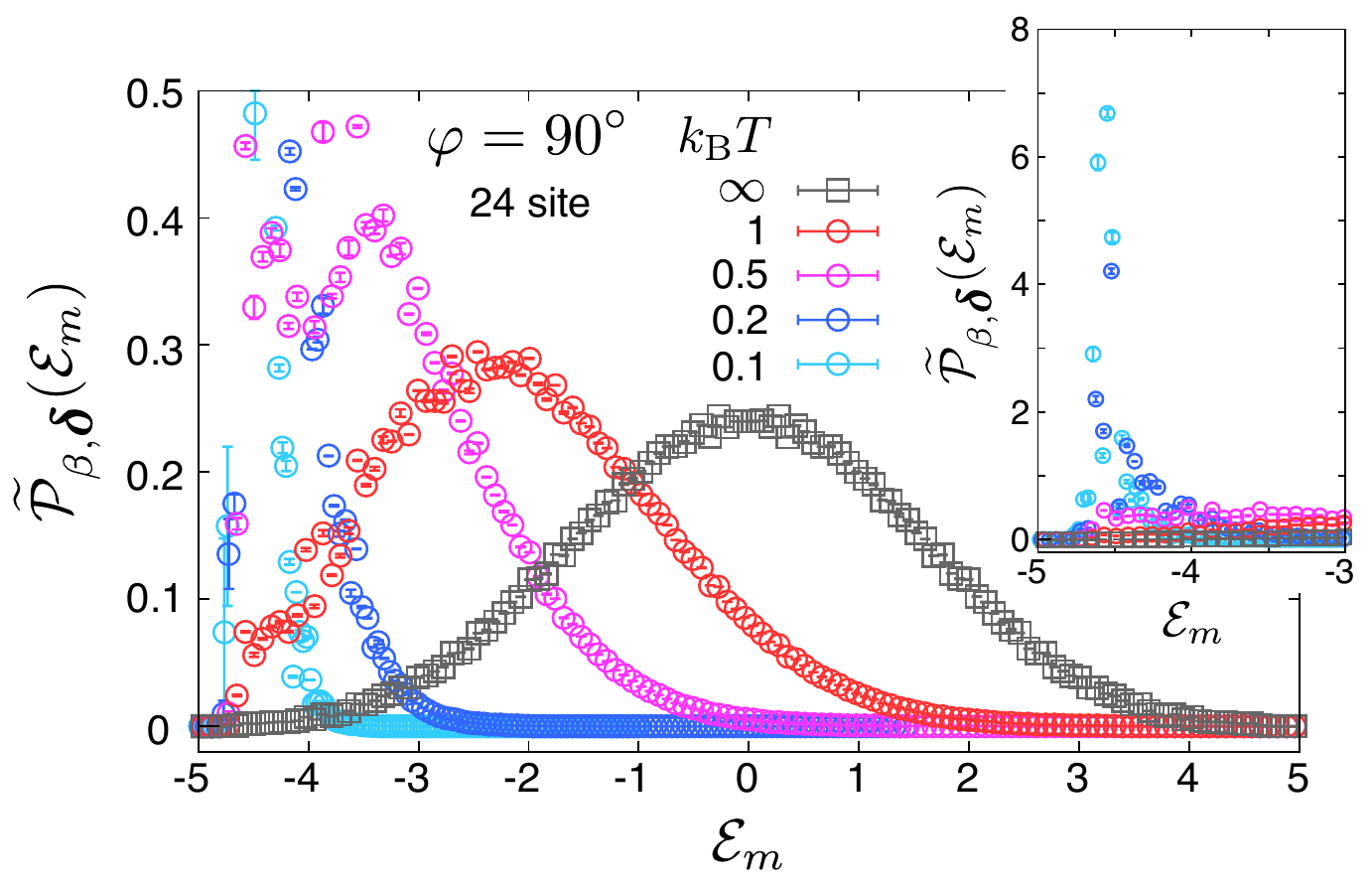}
\caption{(color online):
Probability distribution
in the antiferromagnetic Kitaev model ($\varphi=90^{\circ}$) for the 24 site cluster.
The inset shows the entire peak structures of the probability distribution at low temperatures. 
The variance estimated by two initial random vectors is indicated by vertical bars although
the variance is within the symbol size except for that at $k_{\rm B}T=0.1$.
}
\label{FigProbability}
\end{figure}

As clarified in the literature~\cite{PhysRevE.62.4365,PhysRevLett.111.010401,doi:10.7566/JPSJ.83.094001,PhysRevLett.116.017202}
and the present paper,
there is an upper bound on
variance of finite-temperature physical quantities calculated by the typical pure state approach.
The upper bound has been found to be
proportional to $\exp [-S(\beta^{\ast})/k_{\rm B}]=\exp [-2\beta \{F(2\beta)-F(\beta)\}]$,
regardless of whether physical quantities are static or dynamical.

Here, we note that the square root of the variance (standard deviation) is not an estimate of statistical errors in the physical quantities
due to the distribution of the random initial vectors.
The standard error
proportional to $(\exp [-S(\beta^{\ast})/k_{\rm B}]/N_{\rm s})^{1/2}$
is the estimate of the deviation from physical quantities by canonical ensemble,
when the average is taken over $N_{\rm s}$ initial random vectors~\cite{PhysRevE.62.4365}.

In frustrated quantum spin systems,
sizable entropy
often remains
even at low temperature.
The Kitaev model is an example of
such frustrated magnets.
The Kitaev model has been shown to exhibit a half plateau in temperature dependence of entropy~\cite{PhysRevB.92.115122}.
The 24 site cluster of the Kitaev model employed in the following analysis shows the half plateau in the temperature range $0.02\lesssim k_{\rm B}T \lesssim 0.4$.
As inferred from the plateau, the factor $(\exp [-S(\beta^{\ast})/k_{\rm B}])^{1/2}$ remains as small as 0.01 even below
$k_{\rm B}T\sim\mathcal{O}(J_0/10)$ for $N=24$. 
In Fig.~\ref{fig_expShalf}, temperature dependence of $(\exp [-S(\beta^{\ast})/k_{\rm B}])^{1/2}$
is shown for the antiferromagnetic Kitaev model ($\varphi=90^{\circ}$).
The extensive properties of entropy are reflected in the size dependence of $(\exp [-S(\beta^{\ast})/k_{\rm B}])^{1/2}$,
which is evident in difference between the results for the 24 site and 18 site cluster.

At lower temperature than the plateau region, the standard deviation has been
known to be much smaller than $(\exp [-S(\beta^{\ast})/k_{\rm B}]/N_{\rm s})^{1/2}$
that is a monotonically increasing function of $\beta$~\cite{PhysRevB.93.174425}.
The reduction of the standard deviation originates from prefactors in the variance~\cite{PhysRevLett.111.010401}.

\subsection{Probability distribution}
\begin{figure*}[thb]
\centering
\includegraphics[width=17.0cm]{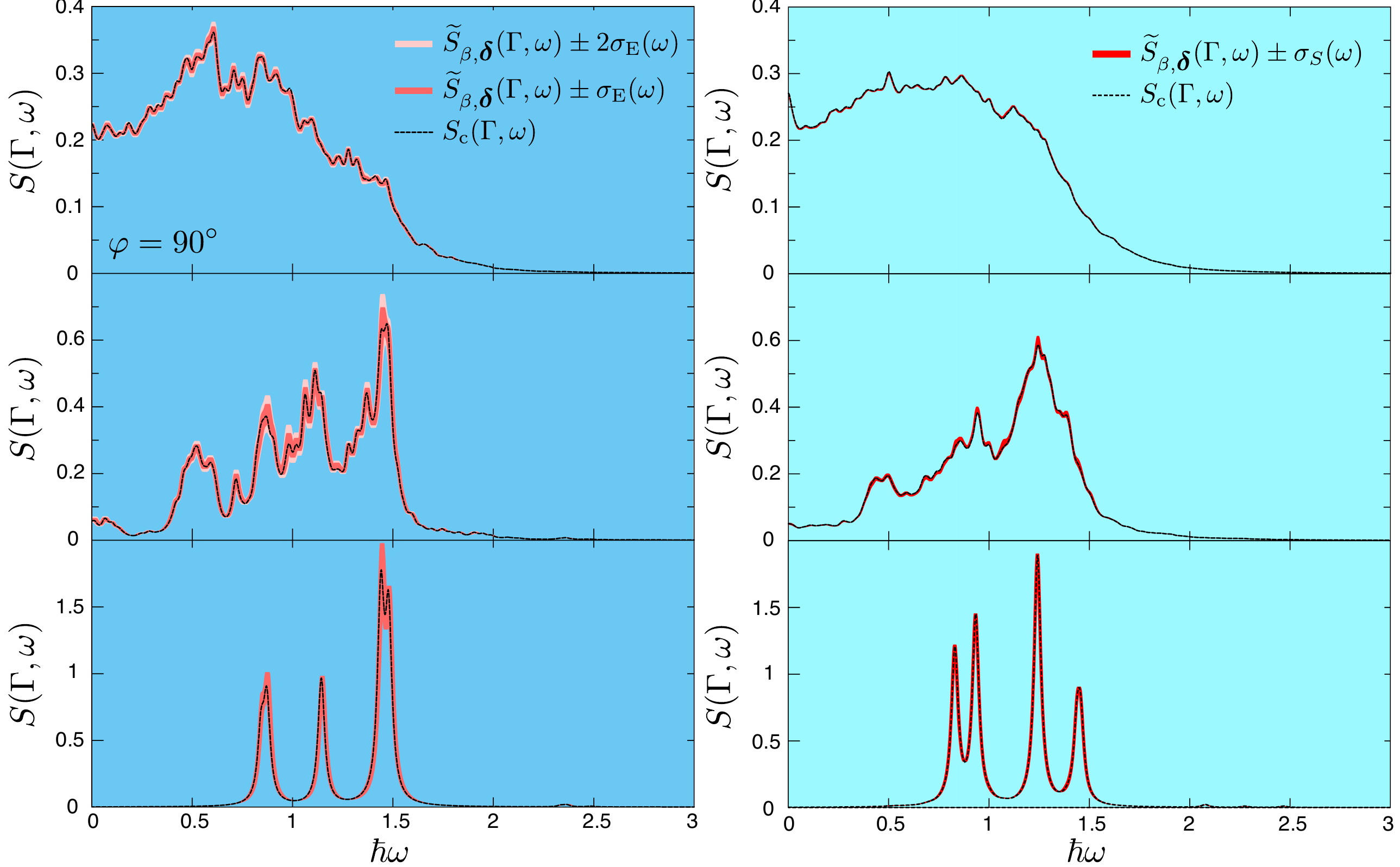}
\caption{(color online):
Comparison between dynamical spin structure factors at the $\Gamma$ point ($\mbox{\boldmath$Q$}=\mbox{\boldmath$0$}$)
obtained by the present algorithm and canonical ensemble for $N=12$ and $18$.
The left column shows the $S(\mbox{\boldmath$Q$}=\Gamma,\omega)$ for $N=12$ and $\varphi=90^{\circ}$.
From the top left to bottom left panel, the dynamical spin structure factors $S(\mbox{\boldmath$Q$}=\Gamma,\omega)$ at $k_{\rm B}T=1$, $0.1$, and $0.01$
are shown.
The broken lines show the canonical ensemble average $S_{\rm c}(\Gamma,\omega)$ and the shaded belts show the FTK$\omega$ results $\widetilde{S}_{\beta,\mbox{\boldmath$\delta$}}(\Gamma,\omega)$.
The vertical width of the red (light red) belts is give by the standard error $\sigma_{\rm E}(\omega)$ ($2\sigma_{\rm E}(\omega)$).
For the left bottom, $2\sigma_{\rm E}(\omega)$ is omitted.
The right column shows the $S(\mbox{\boldmath$Q$}=\Gamma,\omega)$ for $N=18$ and $\varphi=90^{\circ}$.
From the top right to bottom right panel, the dynamical spin structure factors $S(\mbox{\boldmath$Q$}=\Gamma,\omega)$ at $k_{\rm B}T=1$, $0.1$, and $0.01$
are shown.
The vertical width of the red belts is give by the standard deviation $\sigma_{S}(\omega)$.
The horizontal width of the red and light red belts includes discretization errors given by $2\epsilon$.
Here, the broadening factor $\eta$ is set to $0.02$ for both of the canonical ensemble average and the FTK$\omega$ results.
}
\label{FigSQ18}
\end{figure*}

By applying the filter operators to the typical pure sates, we obtain an accurate estimate of probability distribution
for the antiferromagnetic Kitaev model ($\varphi=90^{\circ}$).
The results are shown in Fig.~\ref{FigProbability} for $k_{\rm B}T=0.1$, $0.2$, $0.5$, $1$, and $k_{\rm B}T\rightarrow +\infty$.
At high temperature, $\widetilde{P}_{\beta,\mbox{\boldmath$\delta$}}$ resembles the Gaussian distribution with a width proportional
to $\sqrt{C(\beta)/k_{\rm B}\beta^2}$.
At the high temperature limit,
$\widetilde{P}_{\beta,\mbox{\boldmath$\delta$}}$ becomes nothing but density of states.
The discretization parameter $\mbox{\boldmath$\delta$}=(E_{\rm b},\epsilon,M)$ for the filter operator is chosen as summarized in Appendix \ref{appendix_conv}.

As seen in Fig.~\ref{FigProbability}, the probability distributions for these temperatures overlap each other.
The overlap among them guarantees that the reweighting method interpolates temperature dependence of physical quantities
between the two adjacent temperatures~\cite{PhysRevB.44.5081}.

\subsection{Comparison with canonical ensemble}

To examine accuracy of the present algorithm with a practical choice
of discretization parameters $\mbox{\boldmath$\delta$}$,
we compare the DSFs
obtained by the present FTK$\omega$ algorithm
denoted by $\widetilde{S}_{\beta,\mbox{\boldmath$\delta$}}(\Gamma,\omega)$
with those obtained by canonical ensemble denoted by $S_{\rm c}(\Gamma,\omega)$.
We employ a 12 site and 18 site clusters,
which is illustrated in Fig.~\ref{fig_honeycomb_18_24}(a) and (b).
The 18 site cluster is practically
one of the largest system we can directly take a canonical ensemble average
for the Kitaev-Heisenberg model so far.

In Fig.~\ref{FigSQ18}, we show the DSFs
at $\mbox{\boldmath$Q$}=\Gamma$
obtained by the present algorithm and canonical ensemble average
for the antiferromagnetic Kitaev model ($\varphi=90^{\circ}$).
Here, the broadening factor $\eta$ is fixed at $0.02$.
Convergence of the results is examined by changing the discretization parameter $\mbox{\boldmath$\delta$}=(E_{\rm b},\epsilon,M)$.
From $k_{\rm B}T=1$ to $k_{\rm B}T=0.01$, the present FTK$\omega$ indeed reproduces the canonical ensemble average.

The
DSFs
of the 12 site cluster at $\mbox{\boldmath$Q$}=\Gamma$ are shown in the left panel of Fig.~\ref{FigSQ18}.
By the binning analysis, we estimate standard error $\sigma_{\rm E}$ at each frequency:
First, we prepare 8 sets of 8 samples (8 sets of 16 samples) at $k_{\rm B}T=1$ ($k_{\rm B}T=0.1$ and $0.01$).
Then, we take averages over the samples within each set and 
estimate the standard error $\sigma_{\rm E}$ by the standard deviation of the sets of these averages.
As the result,
we find that
$S_{\rm c}(\Gamma,\omega)$,
and
$\widetilde{S}_{\beta,\mbox{\boldmath$\delta$}}(\Gamma,\omega)$
agree within 1 standard error or, at least, 2 standard error.

Since the simulation for the 18 site cluster is more computationally demanding,
standard deviation $\sigma_{S}$ of 8 samples, instead of the standard error, is estimated.
The number of the available samples is too small to estimate the standard error.
In the right panel of Fig.~\ref{FigSQ18},
we compare $S_{\rm c}(\Gamma,\omega)$ and $\widetilde{S}_{\beta,\mbox{\boldmath$\delta$}}(\Gamma,\omega)$ of the 18 site cluster.
Within 1 standard deviation, $S_{\rm c}(\Gamma,\omega)$ and $\widetilde{S}_{\beta,\mbox{\boldmath$\delta$}}(\Gamma,\omega)$ agree.

\subsection{Temperature evolution of spectra}
\begin{figure}[htb]
\centering
\includegraphics[width=8.5cm]{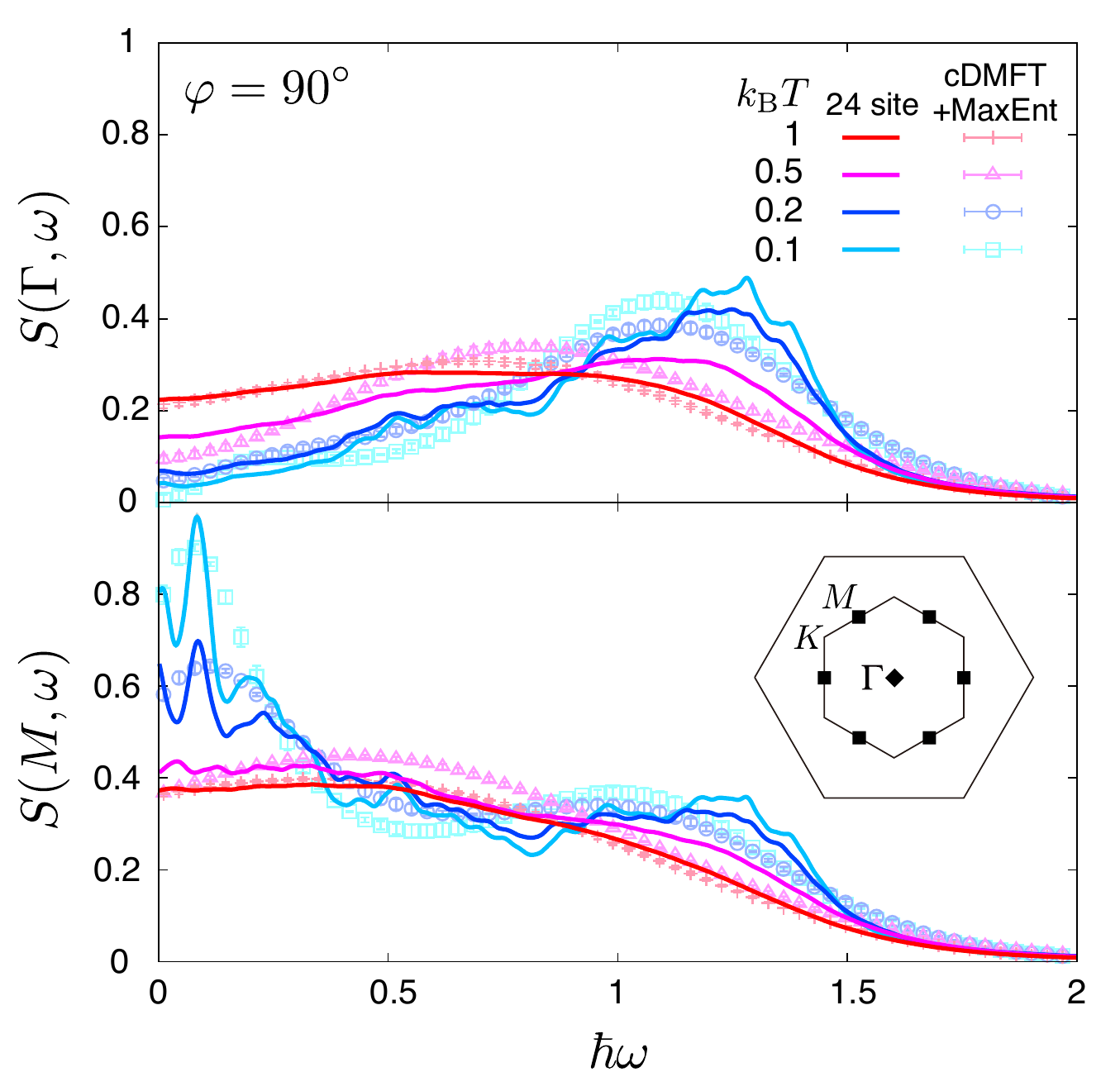}
\caption{(color online):
Dynamical spin structure factors $\widetilde{S}_{\beta,\mbox{\boldmath$\delta$}}(\mbox{\boldmath$Q$},\omega)$
compared with those by the Majorana-fermion cluster dynamical mean-field theory with the maximum entropy method (cDMFT+MaxEnt),
$S_{\rm DMFT}(\mbox{\boldmath$Q$},\omega)$,
from Ref.\onlinecite{PhysRevLett.117.157203}.
The upper (lower) panel shows the dynamical spin structure factors at a typical momentum $\mbox{\boldmath$Q$}=\Gamma$ ($\mbox{\boldmath$Q$}=M$).
The inset of the lower panel illustrates the locations of the symmetric momenta, where
the inner hexagon represents the first Brillouin zone of the honeycomb lattice.
}
\label{FigSQomegaQMC}
\end{figure}

\begin{figure}[thb]
\centering
\includegraphics[width=0.5\textwidth]{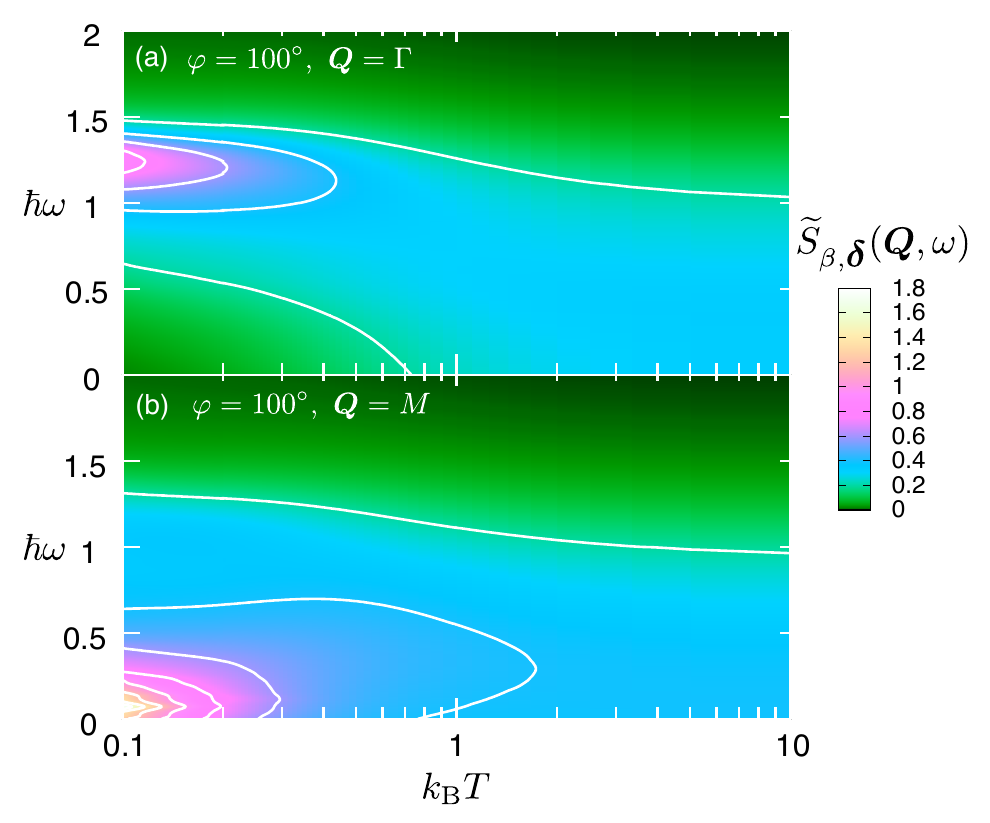}
\caption{(color online):
Temperature dependence of dynamical spin structure factors $\widetilde{S}_{\beta,\mbox{\boldmath$\delta$}}(\mbox{\boldmath$Q$},\omega)$ of the generalized Kitaev-Heisenberg model
for $(\varphi,\mbox{\boldmath$Q$})=(100^{\circ},\Gamma)$ (a)
and $(100^{\circ},M)$ (b). 
The discretization parameters $\mbox{\boldmath$\delta$}$
are the same as $\mbox{\boldmath$\delta$}$ used in Fig.~\ref{FigProbability}.
The broadening factor $\eta$ is set to $0.02$.}
\label{FigSQomega3D}
\end{figure}

\begin{figure*}[thb]
\centering
\includegraphics[width=0.95\textwidth]{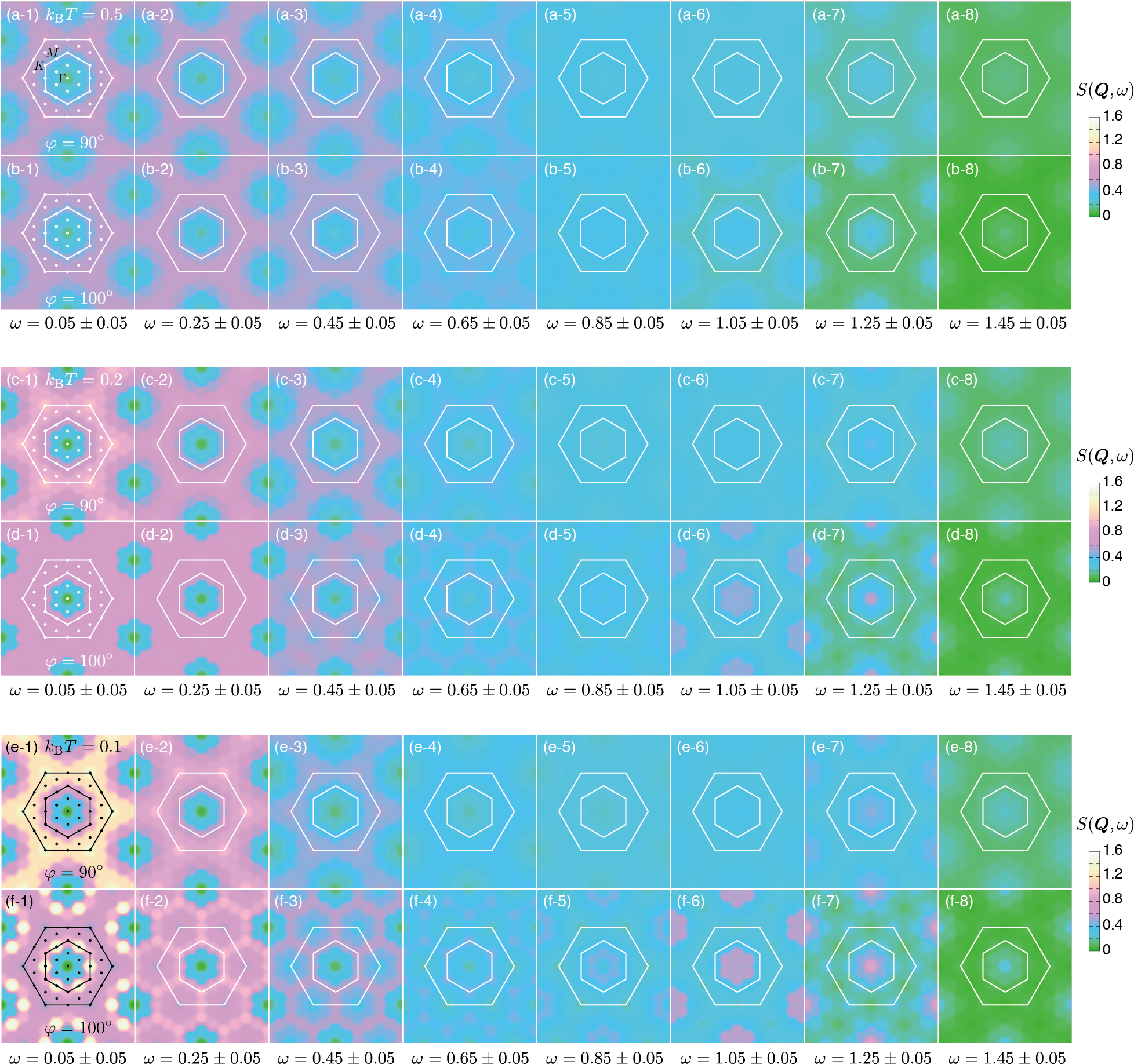}
\caption{(color online):
Equi-energy slices of the dynamical spin structure factors obtained by the FTK$\omega$.
The momentum dependence of the equi-energy slices is shown by changing temperature and frequency.
(a) The equi-energy slices of the dynamical spin structure factor $S(\mbox{\boldmath$Q$},\omega)$ at $k_{\rm B}T=0.5$ for $\varphi=90^{\circ}$
are shown for $\omega=0.05$ (a-1), $\omega=0.25$ (a-2), $\omega=0.45$ (a-3), $\omega=0.65$ (a-4), $\omega=0.85$ (a-5), $\omega=1.05$ (a-6),
$\omega=1.25$ (a-7), and $\omega=1.45$ (a-8).
(b) $S(\mbox{\boldmath$Q$},\omega)$ at $k_{\rm B}T=0.5$ for $\varphi=100^{\circ}$ is shown.
(c) $S(\mbox{\boldmath$Q$},\omega)$ at $k_{\rm B}T=0.2$ for $\varphi=90^{\circ}$ is shown.
(d) $S(\mbox{\boldmath$Q$},\omega)$ at $k_{\rm B}T=0.2$ for $\varphi=100^{\circ}$ is shown.
(e) $S(\mbox{\boldmath$Q$},\omega)$ at $k_{\rm B}T=0.1$ for $\varphi=90^{\circ}$ is shown.
(f) $S(\mbox{\boldmath$Q$},\omega)$ at $k_{\rm B}T=0.1$ for $\varphi=100^{\circ}$ is shown.
The equi-energy slices are prepared by averaging the spectra within an energy window whose width is 0.1.
}
\label{FigSQomegaBZ}
\end{figure*}

\begin{figure}[thb]
\centering
\includegraphics[width=0.49\textwidth]{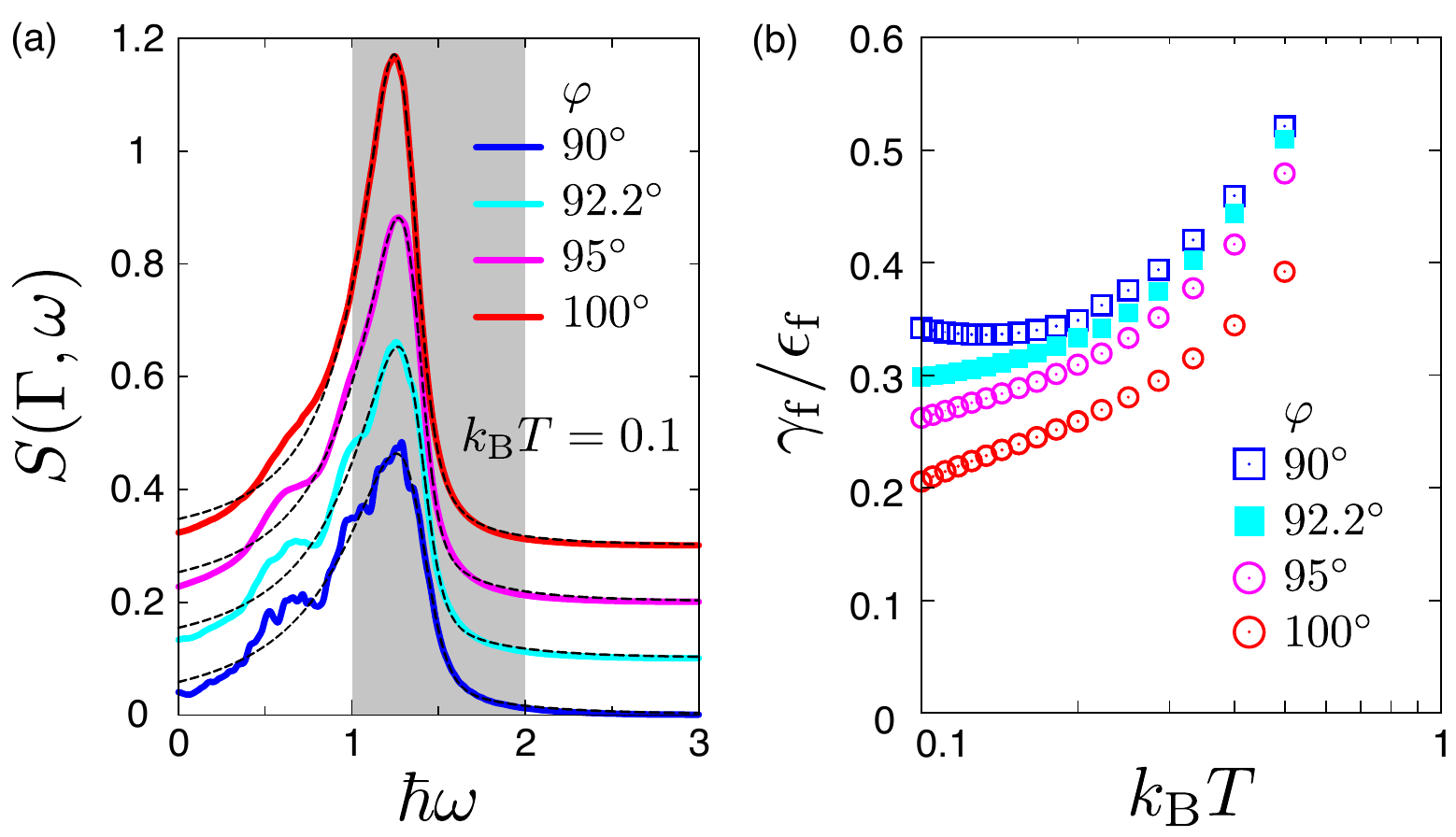}
\caption{(color online):
(a) $S(\Gamma,\omega)$ of the 24 site cluster at $k_{\rm B}T=0.1$
obtained by FTK$\omega$ for $\varphi = 90^{\circ}$, $92.2^{\circ}$,
$95^{\circ}$, and $100^{\circ}$.
The broken curves represent the asymmetric Lorentzian function
$s(\omega)$ defined in Eq.(\ref{asymmetricLorentzian})
fitted to $S(\Gamma,\omega)$ in $1 \leq \hbar\omega \leq 2$,
where $a$, $b$, $c$, $\gamma_{\rm f}$, and $\epsilon_{\rm f}$ are fitting parameters.
These curves are horizontally shifted for visibility.
(b) Temperature dependence of $\gamma_{\rm f}/\epsilon_{\rm f}$.
}
\label{FigGammaET}
\end{figure}

\begin{figure}[thb]
\centering
\includegraphics[width=0.49\textwidth]{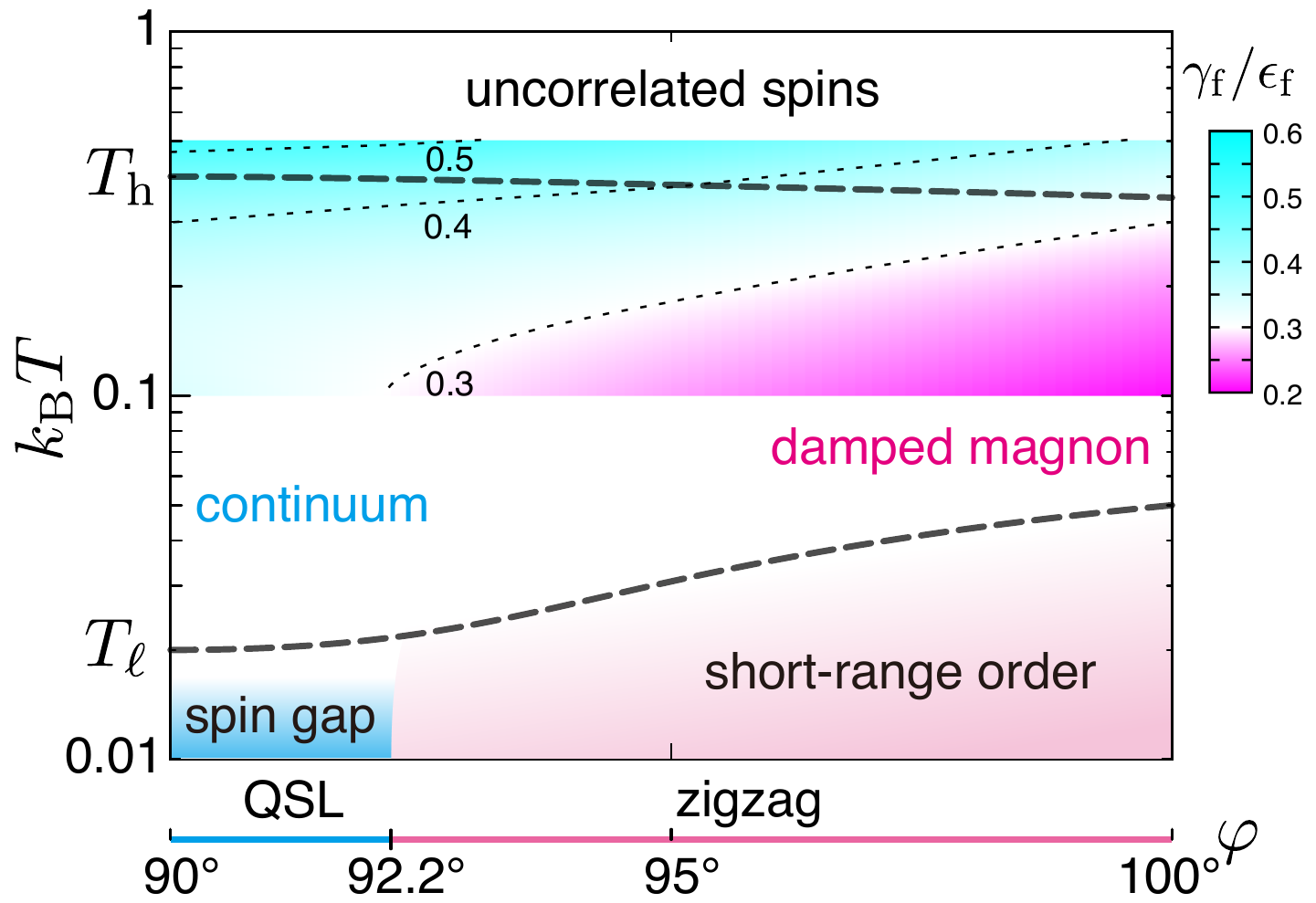}
\caption{(color online):
Summary of temperature dependence of $\gamma_{\rm f}/\epsilon_{\rm f}$
and temperature scales of the Kitaev-Heisenberg model for $90^{\circ}\leq\varphi\leq 100^{\circ}$.
The temperature dependence of $\gamma_{\rm f}/\epsilon_{\rm f}$ for $0.1 \leq k_{\rm B}T \leq 0.5$
is illustrated by
the contour lines (dotted lines) and color plot,
which are obtained by linearly interpolating
the present FTK$\omega$ results shown in Fig.~\ref{FigGammaET}.
The two temperature scales $T_{\rm h}$ and $T_{\rm \ell}$
for the 24 site cluster at which temperature dependence of heat capacity shows peaks
are given in Ref.\onlinecite{PhysRevB.93.174425} at $\varphi=90^{\circ}$ and $100^{\circ}$.
For $90^{\circ}<\varphi<100^{\circ}$, the broken curves are guides for eyes
that interpolate these two temperature scales $T_{\rm h}$ and $T_{\rm \ell}$.
In the shaded region below the low-temperature scale $T_{\rm \ell}$ and for $\varphi\gtrsim 92.2^{\circ}$,
where static spin structure factors become saturated upon cooling,
the long-range zigzag order will develop if there is small but finite magnetic anisotropy or three dimensional couplings. 
}
\label{Figdiagram}
\end{figure}

To examine temperature evolution of the DSF
for finite size clusters,
we compare with $S(\mbox{\boldmath$Q$},\omega)$
of the antiferromagnetic Kitaev model ($\varphi=90^{\circ}$)
obtained by available numerical results 
at the thermodynamic limit.
In Fig.~\ref{FigSQomegaQMC}, the DSFs
at two typical momenta ($\mbox{\boldmath$Q$}=\Gamma, M$)
by FTK$\omega$ for the 24 site cluster
are compared with those at the thermodynamic limit obtained by the Majorana-fermion cluster dynamical mean field theory (cDMFT)
with the maximum entropy method (MaxEnt)
reported in Ref.\onlinecite{PhysRevLett.117.157203} and Ref.\onlinecite{PhysRevB.96.024438}.
Here, the broadening factor $\eta$ is set to $0.02$ for the FTK$\omega$ results.

Even though there is quantitative difference between the spectra of the
finite size cluster and the cDMFT,
the DSF
for the 24 site cluster captures
shifts in spectral weight upon cooling down to $k_{\rm B}T=0.1$, qualitatively.
At the $\Gamma$ point $(\mbox{\boldmath$Q$}=\mbox{\boldmath$0$})$,
the FTK$\omega$ shows
the shift of the spectral weight from low frequency to high frequency ($\hbar\omega\gtrsim 1$),
which is not hampered by the finite-size effect and consistent with the cDMFT.
At the $M$ point, the formation of the low-energy peak and high-energy shoulder upon cooling is obtained by the 24-site simulation,
which is again consistent with the cDMFT.
Although there are detailed peak structures due to the finite-size effect for $k_{\rm B}T\lesssim 0.2$,
the present finite-size simulation presumably captures the temperature evolution of the DSFs.

\subsection{Proximity of Kitaev's spin liquid}
The advantages of the present FTK$\omega$ are its applicability to frustrated systems,
which are hardly tractable by the quantum Monte Carlo methods except special limits,
and its compatibility with the reweighting techniques that enables us to sweep a range of temperatures
with reasonable numerical costs. 
The antiferromagnetic and ferromagnetic Kitaev models ($\varphi=90^{\circ}$ and $270^{\circ}$, respectively)
are the special limits that are tractable by Majorana-fermion
quantum Monte Carlo methods~\cite{PhysRevB.92.115122,PhysRevLett.117.157203}. 
In the section, we demonstrate the advantages and capability of the FTK$\omega$ by simulating the DSFs
of the Kitaev-Heisenberg model and sweeping a range of temperatures
from $k_{\rm B}T=10$ to $0.1$.

Here, we focus on a proximity of the Kitaev's spin liquid.
As clarified in Ref.\onlinecite{PhysRevB.93.174425},
vicinity to the spin liquid is observable as two peak structures in temperature dependence of
heat capacity, even in the magnetically ordered phases.
A concrete example is the Kitaev-Heisenberg model at $\varphi=100^{\circ}$.
For this choice of $\varphi$, temperature dependence of heat capacity
has the two peak structure, which is in a close resemblance to that of the Kitaev model,
although the ground state has been known to show the zigzag order.

Temperature dependence of heat capacity is informative~\cite{PhysRevB.95.144406}.
However, it is not straightforward to extract an electronic contribution
from total heat capacity that may be dominated by the lattice contribution.
Alternative approaches to verifying the emergence or proximity of quantum spin liquids
are highly desirable. 
Spectroscopic measurement is one of the promising approaches.
Especially, the DSFs
have attracted much attention
due to recent inelastic neutron scattering measurements on a Kitaev material $\alpha$-RuCl$_3$~\cite{banerjee2016proximate,banerjee2016neutron,PhysRevLett.118.107203}. 

Before going to the details of the present simulation,
we note that finite size effects are plausibly weak at temperatures above $k_{\rm B}T \sim 0.1$.
As shown in the literature~\cite{PhysRevB.92.115122,PhysRevB.93.174425} on the Kitaev-Heisenberg model,
there are two temperature scales $T_{\rm h}$ and $T_{\rm \ell}$ at which the temperature dependence of
the heat capacity shows local maxima as the function of temperature,
in the proximity of the Kitaev's quantum spin liquid.
In Ref.\onlinecite{PhysRevB.93.174425}, it is clarified that the system size dependence of the heat capacity becomes
negligible for the 24 site or larger clusters
at temperatures well above $T_{\rm \ell}$, where $k_{\rm B}T_{\rm \ell}$ is smaller than $0.1$ for $\varphi \lesssim 100^{\circ}$,
while, at the Kitaev limit, $\varphi=90^{\circ}$, the system size dependence has been shown to be significant around and below the temperature
scale $T_{\rm \ell}$.
Therefore, in the following, we show the DSFs
of the 24 site cluster for $k_{\rm B}T \geq 0.1$
and expect that these results are robust against the finite-size effects.

\subsubsection{Spectral weight evolution at typical momenta}
To capture the proximity of the Kitaev's spin liquid in the Kitaev-Heisenberg model,
we simulate the temperature dependence of the DSFs
of the 24 site cluster.
First, the temperature evolution of the spectra is examined for $\varphi=100^{\circ}$ at typical momenta $\mbox{\boldmath$Q$}=\Gamma$ and $M$.
Then, the finite-temperature spectra at $\varphi=100^{\circ}$ are compared with the spectra at the Kiteav limit ($\varphi =90^{\circ}$).

In Fig.~\ref{FigSQomega3D}, the temperature evolution of $\widetilde{S}_{\beta,\mbox{\boldmath$\delta$}}(\Gamma,\omega)$ and $\widetilde{S}_{\beta,\mbox{\boldmath$\delta$}}(M,\omega)$
for $\varphi=100^{\circ}$ is shown by using the reweighting method.
At the $\Gamma$ point,
the spectral weight shifts from $\hbar\omega\sim 0$ to $\hbar\omega\gtrsim 1$ upon cooling
while the low-energy peak below $\hbar\omega \sim 0.5$ and the high-energy shoulder above $\hbar\omega\sim 1$ develop at the $M$ point at low temperatures below $k_{\rm B}T \sim 0.2$.
These temperature dependences seemingly resemble those of the Kitaev model obtained by the cluster dynamical mean-field theory~\cite{PhysRevLett.117.157203}.
However, as detailed below, there is substantial difference between the spectra at $\varphi=100^{\circ}$ and that at $\varphi=90^{\circ}$. 
Here, the continuous temperature dependence is obtained by the reweighting method.
The filter operators are constructed at $k_{\rm B}T=+\infty$, $1$, $0.5$, and $0.2$. The reweighting method accurately reproduces the spectra for the temperature ranges $1<k_{\rm B}T\leq 10$,
$0.5<k_{\rm B}T\leq 1$, $0.2<k_{\rm B}T\leq 0.5$, and $0.1<k_{\rm B}T\leq 0.2$ by starting with the filtered typical pure states at $k_{\rm B}T=+\infty$, $1$, $0.5$, and $0.2$, respectively.

\subsubsection{Comparison with the Kitaev limit}
To contrast the DSF
for $\varphi=100^{\circ}$
obtained by the FTK$\omega$,
we compare
that 
with $\widetilde{S}_{\beta,\mbox{\boldmath$\delta$}}(\mbox{\boldmath$Q$},\omega)$
for $\varphi=90^{\circ}$ in Fig.~\ref{FigSQomegaBZ}.
The momentum dependence of the equi-energy slices are shown by changing temperature and frequency.
The equi-energy slices are prepared by averaging the spectra within an energy window whose width is 0.1.
The momentum dependence is numerically interpolated for visibility without changing the simulation results
at the discrete momenta $\mbox{\boldmath$Q$}$ compatible with the finite size cluster.  

At $k_{\rm B}T= 0.5$, the DSFs
for $\varphi=90^{\circ}$ and $\varphi=100^{\circ}$
are almost the same, as shown in Figs.~\ref{FigSQomegaBZ}(a) and (b), respectively.
However, below $k_{\rm B}T= 0.2$,
not only the low-energy spectrum at $\mbox{\boldmath$Q$}=M$ but also the high-energy spectrum at $\mbox{\boldmath$Q$}=\Gamma$  
for these two parameters show a stark contrast.
The spectral weight at $\mbox{\boldmath$Q$}=M$ below $\hbar\omega\sim 0.2$ grows significantly for $\varphi=100^{\circ}$.
The growth signals development of the zigzag correlations, which is consistent with the temperature dependence
of the static spin structure factor~\cite{PhysRevB.93.174425} at $\mbox{\boldmath$Q$}=M$\footnote{
We note that the energy unit {\it A} in Ref.\onlinecite{PhysRevB.93.174425} is the half of the present energy unit. The label of the typical momenta is also different: The {\it Y} point in Ref.\onlinecite{PhysRevB.93.174425} is denoted
by {\it M} in the present paper.}.
In addition, the spectral weight at $\mbox{\boldmath$Q$}=\Gamma$ for $1\lesssim\hbar\omega\lesssim 1.5$ grows.

If we recall how small $|J|$ is for $\varphi=100^{\circ}$, 
one may naively wonder why the intensity growth at such high energy region occurs.
Indeed, the onset temperature of the intensity growth is more than twice of $|J|=(1/2)|\cos 100^{\circ}|\sim 0.087$,
and the energy scale $1\lesssim\hbar\omega\lesssim 1.5$ is far beyond the scale of the perturbation $J$. 

\subsubsection{Crossover from continuum to damped magnon mode}
The intensity growth at $\mbox{\boldmath$Q$}=\Gamma$ for $1\lesssim \hbar\omega \lesssim 1.5$
is quantitatively captured by analyzing width of the broad peak in the spectra.
To extract the peak width, we fit the high energy peak by an asymmetric Lorentzian function,
\eqsa{
s(\omega)=
\frac{\gamma_{\rm f}}{\gamma_{\rm f}^2+(\hbar\omega-\epsilon_{\rm f})^2}
\left\{ a + \frac{b}{1+e^{c(\hbar\omega-\epsilon_{\rm f})}}\right\},
\label{asymmetricLorentzian}
}
where $a$, $b$, $c$, $\gamma_{\rm f}$, and $\epsilon_{\rm f}$ are fitting parameters.
For the fitting, we choose a energy window $1 \leq \hbar\omega \leq 2$ to exclude a contribution of low energy tails.
As shown in Fig.~\ref{FigGammaET}(a), the asymmetric Lorentizan function well fit the high energy broad peak for $90^{0}\leq \varphi \leq 100^{\circ}$. 
Then, the dimensionless ratio $\gamma_{\rm f}/\epsilon_{\rm f}$ is a measure of the peak width.
In the standard analysis of the magnon spectrum, the dimensionless measure of the peak width is
given by the ratio of the raw full width at half maximum $\Gamma_{\rm p}$ and the peak energy $E_{\rm p}$, $\Gamma_{\rm p}/E_{\rm p}$.
Here, the full width at half maximum $\Gamma_{\rm p}$ is approximately twice of the imaginary part of the magnon self-energy at $\hbar\omega = E_{\rm p}$.
The present measure $\gamma_{\rm f}/\epsilon_{\rm f}$ is qualitatively similar to $\Gamma_{\rm p}/E_{\rm p}$ although
$\gamma_{\rm f}/\epsilon_{\rm f}$ is always smaller than $\Gamma_{\rm p}/E_{\rm p}$ since
$\epsilon_{\rm f}>E_{\rm p}$ and $\gamma_{\rm f}< \Gamma_{\rm p}$ hold for the fitting function $s(\omega)$.
Thus, the ratio $\gamma_{\rm f}/\epsilon_{\rm f}$ gives a lower bound for $\Gamma_{\rm p}/E_{\rm p}$.

The dimensionless peak width $\gamma_{\rm f}/\epsilon_{\rm f}$ at the Kitaev limit ($\varphi=90^{\circ}$) shows
the temperature dependence distinct from those for $\varphi \geq 92.2^{\circ}$, as shown shown in Fig.~\ref{FigGammaET}(b).
In the quantum spin liquid phase, $\gamma_{\rm f}/\epsilon_{\rm f}$
seems to be always larger than $0.3$.
In contrast, for $\varphi=95^{\circ}$ and $100^{\circ}$,
$\gamma_{\rm f}/\epsilon_{\rm f}$ becomes smaller than $0.3$ around $k_{\rm B}T/\epsilon_{\rm f}\sim 0.1$,
where $\epsilon_{\rm f}\sim 1.4$ for $90^{\circ}\leq \varphi \leq 100^{\circ}$.
The distinct temperature dependence of $\gamma_{\rm f}/\epsilon_{\rm f}$ is coincide with the quantum phase transition
at $\varphi\sim 92.2^{\circ}$ between the Kitaev's quantum spin liquid phase and the magnetically ordered phase.
However, we note that $\gamma_{\rm f}/\epsilon_{\rm f}$ for $\varphi=100^{\circ}$ at $k_{\rm B}T/\epsilon_{\rm f}\sim 0.1$ is
at least twice larger than the observed upper limit of $\Gamma_{\rm p}/E_{\rm p}$
in non-frustrated magnets:
The experimantal and theoretical studies on the magnon peak width of non-frustrated square-lattice antiferromagnets
show $\Gamma_{\rm p}/E_{\rm p} \lesssim 0.1$ at the top of the magnon dispersion for $k_{\rm B}T/E_{\rm p} \lesssim 0.1$~\cite{PhysRevLett.86.5377,Shao2017}.
Thus, the peak width $\gamma_{\rm f}/\epsilon_{\rm f}$ at temperatures around $k_{\rm B}T/\epsilon_{\rm f}\sim 0.1$
is a good measure of
frustration.

The peak narrowing at finite temperatures is associated with
the quantum phase transition from the spin liquid to the zigzag order.
To clarify the relation between the peak narrowing and the quantum phase transition, 
we summarize
the temperature and $\varphi$ dependences of the peak width $\gamma_{\rm f}/\epsilon_{\rm f}$
in a $\varphi$-$T$ phase diagram of the Kitaev-Heisenberg model for $90^{\circ}\leq \varphi \leq 100^{\circ}$, which is shown in Fig.~\ref{Figdiagram}.
The peak width $\gamma_{\rm f}/\epsilon_{\rm f}$ for $k_{\rm B}T \gtrsim 0.1$
reflects the quantum phase transition at zero temperature.
Thus, we attribute the peak narrowing to a finite-temperature crossover
from the spin-excitation continuum at the Kitaev's spin liquid phase ($\varphi=90^{\circ}$)
to the high-energy damped magnon mode that signal the magnetically ordered ground state.

It has already been revealed in the literature that there are two characteristic temperature scales
in the proximity of the Kitaev's spin liquid phase~\cite{PhysRevB.92.115122,PhysRevB.93.174425},
as illustrated in Fig.~\ref{Figdiagram}.
As found in Ref.\onlinecite{PhysRevB.92.115122} for the Kitaev model 
and later for the Kitaev-Heisenberg model~\cite{PhysRevB.93.174425},
there are two temperature scales $T_{\rm h}$ and $T_{\rm \ell}$ at which temperature dependence of heat capacity shows local maxima as the function of temperature.
As clarified for the Kitaev model~\cite{PhysRevB.92.115122},
nearest-neighbor spin-spin correlations develop
upon cooling
around the high-temperature scale $T_{\rm h}$, 
while the spin gap
starts to develop below the low-temperature scale $T_{\rm \ell}$.
In contrast to the Kitaev limit,
in the Kitaev-Heisenberg model, spin-spin correlations start to develop
or long-range magnetic orders appear via an order-by-disorder mechanism~\cite{PhysRevLett.109.187201,PhysRevB.88.024410}
at temperatures below the low-temperature scale $T_{\rm \ell}$.
While the ratio $T_{\rm \ell}/T_{\rm h}$ has been proposed as a measure of distance from the Kitaev's spin liquid phase in Ref.\onlinecite{PhysRevB.93.174425}, 
the peak width $\gamma_{\rm f}/\epsilon_{\rm f}$ at moderately high temperatures far above the low-temperature scale $T_{\rm \ell}$
offers another measure of the closeness to the Kiteav's spin liquid.

We note that 
the classical Kitaev model shows qualitatively similar dynamics
to the quantum counterpart~\cite{PhysRevB.96.134408}.
The semiclassical dynamics of the classical antiferromagnetic Kitaev model
reproduces the high-energy continuum of the quantum Kitaev model at $\varphi=90^{\circ}$
except the difference in the energy scale due to the different spin amplitude,
although development of the spin gap in the quantum Kitaev model
signals the breakdown of the similarity between semiclassical and finite-temperature quantum dynamics~\cite{PhysRevB.96.134408}.
When the finite Heisenberg exchange coupling $J/|K| = -0.1$ is introduced in the classical model,
a crossover from the a high-energy continuum to a high-energy damped magnon mode
is found at $\mbox{\boldmath$Q$}=\Gamma$ upon decreasing temperature~\cite{PhysRevB.96.134408}
across the transition temperature of the order by disorder~\cite{PhysRevLett.109.187201,PhysRevB.88.024410},
which seems to be consistent with the present results for the quantum counterpart.
However, here, we note that there is a significant difference between the semiclassical dynamics
and quantum dynamics of the Kitaev-Heisenberg model if we associate the transition temperature of the order by disorder
in the classical model
with the temperature scale $T_{\rm \ell}$ in the quantum counterpart:
Although the semiclassical dynamics of the Kitaev-Heisenberg model seems to show the continuum as broad as that in the zero-temperature Kitaev limit
above the transition temperature (and below $k_{\rm B}T \sim 1$),
the quantum dynamics shows the high-energy excitation peak at $\mbox{\boldmath$Q$}=\Gamma$ narrower than
that in the Kitaev limit far above the low-temperature scale $T_{\rm \ell}$,
as shown in Fig.~\ref{Figdiagram}.
Detailed comparison between the temperature dependences of the classical and quantum dynamics are left for future studies.

The crossover from the spin-excitation continuum to damped magnon mode at high energy
is plausibly ubiquitous in the proximity of the Kitaev's quantum spin liquid phase.
The intensity growth and line shape narrowing
in high-energy spin excitation spectra
are expected to be independent of specific choice of perturbation that drives the Kitaev's
quantum spin liquid to magnetically ordered states. 
Although the threshold value of $\gamma_{\rm f}/\epsilon_{\rm f}$, which separates the Kitaev's quantum spin liquid
and a frustrated magnet with the magnetically ordered ground state, 
depends on momenta and the Hamiltonian,
the temperature dependence of $\gamma_{\rm f}/\epsilon_{\rm f}$ may offer a common measure of the distance to the Kitaev's
quantum spin liquid.

\section{Summary and discussion}
\label{secVI}
In the present paper, we have proposed an $\mathcal{O}(N_{\rm F})$ algorithm for simulating finite-temperature spectra,
called FTK$\omega$,
by combining the typical pure state approach and the shifted Krylov subspace method.
The present algorithm has advantages over the previous approaches~\cite{PhysRevLett.90.047203,
PhysRevLett.112.120601,doi:10.7566/JPSJ.83.094001,PhysRevB.90.155104,PhysRevB.92.205103,
PhysRevB.68.235106,PhysRevLett.92.067202}.

The present algorithm
enables us to obtain spectra directly in the frequency domain
without the aid of real-time evolution of typical pure states employed in the previous studies~\cite{PhysRevLett.90.047203,
PhysRevLett.112.120601,doi:10.7566/JPSJ.83.094001,PhysRevB.90.155104,PhysRevB.92.205103}.
Probability distribution obtained by utilizing the shifted Krylov subspace method,
which is essential to the present algorithm, makes possible a use of the reweighting method to finely tune temperature.
The reweighting method significantly reduces computational costs to study temperature dependence of the spectra.
From the probability distribution at the set of the discrete temperatures,
the FTK$\omega$ interpolates potentially exact spectra at temperatures
between the two adjacent discrete temperatures with negligible costs.

The present FTK$\omega$ is implemented by using a complete orthonormal basis set of the Fock space in this paper.
As a next step, implementation by a compressed basis set is highly desirable to simulate much larger systems. 
The typical pure state approaches for static observables have already been implemented by variational wave functions~\cite{doi:10.7566/JPSJ.85.034601}.
There have also been several studies on the Krylov subspace method by using various variational basis sets such as tensor-network states~\cite{huang2016generalized}.
The variational bases that are compatible with volume law entanglement will realize the compressed-basis FTK$\omega$.

The capability of the FTK$\omega$ is demonstrated by simulating finite-temperature dynamical spin structure factors
of the Kitaev-Heisenberg model.
We have found that, even though the absolute value of the ratio of the Heisenberg exchange coupling $J$
and the Kitaev couplings, $|J/K|$, is small for $\varphi=100^{\circ}$,
temperature dependence of the dynamical spin structure factor shows substantial deviation from that of the Kitaev model
not only in the low-energy spectrum at the $M$ point, which signals the onset of the zigzag correlation,
but also in the high-energy spectrum at the $\Gamma$ point, even at the temperatures twice larger than $|J|$.
The perturbative approaches fail in explaining the deviation.
The present exact temperature dependence of dynamical spin structure factors
set constraint on approximations,
even though the present results are limited for the finite size clusters.
At least for $k_{\rm B}T\gtrsim 0.1$, the present finite-size simulation
essentially captures the temperature evolution of the spectral weight
that is consistent with that at the thermodynamic limit obtained by the cluster dynamical mean-field theory~\cite{PhysRevLett.117.157203}.

The finite-temperature dynamical spin structure factors shed new light on emergent temperature scales in the proximity of the Kitaev's quantum spin liquid. 
As found
in Ref.\onlinecite{PhysRevB.92.115122} 
for the Kitaev model 
and later for the Kitaev-Heisenberg model~\cite{PhysRevB.93.174425}
in the proximity of the Kitaev quantum spin liquid,
there are two temperature scales $T_{\rm h}$ and $T_{\rm \ell}$ at which temperature dependence of heat capacity shows peak structures.
In the parameter range $90^{\circ} \leq \varphi \leq 100^{\circ}$, there are two peak structures in heat capacity of the 24 site cluster as illustrated in Fig.~\ref{Figdiagram}.
As clarified for the Kitaev model~\cite{PhysRevB.92.115122}, around $T=T_{\rm h}$, the nearest-neighbor spin-spin correlations develop
while the spin gap starts to develop below $T=T_{\rm \ell}$.
In contrast to the Kitaev limit, static structure factors for $\varphi = 100^{\circ}$ grow at the $M$ point around
$T=T_{\rm \ell}\sim 0.05$.
If there is small but finite magnetic anisotropy or a three dimensional coupling, 
or if the order-by-disorder mechanism found in the classical Kitaev-Heisenberg model~\cite{PhysRevLett.109.187201,PhysRevB.88.024410} is
relevant to the quantum model,
spontaneous time-reversal symmetry breakings will occur
below the low-temperature scale.
In Fig.~\ref{Figdiagram},
the parameter region
where we expect the spontaneous symmetry breaking is illustrated as shaded region below $T\sim T_{\ell}$ for $\varphi\gtrsim 92.2^{\circ}$ in the $\varphi$-$T$ phase diagram.
From the measurements on heat capacity and static magnetic orders, these two temperature scales seem to characterize the magnetism in the proximity of the Kitaev's quantum spin liquid.
For example, the ratio $T_{\rm \ell}/T_{\rm h}$ has been proposed as a measure of distance from the Kitaev's spin liquid phase~\cite{PhysRevB.93.174425}.

The present results on the dynamical spin structure factors reveal
that
there is the crossover from the spin-excitation continuum at the Kitaev limit to the damped
magnon modes at high energy.
The dynamical spin structure factors $S(\mbox{\boldmath$Q$},\omega)$ show significant deviations from those at the Kitaev limit,
even at high energy $(1\lesssim \omega \lesssim 1.5)$. 
For $\varphi\gtrsim 92.2^{\circ}$, $S(\mbox{\boldmath$Q$}=\Gamma,\omega)$ deviates from that for $\varphi=90^{\circ}$ below $T\sim T_{\rm h}$,
which is another precursor of the magnetically ordered ground state, in addition to development of the low-energy spectral weight
at the $M$ point due to short-range magnetic correlations.
Although a finite-size cluster is employed in the present simulation,
the high-energy spectra well above the low-temperature scale $T_{\rm \ell}$ are reliable since
the finite-size effects become negligible for $N\geq 24$ above the temperature scale $T_{\rm \ell}$~\cite{PhysRevB.93.174425}.

In addition to thermodynamic measurements such as heat capacity~\cite{PhysRevB.93.174425},
the spectroscopic measurements are found to be useful to measure the distance between a given Kitaev material and the Kitaev limit.  
The present flexible algorithm is also applicable to other linear responses such as thermal conductivity~\cite{PhysRevLett.92.067202,nasu2017thermal},
which will contribute to an understanding of the proximity of not only the Kitaev's spin liquid but also other spin liquid candidates~\cite{PhysRevLett.118.147204,arXiv:1701.07837,arXiv:1706.09908}.

\acknowledgments
Y. Y. gratefully thanks Takeo Hoshi and Tomohiro Sogabe
for continued collaboration on the shifted Krylov subspace method, which leads him to the present study. 
Y. Y. also thanks Naoki Kawashima, Seiji Miyashita, and Hans De Raedt for their enlightening discussions
and letting him know important references of the typical state approaches.
Y. Y. further thanks
Masatoshi Imada for carefully reading the manuscript and helpful comments on it,
Karen Hallberg for the enlightening discussion on numerical approaches for simulating excitation spectra of correlated electron systems,
Synge Todo for his useful comments on statistical treatment in the typical state approaches,
Takahiro Misawa for discussions about Ref.\onlinecite{Takahashi_Umezawa},
and thank 
Tsuyoshi Okubo for bringing his attention to the reweighting method.
We thank Junki Yoshitake
and Yukitoshi Motome 
for providing us their numerical data on finite-temperature dynamical structure factors
of the Kitaev model.
In addition, Y. Y. thanks Junki Yoshitake for discussions on the maximum entropy method.
Yukitoshi Motome for his comment on the temperature scales
of the Kitaev-Heisenberg model.
Y. Y. was supported by JSPS KAKENHI
(Grant Nos. 15K17702 and 16H06345) and was supported
by PRESTO, JST (JPMJPR15NF).
This research was supportd by MEXT as
``Priority Issue on Post-K computer" (Creation of New Functional Devices and High-Performance Materials
to Support Next-Generation Industries) and
``Exploratory Challenge on Post-K computer" (Frontiers of Basic Science: Challengin the Limits).
T. S. was supported by by JSPS KAKENHI
(Grant No. 16K17751).
M. K. acknowledges support by Building of Consortia for the Development of Human Resources in Science and Technology from the MEXT of Japan.
Our numerical calculation was partly carried out at the Supercomputer
Center, Institute for Solid State Physics, University of Tokyo.
The exact diagonalization (ED) calculations are partly double-checked by using
an open-source ED program package $\mathcal{H}\Phi$~\cite{HPhi,Kawamura2017180}. 
The excitation spectra are calculated by employing a numerical library $K\omega$ for 
shifted Krylov subspace methods~\cite{Komega}.

\appendix
\section{Imaginary-time evolution}
\label{appendix_imaginary}
In the present paper, we calculate the typical pure state (or canonical thermal pure quantum state~\cite{PhysRevLett.111.010401})
at inverse temperature $\beta$ by following Ref.\onlinecite{PhysRevLett.111.010401} as
\eqsa{
e^{N\beta \ell/2}\ket{\psi_{\beta}}&=&e^{N\beta (\ell - \hat{h})/2}\ket{\psi_0}
\nonumber\\
&=&\sum_{k=0}^{+\infty} \frac{(N\beta/2)^k}{k!}(\ell-\hat{h})^k \ket{\psi_0},
\label{B1}
}
where $\hat{h}=\hat{H}/N$ is used.
The above formula is not suitable for the numerical simulation,
since terms in the rightmost hand side of Eq.(\ref{B1}) become too large for $\beta\gg 1$
and will introduce cancellation of significant digits.

To avoid the cancellation of significant digits, we split the imaginary time evolution
and divide it into $M_{\rm d}$ steps.
The step size $\beta/M_{\rm d}$ is determined by the following estimation.
First, we estimate the amplitude of the largest term in the rightmost hand side of Eq.(\ref{B1}).
The $k$ th order term is bounded by
\eqsa{
\left\| \frac{(N\beta/2)^k}{k!}(\ell-\hat{h})^k \ket{\psi_0} \right\|
&\leq& \frac{(N\beta/2)^k (\ell+|\epsilon_0|)^k}{k!}
\nonumber\\
&\sim& \frac{(N\beta/2)^k (\ell+|\epsilon_0|)^k}{\sqrt{2\pi k}(k/e)^k}.
\label{lhk}
}
Then, by differentiating the term in the rightmost side of Eq.(\ref{lhk}) with respect to $k$,
we find that, when $k=X-1/2+\mathcal{O}(1/X)$, where $X=(N\beta/2M_{\rm d})(\ell+|\epsilon_0|)$, the $k$ th term
becomes maximum among the series expansion.
For $k=X-1/2+\mathcal{O}(1/X)$, we obtain the asymptotic formula for the extremum as
\eqsa{
\underset{k}{{\rm max}}\left\{\frac{(N\beta/2M_{\rm d})^k (\ell+|\epsilon_0|)^k}{\sqrt{2\pi k}(k/e)^k}\right\}
\sim
\frac{\exp \left(X\right)}{\sqrt{2\pi  X}}.
}
When we set an upper limit $\Lambda$ for $e^{X}/(2\pi X)^{-1/2}$, we can determine an appropriate $M_{\rm d}$
through iteratively solving
\eqsa{
e^{X}/(2\pi X)^{-1/2} = \Lambda.
}
By using
the following recurrence relation for $k\geq 0$ initialized with $X_0=\Lambda$,
\eqsa{
X_{k+1}= \ln \Lambda + \frac{1}{2}\ln (2\pi X_{k}),
}
we obtain $M_{\rm d}$ as
\eqsa{
M_{\rm d}=
\frac{N\beta/2}{\displaystyle \lim_{k\rightarrow + \infty}X_k}(\ell+|\epsilon_0|).
}
\section{Upper bounds of variance}
\label{appendix_bound}
The source of the deviation between $\mathcal{G}_{\beta}^{AB}(\zeta)$ and
$\widetilde{\mathcal{G}}_{\beta,\mbox{\boldmath$\delta$}}^{AB}(\zeta)$ is twofold:
The discretization parameters $\mbox{\boldmath$\delta$}=(E_{\rm b},\epsilon,M)$ and variance of $\{c_n \}$. 
The former source can be examined by changing the set of the discretization parameters $\mbox{\boldmath$\delta$}$.
Therefore, we here focus on the deviation originating from the variance of stochastic variables $\{c_n \}$
and give the upper bounds of the variance of the present $\mathcal{O}(N_{\rm F})$ algorithm 
by following
Refs.\onlinecite{PhysRevE.62.4365} and \onlinecite{PhysRevLett.111.010401}. 

We start with rewriting Eq.(\ref{GAB_TPS}) as
\eqsa{
\widetilde{\mathcal{G}}_{\beta,\mbox{\boldmath$\delta$}}^{AB}(\zeta)
=\frac{\displaystyle\sum_m \sum_{E,E'\in\Delta_m}
e^{-\beta\mathcal{E}_m}c_{E}^{\ast}c_{E'}
\bra{E}
\hat{O}_m (\zeta)
\ket{E'}}
{\displaystyle\sum_n |c_n|^2 e^{-\beta E_n}},
\nonumber\\
}
where we define a equi-energy shell 
as $\Delta_m =[\mathcal{E}_m-\epsilon, \mathcal{E}_m +\epsilon]$
and an operator as
\eqsa{
\hat{O}_m (\zeta)=\hat{A}^{\dagger}(\zeta-\hat{H}+\mathcal{E}_m)^{-1}\hat{B}.
}
For later usage, we introduce the following shorthand for expectation values as 
\eqsa{
\overline{f}_m(\zeta)=e^{-\beta \mathcal{E}_m}\sum_{E\in\Delta_m}\bra{E}\hat{O}_m (\zeta)\ket{E},
}
and
\eqsa{
\overline{g}=Z(\beta),
}
and for stochastic variables as
\eqsa{
&&\overline{f}_m(\zeta)+\delta f_m (\zeta)
\nonumber\\
&&= e^{-\beta \mathcal{E}_m} N_{\rm F} 
\sum_{E,E'\in \Delta_m} c_{E}^{\ast}c_{E'} \bra{E}\hat{O}_m (\zeta)\ket{E'},
}
and
\eqsa{
\overline{g}+\delta g&=& N_{\rm F} \sum_n |c_n|^2 e^{-\beta E_n},
}
where $\widetilde{\mathcal{G}}_{\beta,\mbox{\boldmath$\delta$}}^{AB}(\zeta)=\sum_m [\overline{f}_m(\zeta)+\delta f_m (\zeta)]
/ [\overline{g}+\delta g ] $.

Then, the variance of $\widetilde{\mathcal{G}}_{\beta,\mbox{\boldmath$\delta$}}^{AB}(\zeta)$ is given by
\eqsa{
\sigma^2 (\zeta)&=&\mathbb{E}\left[\left|\sum_m \frac{\overline{f}_m(\zeta)+\delta f_m (\zeta)}{\overline{g}+\delta g}
-\frac{\overline{f}_m(\zeta)}{\overline{g}}\right|^2\right]
\nonumber\\
&\simeq&
\sum_{m,m'}
\frac{1}{\overline{g}^2}\mathbb{E}[\delta f_m (\zeta)^{\dagger} \delta f_{m'} (\zeta)]
\nonumber\\
&&+
\sum_{m,m'}
\frac{\overline{f}_m (\zeta)^{\dagger} \overline{f}_{m'} (\zeta)}{\overline{g}^4}\mathbb{E}[\delta g^2]
\nonumber\\
&&-
2{\rm Re}
\sum_{m,m'}
\frac{\overline{f}_m (\zeta)^{\dagger}}{\overline{g}^3} \mathbb{E}[\delta f_{m'}(\zeta) \delta g].
}
Below, we evaluate $\sigma^2 (\zeta)$ term by term:
By using the following formulae~\cite{ULLAH196465},
\eqsa{
\mathbb{E}[|c_n|^2]&=&1/N_{\rm F},\\
\mathbb{E}[|c_k|^2 |c_{\ell}|^2]&=&1/N_{\rm F}/(N_{\rm F}+1),\\
\mathbb{E}[|c_n|^4]&=&2/N_{\rm F}/(N_{\rm F}+1),
}
we obtain the following expectation values, 
\eqsa{
&&\mathbb{E}[\delta f_m (\zeta)^{\dagger} \delta f_{m'} (\zeta)]
=
\frac{e^{-\beta(\mathcal{E}_m+\mathcal{E}_{m'})}}
{N_{\rm F}+1}
\nonumber\\
&\times&\left[
\delta_{m,m'}
N_{\rm F}\sum_{E,E'\in\Delta_m}\bra{E}\hat{O}_m (\zeta)^{\dagger}\ket{E'}\bra{E'}\hat{O}_m (\zeta)\ket{E}
\right.
\nonumber\\
&&
-
\left(\sum_{E\in\Delta_m}\bra{E}\hat{O}_m (\zeta)^{\dagger}\ket{E}\right)
\nonumber\\
&&\times\left.
\left(\sum_{E'\in\Delta_{m'}}\bra{E'}\hat{O}_{m'}(\zeta)\ket{E'}\right)
\right],
}

\eqsa{
\mathbb{E}[\delta g^2]=\frac{N_{\rm F}}{N_{\rm F}+1}\left[Z(2\beta)-\frac{Z(\beta)^2}{N_{\rm F}}\right],
}
and
\eqsa{
&&\mathbb{E}[\delta f_m (\zeta)^{\dagger} \delta g]
\nonumber\\
&&=e^{-\beta\mathcal{E}_m}
\frac{N_{\rm F}e^{-\beta\mathcal{E}_m}-Z(\beta)}{N_{\rm F}+1}
\sum_{E\in\Delta_m}
\bra{E}\hat{O}_m(\zeta)^{\dagger}\ket{E}.
\nonumber\\
}
After straightforward calculations, we reach the following expression,
\eqsa{
\sigma^2 (\zeta)
&\simeq&
\frac{N_{\rm F}}{N_{\rm F}+1}\frac{Z(2\beta)}{Z(\beta)^2}
\left[
A_{2\beta}(\zeta)
\right.
\nonumber\\
&+&
\left.
|\mathcal{G}_{\beta}^{AB}(\zeta)|^2
-2{\rm Re}\{\mathcal{G}^{AB}_{\beta}(\zeta)^{\dagger}\mathcal{G}^{AB}_{2\beta}(\zeta)\}
\right],\label{sigma2v1}
}
where we introduce the following shorthand, 
\eqsa{
A_{\beta}(\zeta)=
\sum_m \frac{e^{-\beta \mathcal{E}_m }}{Z(\beta)}
\sum_{E,E'\in\Delta_m}
\bra{E}\hat{O}_m (\zeta)^{\dagger}\ket{E'}
\bra{E'}\hat{O}_m (\zeta)\ket{E}.
\nonumber\\
}

Generally, estimate of $A_{\beta}(\zeta)$ in Eq.(\ref{sigma2v1}) is not tractable.
Below, we give an upper bound of $A_{\beta}(\zeta)$.
First, we use the following inequality:
There is a positive constant ${\rm min}\{\eta,\epsilon\}<\overline{\eta}<\mathcal{O}(NJ_0)$ that satisfies
the inequality
\eqsa{
A_{\beta}(\hbar\omega+i\eta)< \frac{\hbar}{\overline{\eta}}\int_{-\infty}^{+\infty}d\omega'\ A_{\beta}(\hbar\omega'+i\eta).
}
Then, we estimate upper bounds of the following integral as
\eqsa{
&&\int_{-\infty}^{+\infty} \hbar d\omega
\sum_{E,E'\in\Delta_m}
\bra{E}\hat{O}_m (\hbar\omega+i\eta)^{\dagger}\ket{E'}
\bra{E'}\hat{O}_m (\hbar\omega+i\eta) \ket{E}
\nonumber\\
&&=
4\pi\hbar\eta
\sum_{E,E'\in\Delta_m}
\sum_{k,\ell} \frac{
\bra{E}\hat{B}^{\dagger}\ket{k}\bra{k}\hat{A}\ket{E'}
\bra{E'}\hat{A}^{\dagger}\ket{\ell}\bra{\ell}\hat{B}\ket{E}
}
{(E_k-E_{\ell})^2+4\eta^2}
\nonumber\\
&&=
\pi \hbar
\sum_{E,E'\in\Delta_m}
\int_{-\infty}^{+\infty}
dt e^{-2\eta |t|/\hbar}
\bra{E}\hat{B}^{\dagger}\hat{A}(t)\ket{E'}
\bra{E'}\hat{A}(t)^{\dagger}\hat{B}\ket{E}
\nonumber\\
&&<
\pi \hbar
\sum_{E\in\Delta_m}
\int_{-\infty}^{+\infty}
dt e^{-2\eta |t|/\hbar}
\bra{E}\hat{B}^{\dagger}\hat{A}(t)
\hat{A}(t)^{\dagger}\hat{B}\ket{E},\label{int_BAAB}
}
where we use the following transformation,
\eqsa{
&&\bra{E}\hat{B}^{\dagger}\hat{A}(t)\ket{E'}
\nonumber\\
&&=
\bra{E}\hat{B}^{\dagger}e^{+i\hat{H}t/\hbar}\hat{A}e^{-i\hat{H}t/\hbar}\ket{E'}
\nonumber\\
&&
=\sum_{k}
e^{-i\mathcal{E}_m t/\hbar +iE_k t/\hbar}\bra{E}\hat{B}^{\dagger}\ket{k}\bra{k}\hat{A}\ket{E'}.
}
The integral in the last line of Eq.(\ref{int_BAAB}) is bounded:
There is a positive constant $\tau$ that satisfies
\eqsa{
&&\int_{-\infty}^{+\infty}dt
e^{-2\eta |t|/\hbar}
\bra{E}\hat{B}^{\dagger}\hat{A}(t)
\hat{A}(t)^{\dagger}\hat{B}\ket{E}
\nonumber\\
&&<
\bra{E}\hat{B}^{\dagger}\hat{A}
\hat{A}^{\dagger}\hat{B}\ket{E} 
\int_{-\infty}^{+\infty}dt
e^{-(2\eta/\hbar + 2/\tau)|t|}
\nonumber\\
&&=
\frac{\bra{E}\hat{B}^{\dagger}\hat{A}
\hat{A}^{\dagger}\hat{B}\ket{E} 
}{\eta/\hbar+/\tau}.
}
The positive constant $\tau$ simply corresponds to a correlation time
that characterizes the correlation function $\bra{E}\hat{B}^{\dagger}\hat{A}(t)
\hat{A}(t)^{\dagger}\hat{B}\ket{E}$.

Finally, we obtain an upper bounds for $\sigma^2 (\omega+i\eta)$ as
\eqsa{
&&\sigma^2 (\hbar\omega+i\eta)
\nonumber\\
&&\lesssim\frac{Z(2\beta)}{Z(\beta)^2}
\Biggl[
\frac{\pi}{\overline{\eta}}\sum_m \frac{e^{-2\beta\mathcal{E}_m}}{Z(2\beta)}
\frac{\bra{E}\hat{B}^{\dagger}\hat{A}
\hat{A}^{\dagger}\hat{B}\ket{E} 
}{\eta+\hbar/\tau}
\Biggr.
\nonumber\\
&&
\Biggl.
+|\mathcal{G}^{AB}_{\beta}(\hbar\omega+i\eta)|^2
-2{\rm Re}\{\mathcal{G}^{AB}_{\beta}(\hbar\omega+i\eta)^{\dagger}\mathcal{G}^{AB}_{2\beta}(\hbar\omega+i\eta)\}
\Biggr]
\nonumber\\
&&
\lesssim 
\frac{Z(2\beta)}{Z(\beta)^2}
\Biggl[
\frac{\pi}{\overline{\eta}}\frac{
\langle\hat{B}^{\dagger}\hat{A}
\hat{A}^{\dagger}\hat{B}\rangle_{2\beta}^{\rm ens}}{\eta + \hbar/\tau}
\Biggr.
\nonumber\\
&&
\Biggl.
+
|\mathcal{G}^{AB}_{\beta}(\hbar\omega+i\eta)
-
\mathcal{G}^{AB}_{2\beta}(\hbar\omega+i\eta)|^2
\Biggr].
}
The factor $Z(2\beta)/Z(\beta)^2$ is known to be exponentially small,
when the system size $N$ grows~\cite{PhysRevLett.111.010401}.
The temperature dependence of $Z(2\beta)/Z(\beta)^2$ is trivially grasped as follows.
First, we take the simple limits $\beta\rightarrow +0$ and $\beta\rightarrow +\infty$ as
\eqsa{
\lim_{\beta\rightarrow+0}\frac{Z(2\beta)}{Z(\beta)^2}=\frac{1}{N_{\rm F}}
}
and
\eqsa{
\lim_{\beta\rightarrow+\infty}\frac{Z(2\beta)}{Z(\beta)^2}=\frac{1}{D},
}
where $D$ is the degeneracy of the ground state.
By taking temperature derivative of $Z(2\beta)/Z(\beta)^2$ as,
\eqsa{
\frac{\partial }{\partial T}\left[\frac{Z(2\beta)}{Z(\beta)^2}\right]
=
2\beta^2
\frac{Z(2\beta)}{Z(\beta)^2}
\left[
\avrg{E}_{2\beta}^{\rm ens}
-
\avrg{E}_{\beta}^{\rm ens}
\right],
}
and remembering that $\avrg{E}_{\beta}^{\rm ens}$ is a monotonically decreasing function of temperature, 
we can prove the following relation, 
\eqsa{
\frac{1}{N_{\rm F}}\leq \frac{Z(2\beta)}{Z(\beta)^2} \leq \frac{1}{D}.
}
If we use standard relations for free energy $F(\beta)$ and entropy $S(\beta)$,
$F(\beta)=-\beta^{-1}\ln Z(\beta)$ and $dF(\beta)/dT = -S (\beta)$,
we obtain the following relation:
There is an inverse temperature that satisfies $\beta \leq \beta^{\ast} \leq 2\beta$
and
\eqsa{
\frac{Z(2\beta)}{Z(\beta)^2}=e^{-2\beta [F(2\beta)-F(\beta)]}=e^{-S(\beta^{\ast})}.
}
At finite temperature, $S(\beta)$ is finite and extensive.
Since the entropy is proportional to $N$, the factor $Z(2\beta)/Z(\beta)^2$ is exponentially small.

\section{Convergence of shifted BiCG method and dependence on discretization}
\label{appendix_conv}
\begin{table}[thb]
\begin{ruledtabular}
\begin{tabular}{l|cccc}
12 site & $E_{\rm b}$ & $\epsilon$ & $L$ & $M$ \\
\hline
$k_{\rm B}T=1$ &    $-3$ & $3.75\times 10^{-3}$ & 800 & 1024 \\
$k_{\rm B}T=0.1$ &  $-3$ & $5\times 10^{-3}$   & 400 & 512 \\
$k_{\rm B}T=0.01$ & $-3$ & $5\times 10^{-3}$   & 400 & 512 \\
\hline
\hline
18 site & $E_{\rm b}$ & $\epsilon$ & $L$ & $M$ \\
\hline
$k_{\rm B}T=1$ &    $-4$ & $3.90625\times 10^{-3}$ & 1024 & 64 \\
$k_{\rm B}T=0.1$ &  $-4$ & $1.953125\times 10^{-3}$ & 1024 & 64 \\
$k_{\rm B}T=0.01$ & $-4$ & $1.953125\times 10^{-3}$  & 1024 & 64 \\
\hline
\hline
24 site & $E_{\rm b}$ & $\epsilon$ & $L$ & $M$ \\
\hline
$k_{\rm B}T\rightarrow+\infty$ &    $-5$ & $3.90625\times 10^{-2}$ & 128 & 16 \\
$k_{\rm B}T=1$ &  $-5$ & $3.90625\times 10^{-2}$ & 128 & 16 \\
$k_{\rm B}T=0.5$ &  $-5$ & $3.90625\times 10^{-2}$ & 128 & 16 \\
$k_{\rm B}T=0.2$ &  $-5$ & $(14\ln 10) \times 0.2 / 128 /2$ & 128 & 16 \\
$k_{\rm B}T=0.1$ &  $-5$ & $1.5625\times 10^{-2}$ & 128 & 16 \\
\end{tabular}
\end{ruledtabular}
\caption{List of discretization parameters $\mbox{\boldmath$\delta$}$ for the filter operator defined in Eq.(\ref{dfilterP})
used in Sec.~\ref{secV}.
To cover the whole energy eigenvalues of $\hat{H}$, we set
the lower bound of energy as $E_{\rm b}$ and the upper bound of energy as $E_{\rm b}+2\epsilon L$.
Here, $\epsilon$ is the radius of the contours that define the filter operators and $L$ is the number of the filter operators.
We take a Riemann sum along the contour with $M$ discrete points.}
\label{table_delta}
\end{table}

The shifted BiCG method~\cite{frommer2003bicgstab} is employed
in the present FTK$\omega$ algorithm to implement
the multiplication of $(\zeta-\hat{H})^{-1}$ and the filter operator.
The convergence of the CG methods is
verified by the 2-norm of the residual vectors.
By setting the upper bound on the 2-norm
\eqsa{
\||\rho_n (\zeta)\rangle\|_2=\sqrt{\langle \rho_n (\zeta)|\rho_n (\zeta)\rangle},
\nonumber
}
we can truncate the CG steps in a controlled fashion.
In the shifted Krylov subspace method that handles a set of shifts or complex numbers $\{\zeta\}$,
we need to choose a residual vector for the truncation.
A choice that guarantees the quality of the convergence is the residual vector with the largest 2-norm,
which is realized by the seed switching method~\cite{doi:10.1143/JPSJ.77.114713,Komega}.

In this section, we examine the CG-step dependence of the
2-norm in the present application.
We choose examples from the calculations for the 18 site and 24 site clusters of the antiferromagnetic Kitaev model ($\varphi=90^{\circ}$)
at $k_{\rm B}T=0.1$.
The examples are chosen from the construction of the filtered typical states
defined in Eq.(\ref{filtered})
since the construction is the most time-consuming part of the present FTK$\omega$ algorithm.

The shifted linear equation $\ket{\chi(\zeta)}=(\zeta - \hat{H})\ket{\phi_{\beta}}$ is solved for the set of shifts $\{\zeta\}$
to construct the $L$ filtered typical states.
The shifted BiCG method is applied to each filter operator separately in the present implementation
and the contour integral in every single filter operator is approximated by the Riemann sum with the $M$ discrete points.

\begin{figure}[thb]
\centering
\includegraphics[width=8.5cm]{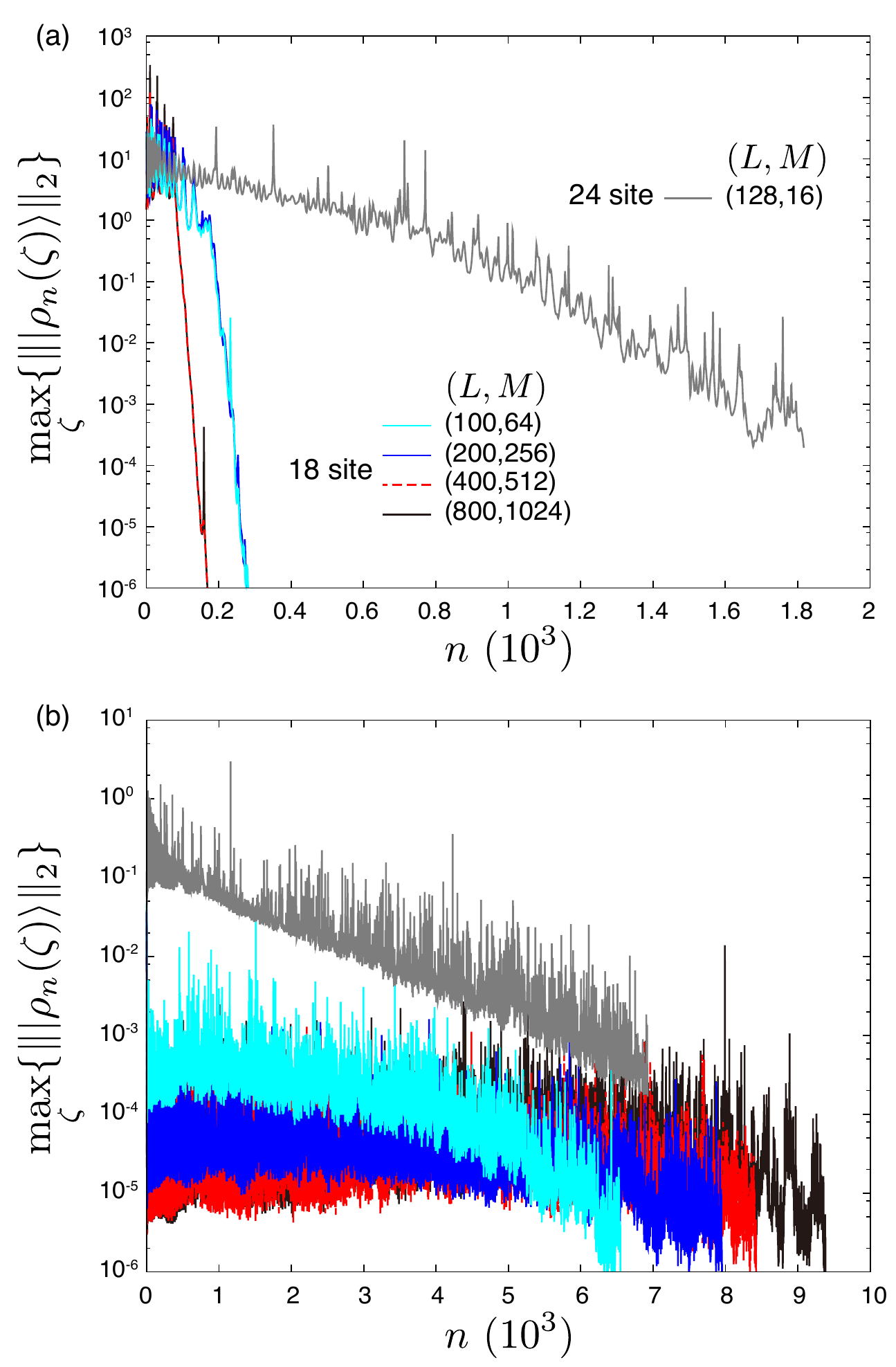}
\caption{(color online):
CG-step dependence of the maximum 2-norm of the residual vectors
for the 18 site and 24 site clusters.
(a) The CG-step dependence is shown for $\mathcal{E}_m$ at which the probability distribution
$\widetilde{\mathcal{P}}_{\beta,\mbox{\boldmath$\delta$}}$ becomes maximum.
(b) The CG-step dependence for $\mathcal{E}_m$ that requires the largest number of the CG steps.
}
\label{Fig_SKloop}
\end{figure}

In Fig.~\ref{Fig_SKloop}, we show typical examples of the CG-step dependence of the maximum 2-norm
in the constructions of the filter operators.
For each filter operator (each $m\in[0,L)$), at each CG step, the maximum 2-norm
$\max_{\zeta} \{\| |\rho_n (\zeta)\rangle\|_2\}$ is chosen
from the set of the $M$ discrete points $\zeta \in \{{\epsilon e^{i\theta_j}+\mathcal{E}_m}\}_{j}$ along the contour illustrated in Fig.~\ref{fig_contour}.
The upper bound of the 2-norm is set to $10^{-4}$ or smaller for the 18 site and 24 site clusters,
which guarantees convergence of expectation values taken by the filtered typical pure states.
For the 18 site cluster, we also show how the CG-step dependence of the 2-norm depends on the discretization $\mbox{\boldmath$\delta$}=(E_{\rm b},\epsilon,M)$.

The CG-step dependence is shown for $\mathcal{E}_m$ at which the probability distribution
$\widetilde{\mathcal{P}}_{\beta,\mbox{\boldmath$\delta$}}$ becomes maximum
in Fig.~\ref{Fig_SKloop}(a).
When we choose $\mathcal{E}_m$ that requires the largest number of the CG steps,
the speed of the convergence also depends on $\min_{\zeta}\{|{\rm Im}\zeta|\}$ for $\zeta \in \{{\epsilon e^{i\theta_j}+\mathcal{E}_m}\}_{j}$.
As shown in Fig.~\ref{Fig_SKloop}(b), larger $L$ and $M$, which decrease $\min_{\zeta}\{|{\rm Im}\zeta|\}$, may require more CG steps.
When both $L$ and $M$ are doubled, which give four times smaller $\min_{\zeta}\{|{\rm Im}\zeta|\}$, the number of the CG steps required for the convergence
increases by around 10 percent.

Here, we note that the number of the required CG steps depends on
the density of states at $\mathcal{E}_m$ while we only show the typical examples in Fig.~\ref{Fig_SKloop}.
This dependence is inferred from
the convergence theorem of the Lanczos method~\cite{saad2011numerical}.
The number of the Lanczos steps required to obtain an eigenstate and eigenvalue becomes larger
as the density of states at the target eigenvalue becomes larger.
At $k_{\rm B}T=0.1$, as shown in Fig.~\ref{FigProbability},
the peak of the probability distribution is located nearby the lower edge of the eigenvalue distribution.
Both the 18 site and 24 site clusters show faster convergences for $\mathcal{E}_m$ close to the edge of the probability distribution
compared with the other choice of $\mathcal{E}_m$.
The sparse density of states nearby the lowest eigenvalue naturally explains the faster convergence.

The system size dependence of the convergence is of practical importance.
As evident in Fig.~\ref{Fig_SKloop}(b), the system size affects the CG-step dependence of the 2-norm.
However, when we lower the upper bound of the 2-norm to ensure the exponential decay of the CG-step dependence of the 2-norm,
we observe that ten thousand CG steps are practically enough to obtain the convergence for any $\mathcal{E}_m$, irrespective of the system size.
Only nearby the edge of the probability distribution, the exponential decay is sensitive to the discretization and the system size.

Then, we examine how the discretization $\mbox{\boldmath$\delta$}$ affects the spectra for the fixed broadening factor $\eta=0.02$.
As formulated in Eq.(\ref{limGAB}), the FTK$\omega$ exactly reproduces the finite-temperature spectra by the canonical ensemble average
after taking the average over the initial random vectors and the two limits, $M\rightarrow +\infty$ and $\epsilon\rightarrow +0$.
The large $M$ limit should be taken before the small $\epsilon$ or large $L$ limit.
Here, we note that the interval of the discrete energy grid, $\epsilon$, is set to be comparable to or smaller than the broadening factor $\eta$.

\begin{figure}[thb]
\centering
\includegraphics[width=8.5cm]{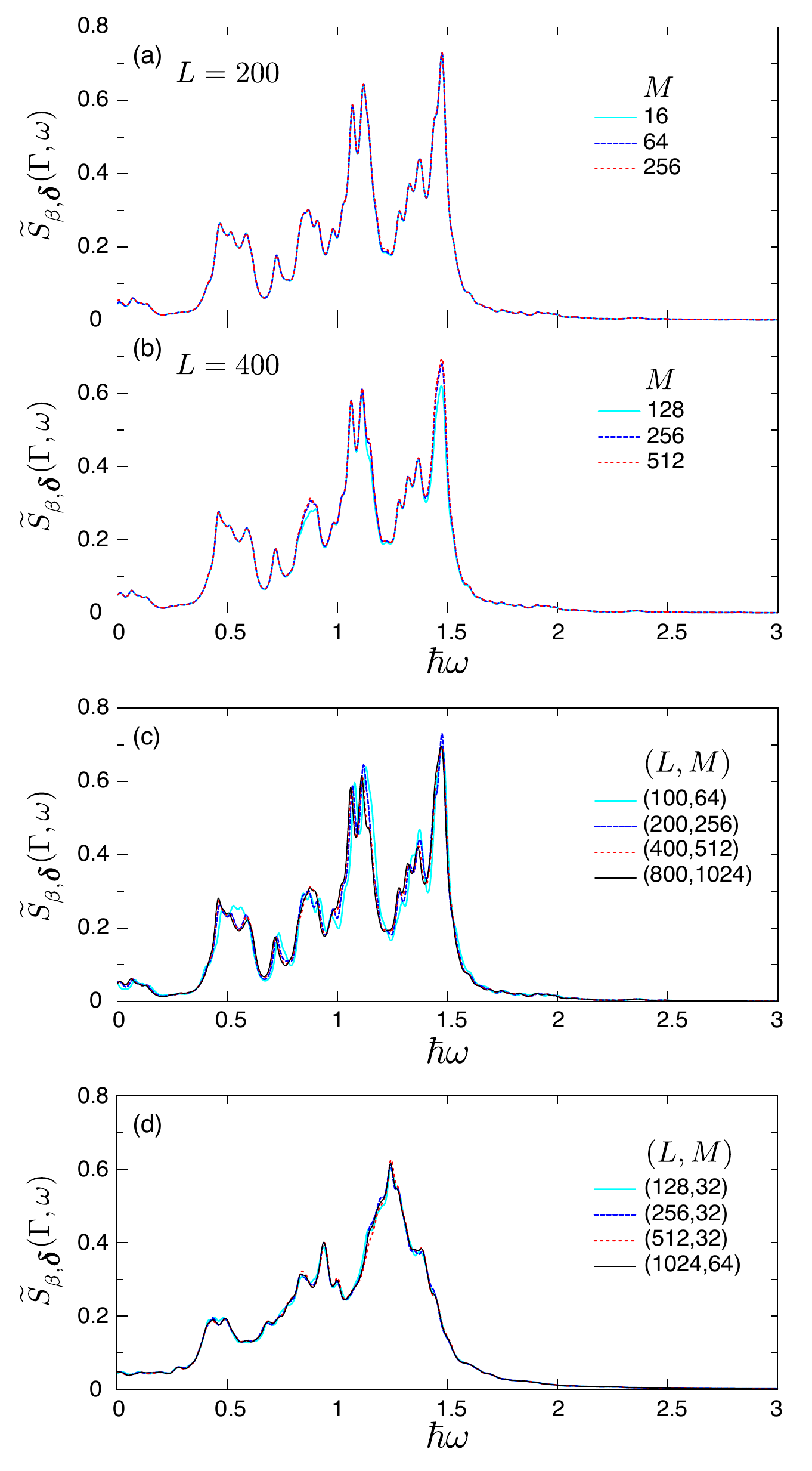}
\caption{(color online):
Discretization dependence of the dynamical structure factor at $\mbox{\boldmath$Q$}=\Gamma$.
(a) The convergence of the structure factor for the 12 site cluster
is examined by changing $M$ form 16 to 256 with $L=200$.
(b) The convergence for the 12 site cluster is examined with fixed $L=400$ when $M$ is increased from 128 to 512.
(c) The convergence for the 12 site cluster is also examined when both $L$ and $M$ are increased.
(d) For the 18 site cluster, the convergence is examined by increasing $L$ and $M$ simultaneously.
}
\label{Fig_convSQomega}
\end{figure}

To find a reasonable choice of $\mbox{\boldmath$\delta$}$,
we examine the $\mbox{\boldmath$\delta$}$ dependence of the dynamical spin structure factor
for the 12 site and 18 site clusters of the antiferromagnetic Kiteav model ($\varphi=90^{\circ}$) at the $\Gamma$ point.
First, we examine the convergence when $M$ is increased.
For $L=200$ and $400$, the $M$ dependence of the spectrum is examined in Figs.~\ref{Fig_convSQomega}(a) and (b)
when $\epsilon L$ is fixed.
The spectrum calculated with larger $L$ requires larger $M$ to converge.
Second, we examine the $L$ dependence of the spectrum with an appropriate $M$ and keeping $\epsilon L$ constant.
As shown in Fig.~\ref{Fig_convSQomega}(c),
we obtain a converged result for the 12 site cluster for $\varphi=90^{\circ}$ by increasing $L$ and choosing an appropriate $M$. 
The system size also affects the convergence.
When the system size is increased from $N=12$ to $18$, the $L$ dependence becomes smaller as shown in Fig.~\ref{Fig_convSQomega}(d).

The discretization parameters used in Sec.~\ref{secV} are summarized in Table \ref{table_delta}.
Here, we choose the discretization parameters to obtain converged results for the 12 site and 18 site clusters.
For the 24 site cluster, to take a balance of accuracy and a numerical cost, a practical parameter set is chosen
based on the $L$ and $M$ dependence of the dynamical spin structure factors for $N=12$ and $18$.
As shown in Fig.~\ref{Fig_convSQomega}(d), the $(L,M)$ dependence is already small for $N=18$, at least, for $L\gtrsim 100$.
Thus, we choose $L=128$ and confirm that $M=16$ is enough to obtain reasonable results.


\bibliography{Kitaev}

\begin{thebibliography}{98}%
\makeatletter
\providecommand \@ifxundefined [1]{%
 \@ifx{#1\undefined}
}%
\providecommand \@ifnum [1]{%
 \ifnum #1\expandafter \@firstoftwo
 \else \expandafter \@secondoftwo
 \fi
}%
\providecommand \@ifx [1]{%
 \ifx #1\expandafter \@firstoftwo
 \else \expandafter \@secondoftwo
 \fi
}%
\providecommand \natexlab [1]{#1}%
\providecommand \enquote  [1]{``#1''}%
\providecommand \bibnamefont  [1]{#1}%
\providecommand \bibfnamefont [1]{#1}%
\providecommand \citenamefont [1]{#1}%
\providecommand \href@noop [0]{\@secondoftwo}%
\providecommand \href [0]{\begingroup \@sanitize@url \@href}%
\providecommand \@href[1]{\@@startlink{#1}\@@href}%
\providecommand \@@href[1]{\endgroup#1\@@endlink}%
\providecommand \@sanitize@url [0]{\catcode `\\12\catcode `\$12\catcode
  `\&12\catcode `\#12\catcode `\^12\catcode `\_12\catcode `\%12\relax}%
\providecommand \@@startlink[1]{}%
\providecommand \@@endlink[0]{}%
\providecommand \url  [0]{\begingroup\@sanitize@url \@url }%
\providecommand \@url [1]{\endgroup\@href {#1}{\urlprefix }}%
\providecommand \urlprefix  [0]{URL }%
\providecommand \Eprint [0]{\href }%
\providecommand \doibase [0]{http://dx.doi.org/}%
\providecommand \selectlanguage [0]{\@gobble}%
\providecommand \bibinfo  [0]{\@secondoftwo}%
\providecommand \bibfield  [0]{\@secondoftwo}%
\providecommand \translation [1]{[#1]}%
\providecommand \BibitemOpen [0]{}%
\providecommand \bibitemStop [0]{}%
\providecommand \bibitemNoStop [0]{.\EOS\space}%
\providecommand \EOS [0]{\spacefactor3000\relax}%
\providecommand \BibitemShut  [1]{\csname bibitem#1\endcsname}%
\let\auto@bib@innerbib\@empty
\bibitem [{\citenamefont {Kohn}(1999)}]{RevModPhys.71.1253}%
  \BibitemOpen
  \bibfield  {author} {\bibinfo {author} {\bibfnamefont {W.}~\bibnamefont
  {Kohn}},\ }\bibfield  {title} {\enquote {\bibinfo {title} {Nobel lecture:
  Electronic structure of matter--{W}ave functions and density functionals},}\
  }\href@noop {} {\bibfield  {journal} {\bibinfo  {journal} {Rev. Mod. Phys.}\
  }\textbf {\bibinfo {volume} {71}},\ \bibinfo {pages} {1253--1266} (\bibinfo
  {year} {1999})}\BibitemShut {NoStop}%
\bibitem [{\citenamefont {Imada}\ and\ \citenamefont
  {Takahashi}(1986)}]{Imada_Takahashi}%
  \BibitemOpen
  \bibfield  {author} {\bibinfo {author} {\bibfnamefont {M.}~\bibnamefont
  {Imada}}\ and\ \bibinfo {author} {\bibfnamefont {M.}~\bibnamefont
  {Takahashi}},\ }\bibfield  {title} {\enquote {\bibinfo {title} {Quantum
  transfer {M}onte {C}arlo method for finite temperature properties and quantum
  molecular dynamics method for dynamical correlation functions},}\ }\href@noop
  {} {\bibfield  {journal} {\bibinfo  {journal} {J. Phys. Soc. Jpn.}\ }\textbf
  {\bibinfo {volume} {55}},\ \bibinfo {pages} {3354} (\bibinfo {year}
  {1986})}\BibitemShut {NoStop}%
\bibitem [{\citenamefont {Skilling}(2013)}]{skilling2013maximum}%
  \BibitemOpen
  \bibfield  {author} {\bibinfo {author} {\bibfnamefont {John}\ \bibnamefont
  {Skilling}},\ }\enquote {\bibinfo {title} {Maximum entropy and bayesian
  methods: Cambridge, england, 1988},}\ \ (\bibinfo  {publisher} {Springer
  Science \& Business Media},\ \bibinfo {year} {2013})\ p.\ \bibinfo {pages}
  {455}\BibitemShut {NoStop}%
\bibitem [{\citenamefont {de~Vries}\ and\ \citenamefont
  {De~Raedt}(1993)}]{PhysRevB.47.7929}%
  \BibitemOpen
  \bibfield  {author} {\bibinfo {author} {\bibfnamefont {Pedro}\ \bibnamefont
  {de~Vries}}\ and\ \bibinfo {author} {\bibfnamefont {Hans}\ \bibnamefont
  {De~Raedt}},\ }\bibfield  {title} {\enquote {\bibinfo {title} {Solution of
  the time-dependent {S}chr\"odinger equation for two-dimensional spin-1/2
  {H}eisenberg systems},}\ }\href@noop {} {\bibfield  {journal} {\bibinfo
  {journal} {Phys. Rev. B}\ }\textbf {\bibinfo {volume} {47}},\ \bibinfo
  {pages} {7929--7937} (\bibinfo {year} {1993})}\BibitemShut {NoStop}%
\bibitem [{\citenamefont {Jakli\ifmmode~\check{c}\else \v{c}\fi{}}\ and\
  \citenamefont {Prelov\ifmmode~\check{s}\else
  \v{s}\fi{}ek}(1994)}]{PhysRevB.49.5065}%
  \BibitemOpen
  \bibfield  {author} {\bibinfo {author} {\bibfnamefont {J.}~\bibnamefont
  {Jakli\ifmmode~\check{c}\else \v{c}\fi{}}}\ and\ \bibinfo {author}
  {\bibfnamefont {P.}~\bibnamefont {Prelov\ifmmode~\check{s}\else
  \v{s}\fi{}ek}},\ }\bibfield  {title} {\enquote {\bibinfo {title} {Lanczos
  method for the calculation of finite-temperature quantities in correlated
  systems},}\ }\href {\doibase 10.1103/PhysRevB.49.5065} {\bibfield  {journal}
  {\bibinfo  {journal} {Phys. Rev. B}\ }\textbf {\bibinfo {volume} {49}},\
  \bibinfo {pages} {5065--5068} (\bibinfo {year} {1994})}\BibitemShut {NoStop}%
\bibitem [{\citenamefont {Hams}\ and\ \citenamefont
  {De~Raedt}(2000)}]{PhysRevE.62.4365}%
  \BibitemOpen
  \bibfield  {author} {\bibinfo {author} {\bibfnamefont {Anthony}\ \bibnamefont
  {Hams}}\ and\ \bibinfo {author} {\bibfnamefont {Hans}\ \bibnamefont
  {De~Raedt}},\ }\bibfield  {title} {\enquote {\bibinfo {title} {Fast algorithm
  for finding the eigenvalue distribution of very large matrices},}\
  }\href@noop {} {\bibfield  {journal} {\bibinfo  {journal} {Phys. Rev. E}\
  }\textbf {\bibinfo {volume} {62}},\ \bibinfo {pages} {4365--4377} (\bibinfo
  {year} {2000})}\BibitemShut {NoStop}%
\bibitem [{Note1()}]{Note1}%
  \BibitemOpen
  \bibinfo {note} {The studies may recall the maximum-entropy approach~\cite
  {mead1984maximum,skilling2013maximum} and the forced oscillator method~\cite
  {PhysRevB.31.4508,PhysRevB.36.8933} as predecessors of them.}\BibitemShut
  {Stop}%
\bibitem [{\citenamefont {Tasaki}(1998)}]{PhysRevLett.80.1373}%
  \BibitemOpen
  \bibfield  {author} {\bibinfo {author} {\bibfnamefont {Hal}\ \bibnamefont
  {Tasaki}},\ }\bibfield  {title} {\enquote {\bibinfo {title} {From quantum
  dynamics to the canonical distribution: General picture and a rigorous
  example},}\ }\href@noop {} {\bibfield  {journal} {\bibinfo  {journal} {Phys.
  Rev. Lett.}\ }\textbf {\bibinfo {volume} {80}},\ \bibinfo {pages}
  {1373--1376} (\bibinfo {year} {1998})}\BibitemShut {NoStop}%
\bibitem [{\citenamefont {Popescu}\ \emph {et~al.}(2006)\citenamefont
  {Popescu}, \citenamefont {Short},\ and\ \citenamefont
  {Winter}}]{popescu2006entanglement}%
  \BibitemOpen
  \bibfield  {author} {\bibinfo {author} {\bibfnamefont {Sandu}\ \bibnamefont
  {Popescu}}, \bibinfo {author} {\bibfnamefont {Anthony~J}\ \bibnamefont
  {Short}}, \ and\ \bibinfo {author} {\bibfnamefont {Andreas}\ \bibnamefont
  {Winter}},\ }\bibfield  {title} {\enquote {\bibinfo {title} {Entanglement and
  the foundations of statistical mechanics},}\ }\href@noop {} {\bibfield
  {journal} {\bibinfo  {journal} {Nature Physics}\ }\textbf {\bibinfo {volume}
  {2}},\ \bibinfo {pages} {754--758} (\bibinfo {year} {2006})}\BibitemShut
  {NoStop}%
\bibitem [{\citenamefont {Goldstein}\ \emph {et~al.}(2006)\citenamefont
  {Goldstein}, \citenamefont {Lebowitz}, \citenamefont {Tumulka},\ and\
  \citenamefont {Zangh\`{\i}}}]{PhysRevLett.96.050403}%
  \BibitemOpen
  \bibfield  {author} {\bibinfo {author} {\bibfnamefont {Sheldon}\ \bibnamefont
  {Goldstein}}, \bibinfo {author} {\bibfnamefont {Joel~L.}\ \bibnamefont
  {Lebowitz}}, \bibinfo {author} {\bibfnamefont {Roderich}\ \bibnamefont
  {Tumulka}}, \ and\ \bibinfo {author} {\bibfnamefont {Nino}\ \bibnamefont
  {Zangh\`{\i}}},\ }\bibfield  {title} {\enquote {\bibinfo {title} {Canonical
  typicality},}\ }\href@noop {} {\bibfield  {journal} {\bibinfo  {journal}
  {Phys. Rev. Lett.}\ }\textbf {\bibinfo {volume} {96}},\ \bibinfo {pages}
  {050403} (\bibinfo {year} {2006})}\BibitemShut {NoStop}%
\bibitem [{\citenamefont {Sugita}(2007)}]{sugita2007basis}%
  \BibitemOpen
  \bibfield  {author} {\bibinfo {author} {\bibfnamefont {A}~\bibnamefont
  {Sugita}},\ }\bibfield  {title} {\enquote {\bibinfo {title} {On the basis of
  quantum statistical mechanics.}}\ }\href@noop {} {\bibfield  {journal}
  {\bibinfo  {journal} {Nonl. Phen. Compl. Sys.}\ }\textbf {\bibinfo {volume}
  {10}},\ \bibinfo {pages} {192--195} (\bibinfo {year} {2007})}\BibitemShut
  {NoStop}%
\bibitem [{\citenamefont {Reimann}(2007)}]{PhysRevLett.99.160404}%
  \BibitemOpen
  \bibfield  {author} {\bibinfo {author} {\bibfnamefont {Peter}\ \bibnamefont
  {Reimann}},\ }\bibfield  {title} {\enquote {\bibinfo {title} {Typicality for
  generalized microcanonical ensembles},}\ }\href@noop {} {\bibfield  {journal}
  {\bibinfo  {journal} {Phys. Rev. Lett.}\ }\textbf {\bibinfo {volume} {99}},\
  \bibinfo {pages} {160404} (\bibinfo {year} {2007})}\BibitemShut {NoStop}%
\bibitem [{\citenamefont {Sugiura}\ and\ \citenamefont
  {Shimizu}(2012)}]{PhysRevLett.108.240401}%
  \BibitemOpen
  \bibfield  {author} {\bibinfo {author} {\bibfnamefont {Sho}\ \bibnamefont
  {Sugiura}}\ and\ \bibinfo {author} {\bibfnamefont {Akira}\ \bibnamefont
  {Shimizu}},\ }\bibfield  {title} {\enquote {\bibinfo {title} {Thermal pure
  quantum states at finite temperature},}\ }\href {\doibase
  10.1103/PhysRevLett.108.240401} {\bibfield  {journal} {\bibinfo  {journal}
  {Phys. Rev. Lett.}\ }\textbf {\bibinfo {volume} {108}},\ \bibinfo {pages}
  {240401} (\bibinfo {year} {2012})}\BibitemShut {NoStop}%
\bibitem [{\citenamefont {Sugiura}\ and\ \citenamefont
  {Shimizu}(2013)}]{PhysRevLett.111.010401}%
  \BibitemOpen
  \bibfield  {author} {\bibinfo {author} {\bibfnamefont {Sho}\ \bibnamefont
  {Sugiura}}\ and\ \bibinfo {author} {\bibfnamefont {Akira}\ \bibnamefont
  {Shimizu}},\ }\bibfield  {title} {\enquote {\bibinfo {title} {Canonical
  thermal pure quantum state},}\ }\href@noop {} {\bibfield  {journal} {\bibinfo
   {journal} {Phys. Rev. Lett.}\ }\textbf {\bibinfo {volume} {111}},\ \bibinfo
  {pages} {010401} (\bibinfo {year} {2013})}\BibitemShut {NoStop}%
\bibitem [{\citenamefont {Takahashi}\ and\ \citenamefont
  {Umezawa}(1975)}]{Takahashi_Umezawa}%
  \BibitemOpen
  \bibfield  {author} {\bibinfo {author} {\bibfnamefont {Y.}~\bibnamefont
  {Takahashi}}\ and\ \bibinfo {author} {\bibfnamefont {H.}~\bibnamefont
  {Umezawa}},\ }\bibfield  {title} {\enquote {\bibinfo {title} {Thermo field
  dynamics},}\ }\href@noop {} {\bibfield  {journal} {\bibinfo  {journal}
  {Collect. Phenom.}\ }\textbf {\bibinfo {volume} {2}},\ \bibinfo {pages} {55}
  (\bibinfo {year} {1975})}\BibitemShut {NoStop}%
\bibitem [{\citenamefont {Iitaka}\ and\ \citenamefont
  {Ebisuzaki}(2003)}]{PhysRevLett.90.047203}%
  \BibitemOpen
  \bibfield  {author} {\bibinfo {author} {\bibfnamefont {Toshiaki}\
  \bibnamefont {Iitaka}}\ and\ \bibinfo {author} {\bibfnamefont {Toshikazu}\
  \bibnamefont {Ebisuzaki}},\ }\bibfield  {title} {\enquote {\bibinfo {title}
  {Algorithm for linear response functions at finite temperatures: Application
  to {ESR} spectrum of ${S}=\frac{1}{2}$ antiferromagnet {C}u benzoate},}\
  }\href@noop {} {\bibfield  {journal} {\bibinfo  {journal} {Phys. Rev. Lett.}\
  }\textbf {\bibinfo {volume} {90}},\ \bibinfo {pages} {047203} (\bibinfo
  {year} {2003})}\BibitemShut {NoStop}%
\bibitem [{\citenamefont {Bartsch}\ and\ \citenamefont
  {Gemmer}(2009)}]{PhysRevLett.102.110403}%
  \BibitemOpen
  \bibfield  {author} {\bibinfo {author} {\bibfnamefont {Christian}\
  \bibnamefont {Bartsch}}\ and\ \bibinfo {author} {\bibfnamefont {Jochen}\
  \bibnamefont {Gemmer}},\ }\bibfield  {title} {\enquote {\bibinfo {title}
  {Dynamical typicality of quantum expectation values},}\ }\href@noop {}
  {\bibfield  {journal} {\bibinfo  {journal} {Phys. Rev. Lett.}\ }\textbf
  {\bibinfo {volume} {102}},\ \bibinfo {pages} {110403} (\bibinfo {year}
  {2009})}\BibitemShut {NoStop}%
\bibitem [{\citenamefont {Elsayed}\ and\ \citenamefont
  {Fine}(2013)}]{PhysRevLett.110.070404}%
  \BibitemOpen
  \bibfield  {author} {\bibinfo {author} {\bibfnamefont {Tarek~A.}\
  \bibnamefont {Elsayed}}\ and\ \bibinfo {author} {\bibfnamefont {Boris~V.}\
  \bibnamefont {Fine}},\ }\bibfield  {title} {\enquote {\bibinfo {title}
  {Regression relation for pure quantum states and its implications for
  efficient computing},}\ }\href@noop {} {\bibfield  {journal} {\bibinfo
  {journal} {Phys. Rev. Lett.}\ }\textbf {\bibinfo {volume} {110}},\ \bibinfo
  {pages} {070404} (\bibinfo {year} {2013})}\BibitemShut {NoStop}%
\bibitem [{\citenamefont {Steinigeweg}\ \emph
  {et~al.}(2014{\natexlab{a}})\citenamefont {Steinigeweg}, \citenamefont
  {Gemmer},\ and\ \citenamefont {Brenig}}]{PhysRevLett.112.120601}%
  \BibitemOpen
  \bibfield  {author} {\bibinfo {author} {\bibfnamefont {Robin}\ \bibnamefont
  {Steinigeweg}}, \bibinfo {author} {\bibfnamefont {Jochen}\ \bibnamefont
  {Gemmer}}, \ and\ \bibinfo {author} {\bibfnamefont {Wolfram}\ \bibnamefont
  {Brenig}},\ }\bibfield  {title} {\enquote {\bibinfo {title} {Spin-current
  autocorrelations from single pure-state propagation},}\ }\href@noop {}
  {\bibfield  {journal} {\bibinfo  {journal} {Phys. Rev. Lett.}\ }\textbf
  {\bibinfo {volume} {112}},\ \bibinfo {pages} {120601} (\bibinfo {year}
  {2014}{\natexlab{a}})}\BibitemShut {NoStop}%
\bibitem [{\citenamefont {Steinigeweg}\ \emph
  {et~al.}(2014{\natexlab{b}})\citenamefont {Steinigeweg}, \citenamefont
  {Khodja}, \citenamefont {Niemeyer}, \citenamefont {Gogolin},\ and\
  \citenamefont {Gemmer}}]{PhysRevLett.112.130403}%
  \BibitemOpen
  \bibfield  {author} {\bibinfo {author} {\bibfnamefont {R.}~\bibnamefont
  {Steinigeweg}}, \bibinfo {author} {\bibfnamefont {A.}~\bibnamefont {Khodja}},
  \bibinfo {author} {\bibfnamefont {H.}~\bibnamefont {Niemeyer}}, \bibinfo
  {author} {\bibfnamefont {C.}~\bibnamefont {Gogolin}}, \ and\ \bibinfo
  {author} {\bibfnamefont {J.}~\bibnamefont {Gemmer}},\ }\bibfield  {title}
  {\enquote {\bibinfo {title} {Pushing the limits of the eigenstate
  thermalization hypothesis towards mesoscopic quantum systems},}\ }\href@noop
  {} {\bibfield  {journal} {\bibinfo  {journal} {Phys. Rev. Lett.}\ }\textbf
  {\bibinfo {volume} {112}},\ \bibinfo {pages} {130403} (\bibinfo {year}
  {2014}{\natexlab{b}})}\BibitemShut {NoStop}%
\bibitem [{\citenamefont {Monnai}\ and\ \citenamefont
  {Sugita}(2014)}]{doi:10.7566/JPSJ.83.094001}%
  \BibitemOpen
  \bibfield  {author} {\bibinfo {author} {\bibfnamefont {Takaaki}\ \bibnamefont
  {Monnai}}\ and\ \bibinfo {author} {\bibfnamefont {Ayumu}\ \bibnamefont
  {Sugita}},\ }\bibfield  {title} {\enquote {\bibinfo {title} {Typical pure
  states and nonequilibrium processes in quantum many-body systems},}\
  }\href@noop {} {\bibfield  {journal} {\bibinfo  {journal} {J. Phys. Soc.
  Jpn.}\ }\textbf {\bibinfo {volume} {83}},\ \bibinfo {pages} {094001}
  (\bibinfo {year} {2014})}\BibitemShut {NoStop}%
\bibitem [{\citenamefont {Karrasch}\ \emph {et~al.}(2014)\citenamefont
  {Karrasch}, \citenamefont {Kennes},\ and\ \citenamefont
  {Moore}}]{PhysRevB.90.155104}%
  \BibitemOpen
  \bibfield  {author} {\bibinfo {author} {\bibfnamefont {C.}~\bibnamefont
  {Karrasch}}, \bibinfo {author} {\bibfnamefont {D.~M.}\ \bibnamefont
  {Kennes}}, \ and\ \bibinfo {author} {\bibfnamefont {J.~E.}\ \bibnamefont
  {Moore}},\ }\bibfield  {title} {\enquote {\bibinfo {title} {Transport
  properties of the one-dimensional {H}ubbard model at finite temperature},}\
  }\href@noop {} {\bibfield  {journal} {\bibinfo  {journal} {Phys. Rev. B}\
  }\textbf {\bibinfo {volume} {90}},\ \bibinfo {pages} {155104} (\bibinfo
  {year} {2014})}\BibitemShut {NoStop}%
\bibitem [{\citenamefont {Jin}\ \emph {et~al.}(2015)\citenamefont {Jin},
  \citenamefont {Steinigeweg}, \citenamefont {Heidrich-Meisner}, \citenamefont
  {Michielsen},\ and\ \citenamefont {De~Raedt}}]{PhysRevB.92.205103}%
  \BibitemOpen
  \bibfield  {author} {\bibinfo {author} {\bibfnamefont {F.}~\bibnamefont
  {Jin}}, \bibinfo {author} {\bibfnamefont {R.}~\bibnamefont {Steinigeweg}},
  \bibinfo {author} {\bibfnamefont {F.}~\bibnamefont {Heidrich-Meisner}},
  \bibinfo {author} {\bibfnamefont {K.}~\bibnamefont {Michielsen}}, \ and\
  \bibinfo {author} {\bibfnamefont {H.}~\bibnamefont {De~Raedt}},\ }\bibfield
  {title} {\enquote {\bibinfo {title} {Finite-temperature charge transport in
  the one-dimensional {H}ubbard model},}\ }\href@noop {} {\bibfield  {journal}
  {\bibinfo  {journal} {Phys. Rev. B}\ }\textbf {\bibinfo {volume} {92}},\
  \bibinfo {pages} {205103} (\bibinfo {year} {2015})}\BibitemShut {NoStop}%
\bibitem [{\citenamefont {Long}\ \emph {et~al.}(2003)\citenamefont {Long},
  \citenamefont {Prelov\ifmmode~\check{s}\else \v{s}\fi{}ek}, \citenamefont
  {El~Shawish}, \citenamefont {Karadamoglou},\ and\ \citenamefont
  {Zotos}}]{PhysRevB.68.235106}%
  \BibitemOpen
  \bibfield  {author} {\bibinfo {author} {\bibfnamefont {M.~W.}\ \bibnamefont
  {Long}}, \bibinfo {author} {\bibfnamefont {P.}~\bibnamefont
  {Prelov\ifmmode~\check{s}\else \v{s}\fi{}ek}}, \bibinfo {author}
  {\bibfnamefont {S.}~\bibnamefont {El~Shawish}}, \bibinfo {author}
  {\bibfnamefont {J.}~\bibnamefont {Karadamoglou}}, \ and\ \bibinfo {author}
  {\bibfnamefont {X.}~\bibnamefont {Zotos}},\ }\bibfield  {title} {\enquote
  {\bibinfo {title} {Finite-temperature dynamical correlations using the
  microcanonical ensemble and the {L}anczos algorithm},}\ }\href@noop {}
  {\bibfield  {journal} {\bibinfo  {journal} {Phys. Rev. B}\ }\textbf {\bibinfo
  {volume} {68}},\ \bibinfo {pages} {235106} (\bibinfo {year}
  {2003})}\BibitemShut {NoStop}%
\bibitem [{\citenamefont {Zotos}(2004)}]{PhysRevLett.92.067202}%
  \BibitemOpen
  \bibfield  {author} {\bibinfo {author} {\bibfnamefont {X.}~\bibnamefont
  {Zotos}},\ }\bibfield  {title} {\enquote {\bibinfo {title} {High temperature
  thermal conductivity of two-leg spin-$1/2$ ladders},}\ }\href@noop {}
  {\bibfield  {journal} {\bibinfo  {journal} {Phys. Rev. Lett.}\ }\textbf
  {\bibinfo {volume} {92}},\ \bibinfo {pages} {067202} (\bibinfo {year}
  {2004})}\BibitemShut {NoStop}%
\bibitem [{\citenamefont {Kubo}\ and\ \citenamefont
  {Tomita}(1954)}]{doi:10.1143/JPSJ.9.888}%
  \BibitemOpen
  \bibfield  {author} {\bibinfo {author} {\bibfnamefont {Ryogo}\ \bibnamefont
  {Kubo}}\ and\ \bibinfo {author} {\bibfnamefont {Kazuhisa}\ \bibnamefont
  {Tomita}},\ }\bibfield  {title} {\enquote {\bibinfo {title} {A general theory
  of magnetic resonance absorption},}\ }\href@noop {} {\bibfield  {journal}
  {\bibinfo  {journal} {J. Phys. Soc. Jpn.}\ }\textbf {\bibinfo {volume} {9}},\
  \bibinfo {pages} {888--919} (\bibinfo {year} {1954})}\BibitemShut {NoStop}%
\bibitem [{\citenamefont {Mori}\ and\ \citenamefont
  {Kawasaki}(1962)}]{doi:10.1143/PTP.27.529}%
  \BibitemOpen
  \bibfield  {author} {\bibinfo {author} {\bibfnamefont {Hazime}\ \bibnamefont
  {Mori}}\ and\ \bibinfo {author} {\bibfnamefont {Kyozi}\ \bibnamefont
  {Kawasaki}},\ }\bibfield  {title} {\enquote {\bibinfo {title} {Theory of
  dynamical behaviors of ferromagnetic spins},}\ }\href@noop {} {\bibfield
  {journal} {\bibinfo  {journal} {Progress of Theoretical Physics}\ }\textbf
  {\bibinfo {volume} {27}},\ \bibinfo {pages} {529} (\bibinfo {year}
  {1962})}\BibitemShut {NoStop}%
\bibitem [{\citenamefont {Dietz}\ \emph {et~al.}(1971)\citenamefont {Dietz},
  \citenamefont {Merritt}, \citenamefont {Dingle}, \citenamefont {Hone},
  \citenamefont {Silbernagel},\ and\ \citenamefont
  {Richards}}]{PhysRevLett.26.1186}%
  \BibitemOpen
  \bibfield  {author} {\bibinfo {author} {\bibfnamefont {R.~E.}\ \bibnamefont
  {Dietz}}, \bibinfo {author} {\bibfnamefont {F.~R.}\ \bibnamefont {Merritt}},
  \bibinfo {author} {\bibfnamefont {R.}~\bibnamefont {Dingle}}, \bibinfo
  {author} {\bibfnamefont {Daniel}\ \bibnamefont {Hone}}, \bibinfo {author}
  {\bibfnamefont {B.~G.}\ \bibnamefont {Silbernagel}}, \ and\ \bibinfo {author}
  {\bibfnamefont {Peter~M.}\ \bibnamefont {Richards}},\ }\bibfield  {title}
  {\enquote {\bibinfo {title} {Exchange narrowing in one-dimensional
  systems},}\ }\href@noop {} {\bibfield  {journal} {\bibinfo  {journal} {Phys.
  Rev. Lett.}\ }\textbf {\bibinfo {volume} {26}},\ \bibinfo {pages}
  {1186--1188} (\bibinfo {year} {1971})}\BibitemShut {NoStop}%
\bibitem [{\citenamefont {Oshikawa}\ and\ \citenamefont
  {Affleck}(2002)}]{PhysRevB.65.134410}%
  \BibitemOpen
  \bibfield  {author} {\bibinfo {author} {\bibfnamefont {Masaki}\ \bibnamefont
  {Oshikawa}}\ and\ \bibinfo {author} {\bibfnamefont {Ian}\ \bibnamefont
  {Affleck}},\ }\bibfield  {title} {\enquote {\bibinfo {title} {Electron spin
  resonance in ${S}=\frac{1}{2}$ antiferromagnetic chains},}\ }\href@noop {}
  {\bibfield  {journal} {\bibinfo  {journal} {Phys. Rev. B}\ }\textbf {\bibinfo
  {volume} {65}},\ \bibinfo {pages} {134410} (\bibinfo {year}
  {2002})}\BibitemShut {NoStop}%
\bibitem [{\citenamefont {Machida}\ \emph {et~al.}(2012)\citenamefont
  {Machida}, \citenamefont {Iitaka},\ and\ \citenamefont
  {Miyashita}}]{PhysRevB.86.224412}%
  \BibitemOpen
  \bibfield  {author} {\bibinfo {author} {\bibfnamefont {Manabu}\ \bibnamefont
  {Machida}}, \bibinfo {author} {\bibfnamefont {Toshiaki}\ \bibnamefont
  {Iitaka}}, \ and\ \bibinfo {author} {\bibfnamefont {Seiji}\ \bibnamefont
  {Miyashita}},\ }\bibfield  {title} {\enquote {\bibinfo {title} {{ESR}
  intensity and the {D}zyaloshinsky-{M}oriya interaction of the nanoscale
  molecular magnet {V}${}_{15}$},}\ }\href@noop {} {\bibfield  {journal}
  {\bibinfo  {journal} {Phys. Rev. B}\ }\textbf {\bibinfo {volume} {86}},\
  \bibinfo {pages} {224412} (\bibinfo {year} {2012})}\BibitemShut {NoStop}%
\bibitem [{\citenamefont {El~Shawish}\ \emph {et~al.}(2010)\citenamefont
  {El~Shawish}, \citenamefont {C\'epas},\ and\ \citenamefont
  {Miyashita}}]{PhysRevB.81.224421}%
  \BibitemOpen
  \bibfield  {author} {\bibinfo {author} {\bibfnamefont {S.}~\bibnamefont
  {El~Shawish}}, \bibinfo {author} {\bibfnamefont {O.}~\bibnamefont {C\'epas}},
  \ and\ \bibinfo {author} {\bibfnamefont {S.}~\bibnamefont {Miyashita}},\
  }\bibfield  {title} {\enquote {\bibinfo {title} {Electron spin resonance in
  ${S}=\frac{1}{2}$ antiferromagnets at high temperature},}\ }\href@noop {}
  {\bibfield  {journal} {\bibinfo  {journal} {Phys. Rev. B}\ }\textbf {\bibinfo
  {volume} {81}},\ \bibinfo {pages} {224421} (\bibinfo {year}
  {2010})}\BibitemShut {NoStop}%
\bibitem [{Not()}]{Note_Kitaev_materials}%
  \BibitemOpen
  \href@noop {} {}\bibinfo {note} {As reviews, see
  Refs.\onlinecite{gegenwart2015spin} and
  \onlinecite{trebst2017kitaev}}\BibitemShut {NoStop}%
\bibitem [{\citenamefont {Sandilands}\ \emph {et~al.}(2015)\citenamefont
  {Sandilands}, \citenamefont {Tian}, \citenamefont {Plumb}, \citenamefont
  {Kim},\ and\ \citenamefont {Burch}}]{PhysRevLett.114.147201}%
  \BibitemOpen
  \bibfield  {author} {\bibinfo {author} {\bibfnamefont {Luke~J.}\ \bibnamefont
  {Sandilands}}, \bibinfo {author} {\bibfnamefont {Yao}\ \bibnamefont {Tian}},
  \bibinfo {author} {\bibfnamefont {Kemp~W.}\ \bibnamefont {Plumb}}, \bibinfo
  {author} {\bibfnamefont {Young-June}\ \bibnamefont {Kim}}, \ and\ \bibinfo
  {author} {\bibfnamefont {Kenneth~S.}\ \bibnamefont {Burch}},\ }\bibfield
  {title} {\enquote {\bibinfo {title} {Scattering continuum and possible
  fractionalized excitations in
  $\ensuremath{\alpha}\text{\ensuremath{-}}{\mathrm{{r}u{c}l}}_{3}$},}\
  }\href@noop {} {\bibfield  {journal} {\bibinfo  {journal} {Phys. Rev. Lett.}\
  }\textbf {\bibinfo {volume} {114}},\ \bibinfo {pages} {147201} (\bibinfo
  {year} {2015})}\BibitemShut {NoStop}%
\bibitem [{\citenamefont {Banerjee}\ \emph {et~al.}(2016)\citenamefont
  {Banerjee}, \citenamefont {Bridges}, \citenamefont {Yan}, \citenamefont
  {Aczel}, \citenamefont {Li}, \citenamefont {Stone}, \citenamefont {Granroth},
  \citenamefont {Lumsden}, \citenamefont {Yiu}, \citenamefont {Knolle} \emph
  {et~al.}}]{banerjee2016proximate}%
  \BibitemOpen
  \bibfield  {author} {\bibinfo {author} {\bibfnamefont {A}~\bibnamefont
  {Banerjee}}, \bibinfo {author} {\bibfnamefont {CA}~\bibnamefont {Bridges}},
  \bibinfo {author} {\bibfnamefont {J-Q}\ \bibnamefont {Yan}}, \bibinfo
  {author} {\bibfnamefont {AA}~\bibnamefont {Aczel}}, \bibinfo {author}
  {\bibfnamefont {L}~\bibnamefont {Li}}, \bibinfo {author} {\bibfnamefont
  {MB}~\bibnamefont {Stone}}, \bibinfo {author} {\bibfnamefont
  {GE}~\bibnamefont {Granroth}}, \bibinfo {author} {\bibfnamefont
  {MD}~\bibnamefont {Lumsden}}, \bibinfo {author} {\bibfnamefont
  {Y}~\bibnamefont {Yiu}}, \bibinfo {author} {\bibfnamefont {J}~\bibnamefont
  {Knolle}},  \emph {et~al.},\ }\bibfield  {title} {\enquote {\bibinfo {title}
  {Proximate {K}itaev quantum spin liquid behaviour in a honeycomb magnet},}\
  }\href@noop {} {\bibfield  {journal} {\bibinfo  {journal} {Nature materials}\
  }\textbf {\bibinfo {volume} {15}},\ \bibinfo {pages} {733--740} (\bibinfo
  {year} {2016})}\BibitemShut {NoStop}%
\bibitem [{\citenamefont {Banerjee}\ \emph {et~al.}(2017)\citenamefont
  {Banerjee}, \citenamefont {Yan}, \citenamefont {Knolle}, \citenamefont
  {Bridges}, \citenamefont {Stone}, \citenamefont {Lumsden}, \citenamefont
  {Mandrus}, \citenamefont {Tennant}, \citenamefont {Moessner},\ and\
  \citenamefont {Nagler}}]{banerjee2016neutron}%
  \BibitemOpen
  \bibfield  {author} {\bibinfo {author} {\bibfnamefont {Arnab}\ \bibnamefont
  {Banerjee}}, \bibinfo {author} {\bibfnamefont {Jiaqiang}\ \bibnamefont
  {Yan}}, \bibinfo {author} {\bibfnamefont {Johannes}\ \bibnamefont {Knolle}},
  \bibinfo {author} {\bibfnamefont {Craig~A.}\ \bibnamefont {Bridges}},
  \bibinfo {author} {\bibfnamefont {Matthew~B.}\ \bibnamefont {Stone}},
  \bibinfo {author} {\bibfnamefont {Mark~D.}\ \bibnamefont {Lumsden}}, \bibinfo
  {author} {\bibfnamefont {David~G.}\ \bibnamefont {Mandrus}}, \bibinfo
  {author} {\bibfnamefont {David~A.}\ \bibnamefont {Tennant}}, \bibinfo
  {author} {\bibfnamefont {Roderich}\ \bibnamefont {Moessner}}, \ and\ \bibinfo
  {author} {\bibfnamefont {Stephen~E.}\ \bibnamefont {Nagler}},\ }\bibfield
  {title} {\enquote {\bibinfo {title} {Neutron scattering in the proximate
  quantum spin liquid $\alpha$-{R}u{C}l$_3$},}\ }\href@noop {} {\bibfield
  {journal} {\bibinfo  {journal} {Science}\ }\textbf {\bibinfo {volume}
  {356}},\ \bibinfo {pages} {1055--1059} (\bibinfo {year} {2017})}\BibitemShut
  {NoStop}%
\bibitem [{\citenamefont {Kitaev}(2006)}]{AnnalsofPhysics321.2}%
  \BibitemOpen
  \bibfield  {author} {\bibinfo {author} {\bibfnamefont {A.}~\bibnamefont
  {Kitaev}},\ }\bibfield  {title} {\enquote {\bibinfo {title} {Anyons in an
  exactly solved model and beyond},}\ }\href@noop {} {\bibfield  {journal}
  {\bibinfo  {journal} {Annals Phys.}\ }\textbf {\bibinfo {volume} {321}},\
  \bibinfo {pages} {2} (\bibinfo {year} {2006})}\BibitemShut {NoStop}%
\bibitem [{\citenamefont {Mandal}\ and\ \citenamefont
  {Surendran}(2009)}]{PhysRevB.79.024426}%
  \BibitemOpen
  \bibfield  {author} {\bibinfo {author} {\bibfnamefont {Saptarshi}\
  \bibnamefont {Mandal}}\ and\ \bibinfo {author} {\bibfnamefont {Naveen}\
  \bibnamefont {Surendran}},\ }\bibfield  {title} {\enquote {\bibinfo {title}
  {Exactly solvable {K}itaev model in three dimensions},}\ }\href@noop {}
  {\bibfield  {journal} {\bibinfo  {journal} {Phys. Rev. B}\ }\textbf {\bibinfo
  {volume} {79}},\ \bibinfo {pages} {024426} (\bibinfo {year}
  {2009})}\BibitemShut {NoStop}%
\bibitem [{rev()}]{review_Balents}%
  \BibitemOpen
  \href@noop {} {}\bibinfo {note} {As a review on quantum spin liquids, see
  Ref.\onlinecite{balents2010spin}.}\BibitemShut {Stop}%
\bibitem [{\citenamefont {Jackeli}\ and\ \citenamefont
  {Khaliullin}(2009)}]{Jackeli}%
  \BibitemOpen
  \bibfield  {author} {\bibinfo {author} {\bibfnamefont {G.}~\bibnamefont
  {Jackeli}}\ and\ \bibinfo {author} {\bibfnamefont {G.}~\bibnamefont
  {Khaliullin}},\ }\bibfield  {title} {\enquote {\bibinfo {title} {Mott
  insulators in the strong spin-orbit coupling limit: From {H}eisenberg to a
  quantum compass and {K}itaev models},}\ }\href {\doibase
  10.1103/PhysRevLett.102.017205} {\bibfield  {journal} {\bibinfo  {journal}
  {Phys. Rev. Lett.}\ }\textbf {\bibinfo {volume} {102}},\ \bibinfo {pages}
  {017205} (\bibinfo {year} {2009})}\BibitemShut {NoStop}%
\bibitem [{\citenamefont {Chaloupka}\ \emph {et~al.}(2010)\citenamefont
  {Chaloupka}, \citenamefont {Jackeli},\ and\ \citenamefont
  {Khaliullin}}]{PhysRevLett.105.027204}%
  \BibitemOpen
  \bibfield  {author} {\bibinfo {author} {\bibfnamefont {Jiri}\ \bibnamefont
  {Chaloupka}}, \bibinfo {author} {\bibfnamefont {George}\ \bibnamefont
  {Jackeli}}, \ and\ \bibinfo {author} {\bibfnamefont {Giniyat}\ \bibnamefont
  {Khaliullin}},\ }\bibfield  {title} {\enquote {\bibinfo {title}
  {Kitaev-{H}eisenberg model on a honeycomb lattice: Possible exotic phases in
  iridium oxides ${{A}}_{2}${I}r{O}$_{3}$},}\ }\href {\doibase
  10.1103/PhysRevLett.105.027204} {\bibfield  {journal} {\bibinfo  {journal}
  {Phys. Rev. Lett.}\ }\textbf {\bibinfo {volume} {105}},\ \bibinfo {pages}
  {027204} (\bibinfo {year} {2010})}\BibitemShut {NoStop}%
\bibitem [{\citenamefont {Shitade}\ \emph {et~al.}(2009)\citenamefont
  {Shitade}, \citenamefont {Katsura}, \citenamefont
  {Kune\ifmmode~\check{s}\else \v{s}\fi{}}, \citenamefont {Qi}, \citenamefont
  {Zhang},\ and\ \citenamefont {Nagaosa}}]{PhysRevLett.102.256403}%
  \BibitemOpen
  \bibfield  {author} {\bibinfo {author} {\bibfnamefont {Atsuo}\ \bibnamefont
  {Shitade}}, \bibinfo {author} {\bibfnamefont {Hosho}\ \bibnamefont
  {Katsura}}, \bibinfo {author} {\bibfnamefont {Jan}\ \bibnamefont
  {Kune\ifmmode~\check{s}\else \v{s}\fi{}}}, \bibinfo {author} {\bibfnamefont
  {Xiao-Liang}\ \bibnamefont {Qi}}, \bibinfo {author} {\bibfnamefont
  {Shou-Cheng}\ \bibnamefont {Zhang}}, \ and\ \bibinfo {author} {\bibfnamefont
  {Naoto}\ \bibnamefont {Nagaosa}},\ }\bibfield  {title} {\enquote {\bibinfo
  {title} {Quantum spin {H}all effect in a transition metal oxide
  {N}a$_{2}${I}r{O}$_{3}$},}\ }\href@noop {} {\bibfield  {journal} {\bibinfo
  {journal} {Phys. Rev. Lett.}\ }\textbf {\bibinfo {volume} {102}},\ \bibinfo
  {pages} {256403} (\bibinfo {year} {2009})}\BibitemShut {NoStop}%
\bibitem [{\citenamefont {Singh}\ and\ \citenamefont
  {Gegenwart}(2010)}]{Singh_Gegenwart}%
  \BibitemOpen
  \bibfield  {author} {\bibinfo {author} {\bibfnamefont {Yogesh}\ \bibnamefont
  {Singh}}\ and\ \bibinfo {author} {\bibfnamefont {P.}~\bibnamefont
  {Gegenwart}},\ }\bibfield  {title} {\enquote {\bibinfo {title}
  {Antiferromagnetic mott insulating state in single crystals of the honeycomb
  lattice material {N}a$_{2}${I}r{O}$_{3}$},}\ }\href@noop {} {\bibfield
  {journal} {\bibinfo  {journal} {Phys. Rev. B}\ }\textbf {\bibinfo {volume}
  {82}},\ \bibinfo {pages} {064412} (\bibinfo {year} {2010})}\BibitemShut
  {NoStop}%
\bibitem [{\citenamefont {Takayama}\ \emph {et~al.}(2015)\citenamefont
  {Takayama}, \citenamefont {Kato}, \citenamefont {Dinnebier}, \citenamefont
  {Nuss}, \citenamefont {Kono}, \citenamefont {Veiga}, \citenamefont {Fabbris},
  \citenamefont {Haskel},\ and\ \citenamefont
  {Takagi}}]{PhysRevLett.114.077202}%
  \BibitemOpen
  \bibfield  {author} {\bibinfo {author} {\bibfnamefont {T.}~\bibnamefont
  {Takayama}}, \bibinfo {author} {\bibfnamefont {A.}~\bibnamefont {Kato}},
  \bibinfo {author} {\bibfnamefont {R.}~\bibnamefont {Dinnebier}}, \bibinfo
  {author} {\bibfnamefont {J.}~\bibnamefont {Nuss}}, \bibinfo {author}
  {\bibfnamefont {H.}~\bibnamefont {Kono}}, \bibinfo {author} {\bibfnamefont
  {L.~S.~I.}\ \bibnamefont {Veiga}}, \bibinfo {author} {\bibfnamefont
  {G.}~\bibnamefont {Fabbris}}, \bibinfo {author} {\bibfnamefont
  {D.}~\bibnamefont {Haskel}}, \ and\ \bibinfo {author} {\bibfnamefont
  {H.}~\bibnamefont {Takagi}},\ }\bibfield  {title} {\enquote {\bibinfo {title}
  {Hyperhoneycomb iridate $\beta$-{Li}$_{2}${I}r{O}$_{3}$ as a platform for
  {K}itaev magnetism},}\ }\href@noop {} {\bibfield  {journal} {\bibinfo
  {journal} {Phys. Rev. Lett.}\ }\textbf {\bibinfo {volume} {114}},\ \bibinfo
  {pages} {077202} (\bibinfo {year} {2015})}\BibitemShut {NoStop}%
\bibitem [{\citenamefont {Modic}\ \emph {et~al.}(2014)\citenamefont {Modic},
  \citenamefont {Smidt}, \citenamefont {Kimchi}, \citenamefont {Breznay},
  \citenamefont {Biffin}, \citenamefont {Choi}, \citenamefont {Johnson},
  \citenamefont {Coldea}, \citenamefont {Watkins-Curry}, \citenamefont
  {McCandless}, \citenamefont {Chan}, \citenamefont {Gandara}, \citenamefont
  {Islam}, \citenamefont {Vishwanath}, \citenamefont {Shekhter}, \citenamefont
  {McDonald},\ and\ \citenamefont {Analytis}}]{modic2014new}%
  \BibitemOpen
  \bibfield  {author} {\bibinfo {author} {\bibfnamefont {Kimberly~A.}\
  \bibnamefont {Modic}}, \bibinfo {author} {\bibfnamefont {Tess~E.}\
  \bibnamefont {Smidt}}, \bibinfo {author} {\bibfnamefont {Itamar}\
  \bibnamefont {Kimchi}}, \bibinfo {author} {\bibfnamefont {Nicholas~P.}\
  \bibnamefont {Breznay}}, \bibinfo {author} {\bibfnamefont {Alun}\
  \bibnamefont {Biffin}}, \bibinfo {author} {\bibfnamefont {Sungkyun}\
  \bibnamefont {Choi}}, \bibinfo {author} {\bibfnamefont {Roger~D.}\
  \bibnamefont {Johnson}}, \bibinfo {author} {\bibfnamefont {Radu}\
  \bibnamefont {Coldea}}, \bibinfo {author} {\bibfnamefont {Pilanda}\
  \bibnamefont {Watkins-Curry}}, \bibinfo {author} {\bibfnamefont {Gregory~T.}\
  \bibnamefont {McCandless}}, \bibinfo {author} {\bibfnamefont {Julia~Y.}\
  \bibnamefont {Chan}}, \bibinfo {author} {\bibfnamefont {Felipe}\ \bibnamefont
  {Gandara}}, \bibinfo {author} {\bibfnamefont {Z.}~\bibnamefont {Islam}},
  \bibinfo {author} {\bibfnamefont {Ashvin}\ \bibnamefont {Vishwanath}},
  \bibinfo {author} {\bibfnamefont {Arkady}\ \bibnamefont {Shekhter}}, \bibinfo
  {author} {\bibfnamefont {Ross~D.}\ \bibnamefont {McDonald}}, \ and\ \bibinfo
  {author} {\bibfnamefont {James~G.}\ \bibnamefont {Analytis}},\ }\bibfield
  {title} {\enquote {\bibinfo {title} {Realization of a three-dimensional
  spin-anisotropic harmonic honeycomb iridate},}\ }\href@noop {} {\bibfield
  {journal} {\bibinfo  {journal} {Nature commun.}\ }\textbf {\bibinfo {volume}
  {5}},\ \bibinfo {pages} {4203} (\bibinfo {year} {2014})}\BibitemShut
  {NoStop}%
\bibitem [{\citenamefont {Knolle}\ \emph
  {et~al.}(2014{\natexlab{a}})\citenamefont {Knolle}, \citenamefont
  {Kovrizhin}, \citenamefont {Chalker},\ and\ \citenamefont
  {Moessner}}]{PhysRevLett.112.207203}%
  \BibitemOpen
  \bibfield  {author} {\bibinfo {author} {\bibfnamefont {J.}~\bibnamefont
  {Knolle}}, \bibinfo {author} {\bibfnamefont {D.~L.}\ \bibnamefont
  {Kovrizhin}}, \bibinfo {author} {\bibfnamefont {J.~T.}\ \bibnamefont
  {Chalker}}, \ and\ \bibinfo {author} {\bibfnamefont {R.}~\bibnamefont
  {Moessner}},\ }\bibfield  {title} {\enquote {\bibinfo {title} {Dynamics of a
  two-dimensional quantum spin liquid: Signatures of emergent {M}ajorana
  fermions and fluxes},}\ }\href@noop {} {\bibfield  {journal} {\bibinfo
  {journal} {Phys. Rev. Lett.}\ }\textbf {\bibinfo {volume} {112}},\ \bibinfo
  {pages} {207203} (\bibinfo {year} {2014}{\natexlab{a}})}\BibitemShut
  {NoStop}%
\bibitem [{\citenamefont {Knolle}\ \emph
  {et~al.}(2014{\natexlab{b}})\citenamefont {Knolle}, \citenamefont {Chern},
  \citenamefont {Kovrizhin}, \citenamefont {Moessner},\ and\ \citenamefont
  {Perkins}}]{PhysRevLett.113.187201}%
  \BibitemOpen
  \bibfield  {author} {\bibinfo {author} {\bibfnamefont {J.}~\bibnamefont
  {Knolle}}, \bibinfo {author} {\bibfnamefont {Gia-Wei}\ \bibnamefont {Chern}},
  \bibinfo {author} {\bibfnamefont {D.~L.}\ \bibnamefont {Kovrizhin}}, \bibinfo
  {author} {\bibfnamefont {R.}~\bibnamefont {Moessner}}, \ and\ \bibinfo
  {author} {\bibfnamefont {N.~B.}\ \bibnamefont {Perkins}},\ }\bibfield
  {title} {\enquote {\bibinfo {title} {Raman scattering signatures of {K}itaev
  spin liquids in ${A}_{2}${I}r{O}$_{3}$ iridates with ${A}$={N}a or {L}i},}\
  }\href@noop {} {\bibfield  {journal} {\bibinfo  {journal} {Phys. Rev. Lett.}\
  }\textbf {\bibinfo {volume} {113}},\ \bibinfo {pages} {187201} (\bibinfo
  {year} {2014}{\natexlab{b}})}\BibitemShut {NoStop}%
\bibitem [{\citenamefont {Nasu}\ \emph {et~al.}(2016)\citenamefont {Nasu},
  \citenamefont {Knolle}, \citenamefont {Kovrizhin}, \citenamefont {Motome},\
  and\ \citenamefont {Moessner}}]{nasu2016fermionic}%
  \BibitemOpen
  \bibfield  {author} {\bibinfo {author} {\bibfnamefont {J}~\bibnamefont
  {Nasu}}, \bibinfo {author} {\bibfnamefont {J}~\bibnamefont {Knolle}},
  \bibinfo {author} {\bibfnamefont {DL}~\bibnamefont {Kovrizhin}}, \bibinfo
  {author} {\bibfnamefont {Y}~\bibnamefont {Motome}}, \ and\ \bibinfo {author}
  {\bibfnamefont {Roderich}\ \bibnamefont {Moessner}},\ }\bibfield  {title}
  {\enquote {\bibinfo {title} {Fermionic response from fractionalization in an
  insulating two-dimensional magnet},}\ }\href@noop {} {\bibfield  {journal}
  {\bibinfo  {journal} {Nature Physics}\ }\textbf {\bibinfo {volume} {12}},\
  \bibinfo {pages} {912--915} (\bibinfo {year} {2016})}\BibitemShut {NoStop}%
\bibitem [{\citenamefont {Yoshitake}\ \emph {et~al.}(2016)\citenamefont
  {Yoshitake}, \citenamefont {Nasu},\ and\ \citenamefont
  {Motome}}]{PhysRevLett.117.157203}%
  \BibitemOpen
  \bibfield  {author} {\bibinfo {author} {\bibfnamefont {Junki}\ \bibnamefont
  {Yoshitake}}, \bibinfo {author} {\bibfnamefont {Joji}\ \bibnamefont {Nasu}},
  \ and\ \bibinfo {author} {\bibfnamefont {Yukitoshi}\ \bibnamefont {Motome}},\
  }\bibfield  {title} {\enquote {\bibinfo {title} {Fractional spin fluctuations
  as a precursor of quantum spin liquids: {M}ajorana dynamical mean-field study
  for the {K}itaev model},}\ }\href@noop {} {\bibfield  {journal} {\bibinfo
  {journal} {Phys. Rev. Lett.}\ }\textbf {\bibinfo {volume} {117}},\ \bibinfo
  {pages} {157203} (\bibinfo {year} {2016})}\BibitemShut {NoStop}%
\bibitem [{\citenamefont {Hermanns}\ and\ \citenamefont
  {Trebst}(2014)}]{PhysRevB.89.235102}%
  \BibitemOpen
  \bibfield  {author} {\bibinfo {author} {\bibfnamefont {M.}~\bibnamefont
  {Hermanns}}\ and\ \bibinfo {author} {\bibfnamefont {S.}~\bibnamefont
  {Trebst}},\ }\bibfield  {title} {\enquote {\bibinfo {title} {Quantum spin
  liquid with a {M}ajorana fermi surface on the three-dimensional hyperoctagon
  lattice},}\ }\href@noop {} {\bibfield  {journal} {\bibinfo  {journal} {Phys.
  Rev. B}\ }\textbf {\bibinfo {volume} {89}},\ \bibinfo {pages} {235102}
  (\bibinfo {year} {2014})}\BibitemShut {NoStop}%
\bibitem [{\citenamefont {Chaloupka}\ \emph {et~al.}(2013)\citenamefont
  {Chaloupka}, \citenamefont {Jackeli},\ and\ \citenamefont
  {Khaliullin}}]{PhysRevLett.110.097204}%
  \BibitemOpen
  \bibfield  {author} {\bibinfo {author} {\bibfnamefont {Jiri}\ \bibnamefont
  {Chaloupka}}, \bibinfo {author} {\bibfnamefont {George}\ \bibnamefont
  {Jackeli}}, \ and\ \bibinfo {author} {\bibfnamefont {Giniyat}\ \bibnamefont
  {Khaliullin}},\ }\bibfield  {title} {\enquote {\bibinfo {title} {Zigzag
  magnetic order in the iridium oxide ${\mathrm{na}}_{2}{\mathrm{iro}}_{3}$},}\
  }\href {\doibase 10.1103/PhysRevLett.110.097204} {\bibfield  {journal}
  {\bibinfo  {journal} {Phys. Rev. Lett.}\ }\textbf {\bibinfo {volume} {110}},\
  \bibinfo {pages} {097204} (\bibinfo {year} {2013})}\BibitemShut {NoStop}%
\bibitem [{\citenamefont {Katukuri}\ \emph {et~al.}(2014)\citenamefont
  {Katukuri}, \citenamefont {Nishimoto}, \citenamefont {Yushankhai},
  \citenamefont {Stoyanova}, \citenamefont {Kandpal}, \citenamefont {Sungkyun},
  \citenamefont {Coldea}, \citenamefont {Rousochatzakis}, \citenamefont
  {Hozoi},\ and\ \citenamefont {van~den Brink}}]{arXiv:1312.7437}%
  \BibitemOpen
  \bibfield  {author} {\bibinfo {author} {\bibfnamefont {V.~K.}\ \bibnamefont
  {Katukuri}}, \bibinfo {author} {\bibfnamefont {S.}~\bibnamefont {Nishimoto}},
  \bibinfo {author} {\bibfnamefont {V.}~\bibnamefont {Yushankhai}}, \bibinfo
  {author} {\bibfnamefont {A.}~\bibnamefont {Stoyanova}}, \bibinfo {author}
  {\bibfnamefont {H.}~\bibnamefont {Kandpal}}, \bibinfo {author} {\bibfnamefont
  {C.}~\bibnamefont {Sungkyun}}, \bibinfo {author} {\bibfnamefont
  {R.}~\bibnamefont {Coldea}}, \bibinfo {author} {\bibfnamefont
  {I.}~\bibnamefont {Rousochatzakis}}, \bibinfo {author} {\bibfnamefont
  {L.}~\bibnamefont {Hozoi}}, \ and\ \bibinfo {author} {\bibfnamefont
  {J.}~\bibnamefont {van~den Brink}},\ }\bibfield  {title} {\enquote {\bibinfo
  {title} {Kitaev interactions between $j$=1/2 moments in honeycomb
  {N}a$_2${I}r{O}$_3$ are large and ferromagnetic: insights from ab initio
  quantum chemistry calculations},}\ }\href@noop {} {\bibfield  {journal}
  {\bibinfo  {journal} {New J. Phys.}\ }\textbf {\bibinfo {volume} {16}},\
  \bibinfo {pages} {013056} (\bibinfo {year} {2014})}\BibitemShut {NoStop}%
\bibitem [{\citenamefont {Rau}\ \emph {et~al.}(2014)\citenamefont {Rau},
  \citenamefont {Lee},\ and\ \citenamefont {Kee}}]{PhysRevLett.112.077204}%
  \BibitemOpen
  \bibfield  {author} {\bibinfo {author} {\bibfnamefont {Jeffrey~G.}\
  \bibnamefont {Rau}}, \bibinfo {author} {\bibfnamefont {Eric Kin-Ho}\
  \bibnamefont {Lee}}, \ and\ \bibinfo {author} {\bibfnamefont {Hae-Young}\
  \bibnamefont {Kee}},\ }\bibfield  {title} {\enquote {\bibinfo {title}
  {Generic spin model for the honeycomb iridates beyond the {K}itaev limit},}\
  }\href {\doibase 10.1103/PhysRevLett.112.077204} {\bibfield  {journal}
  {\bibinfo  {journal} {Phys. Rev. Lett.}\ }\textbf {\bibinfo {volume} {112}},\
  \bibinfo {pages} {077204} (\bibinfo {year} {2014})}\BibitemShut {NoStop}%
\bibitem [{\citenamefont {Yamaji}\ \emph {et~al.}(2014)\citenamefont {Yamaji},
  \citenamefont {Nomura}, \citenamefont {Kurita}, \citenamefont {Arita},\ and\
  \citenamefont {Imada}}]{PhysRevLett.113.107201}%
  \BibitemOpen
  \bibfield  {author} {\bibinfo {author} {\bibfnamefont {Youhei}\ \bibnamefont
  {Yamaji}}, \bibinfo {author} {\bibfnamefont {Yusuke}\ \bibnamefont {Nomura}},
  \bibinfo {author} {\bibfnamefont {Moyuru}\ \bibnamefont {Kurita}}, \bibinfo
  {author} {\bibfnamefont {Ryotaro}\ \bibnamefont {Arita}}, \ and\ \bibinfo
  {author} {\bibfnamefont {Masatoshi}\ \bibnamefont {Imada}},\ }\bibfield
  {title} {\enquote {\bibinfo {title} {First-principles study of the
  honeycomb-lattice iridates {N}a$_{2}${I}r{O}$_{3}$ in the presence of strong
  spin-orbit interaction and electron correlations},}\ }\href@noop {}
  {\bibfield  {journal} {\bibinfo  {journal} {Phys. Rev. Lett.}\ }\textbf
  {\bibinfo {volume} {113}},\ \bibinfo {pages} {107201} (\bibinfo {year}
  {2014})}\BibitemShut {NoStop}%
\bibitem [{\citenamefont {Frommer}(2003)}]{frommer2003bicgstab}%
  \BibitemOpen
  \bibfield  {author} {\bibinfo {author} {\bibfnamefont {Andreas}\ \bibnamefont
  {Frommer}},\ }\bibfield  {title} {\enquote {\bibinfo {title} {{B}i{CG}stab
  ($\ell$) for families of shifted linear systems},}\ }\href@noop {} {\bibfield
   {journal} {\bibinfo  {journal} {Computing}\ }\textbf {\bibinfo {volume}
  {70}},\ \bibinfo {pages} {87--109} (\bibinfo {year} {2003})}\BibitemShut
  {NoStop}%
\bibitem [{Note2()}]{Note2}%
  \BibitemOpen
  \bibinfo {note} {The construction of the microcanonical shell may remind the
  readers of the microcanonical thermal pure quantum (TPQ) state proposed in
  Ref.\protect \rev@citealpnum {PhysRevLett.108.240401}. To avoid possible
  confusion, we note that the microcanonical TPQ state does not correspond to a
  microcannonical shell. As proven in Ref.\protect \rev@citealpnum
  {PhysRevLett.111.010401}, the microcanonical thermal pure quantum state
  reproduces probability distribution of the canonical ensemble}\BibitemShut
  {NoStop}%
\bibitem [{\citenamefont {Ferrenberg}\ and\ \citenamefont
  {Landau}(1991)}]{PhysRevB.44.5081}%
  \BibitemOpen
  \bibfield  {author} {\bibinfo {author} {\bibfnamefont {Alan~M.}\ \bibnamefont
  {Ferrenberg}}\ and\ \bibinfo {author} {\bibfnamefont {D.~P.}\ \bibnamefont
  {Landau}},\ }\bibfield  {title} {\enquote {\bibinfo {title} {Critical
  behavior of the three-dimensional {I}sing model: A high-resolution {M}onte
  {C}arlo study},}\ }\href@noop {} {\bibfield  {journal} {\bibinfo  {journal}
  {Phys. Rev. B}\ }\textbf {\bibinfo {volume} {44}},\ \bibinfo {pages}
  {5081--5091} (\bibinfo {year} {1991})}\BibitemShut {NoStop}%
\bibitem [{\citenamefont {Jiang}\ \emph {et~al.}(2011)\citenamefont {Jiang},
  \citenamefont {Gu}, \citenamefont {Qi},\ and\ \citenamefont
  {Trebst}}]{PhysRevB.83.245104}%
  \BibitemOpen
  \bibfield  {author} {\bibinfo {author} {\bibfnamefont {Hong-Chen}\
  \bibnamefont {Jiang}}, \bibinfo {author} {\bibfnamefont {Zheng-Cheng}\
  \bibnamefont {Gu}}, \bibinfo {author} {\bibfnamefont {Xiao-Liang}\
  \bibnamefont {Qi}}, \ and\ \bibinfo {author} {\bibfnamefont {Simon}\
  \bibnamefont {Trebst}},\ }\bibfield  {title} {\enquote {\bibinfo {title}
  {Possible proximity of the mott insulating iridate na${}_{2}$iro${}_{3}$ to a
  topological phase: Phase diagram of the heisenberg-kitaev model in a magnetic
  field},}\ }\href@noop {} {\bibfield  {journal} {\bibinfo  {journal} {Phys.
  Rev. B}\ }\textbf {\bibinfo {volume} {83}},\ \bibinfo {pages} {245104}
  (\bibinfo {year} {2011})}\BibitemShut {NoStop}%
\bibitem [{\citenamefont {Osorio~Iregui}\ \emph {et~al.}(2014)\citenamefont
  {Osorio~Iregui}, \citenamefont {Corboz},\ and\ \citenamefont
  {Troyer}}]{PhysRevB.90.195102}%
  \BibitemOpen
  \bibfield  {author} {\bibinfo {author} {\bibfnamefont {Juan}\ \bibnamefont
  {Osorio~Iregui}}, \bibinfo {author} {\bibfnamefont {Philippe}\ \bibnamefont
  {Corboz}}, \ and\ \bibinfo {author} {\bibfnamefont {Matthias}\ \bibnamefont
  {Troyer}},\ }\bibfield  {title} {\enquote {\bibinfo {title} {Probing the
  stability of the spin-liquid phases in the {K}itaev-{H}eisenberg model using
  tensor network algorithms},}\ }\href@noop {} {\bibfield  {journal} {\bibinfo
  {journal} {Phys. Rev. B}\ }\textbf {\bibinfo {volume} {90}},\ \bibinfo
  {pages} {195102} (\bibinfo {year} {2014})}\BibitemShut {NoStop}%
\bibitem [{\citenamefont {Nasu}\ \emph {et~al.}(2015)\citenamefont {Nasu},
  \citenamefont {Udagawa},\ and\ \citenamefont {Motome}}]{PhysRevB.92.115122}%
  \BibitemOpen
  \bibfield  {author} {\bibinfo {author} {\bibfnamefont {Joji}\ \bibnamefont
  {Nasu}}, \bibinfo {author} {\bibfnamefont {Masafumi}\ \bibnamefont
  {Udagawa}}, \ and\ \bibinfo {author} {\bibfnamefont {Yukitoshi}\ \bibnamefont
  {Motome}},\ }\bibfield  {title} {\enquote {\bibinfo {title} {Thermal
  fractionalization of quantum spins in a kitaev model: Temperature-linear
  specific heat and coherent transport of majorana fermions},}\ }\href@noop {}
  {\bibfield  {journal} {\bibinfo  {journal} {Phys. Rev. B}\ }\textbf {\bibinfo
  {volume} {92}},\ \bibinfo {pages} {115122} (\bibinfo {year}
  {2015})}\BibitemShut {NoStop}%
\bibitem [{\citenamefont {Yamaji}\ \emph {et~al.}(2016)\citenamefont {Yamaji},
  \citenamefont {Suzuki}, \citenamefont {Yamada}, \citenamefont {Suga},
  \citenamefont {Kawashima},\ and\ \citenamefont {Imada}}]{PhysRevB.93.174425}%
  \BibitemOpen
  \bibfield  {author} {\bibinfo {author} {\bibfnamefont {Youhei}\ \bibnamefont
  {Yamaji}}, \bibinfo {author} {\bibfnamefont {Takafumi}\ \bibnamefont
  {Suzuki}}, \bibinfo {author} {\bibfnamefont {Takuto}\ \bibnamefont {Yamada}},
  \bibinfo {author} {\bibfnamefont {Sei-ichiro}\ \bibnamefont {Suga}}, \bibinfo
  {author} {\bibfnamefont {Naoki}\ \bibnamefont {Kawashima}}, \ and\ \bibinfo
  {author} {\bibfnamefont {Masatoshi}\ \bibnamefont {Imada}},\ }\bibfield
  {title} {\enquote {\bibinfo {title} {Clues and criteria for designing a
  {K}itaev spin liquid revealed by thermal and spin excitations of the
  honeycomb iridate {N}a$_{2}${I}r{O}$_{3}$},}\ }\href@noop {} {\bibfield
  {journal} {\bibinfo  {journal} {Phys. Rev. B}\ }\textbf {\bibinfo {volume}
  {93}},\ \bibinfo {pages} {174425} (\bibinfo {year} {2016})}\BibitemShut
  {NoStop}%
\bibitem [{\citenamefont {Ullah}(1964)}]{ULLAH196465}%
  \BibitemOpen
  \bibfield  {author} {\bibinfo {author} {\bibfnamefont {Nazakat}\ \bibnamefont
  {Ullah}},\ }\bibfield  {title} {\enquote {\bibinfo {title} {Invariance
  hypothesis and higher correlations of hamiltonian matrix elements},}\
  }\href@noop {} {\bibfield  {journal} {\bibinfo  {journal} {Nuclear Physics}\
  }\textbf {\bibinfo {volume} {58}},\ \bibinfo {pages} {65 -- 71} (\bibinfo
  {year} {1964})}\BibitemShut {NoStop}%
\bibitem [{Miy()}]{Miyashita_DeRaedt}%
  \BibitemOpen
  \href@noop {} {}\bibinfo {note} {S. Miyashita and H. De Raedt, private
  communication.}\BibitemShut {Stop}%
\bibitem [{\citenamefont {Gagliano}\ and\ \citenamefont
  {Balseiro}(1987)}]{PhysRevLett.59.2999}%
  \BibitemOpen
  \bibfield  {author} {\bibinfo {author} {\bibfnamefont {E.~R.}\ \bibnamefont
  {Gagliano}}\ and\ \bibinfo {author} {\bibfnamefont {C.~A.}\ \bibnamefont
  {Balseiro}},\ }\bibfield  {title} {\enquote {\bibinfo {title} {Dynamical
  properties of quantum many-body systems at zero temperature},}\ }\href@noop
  {} {\bibfield  {journal} {\bibinfo  {journal} {Phys. Rev. Lett.}\ }\textbf
  {\bibinfo {volume} {59}},\ \bibinfo {pages} {2999--3002} (\bibinfo {year}
  {1987})}\BibitemShut {NoStop}%
\bibitem [{Kom()}]{Komega}%
  \BibitemOpen
  \href@noop {} {}\bibinfo {note} {A numerical library for the shifted Krylov
  methods, $K\omega$, developed by M. Kawamura based on
  Refs.\onlinecite{sogabe2007numerical} and
  \onlinecite{doi:10.1143/JPSJ.77.114713}, is available through
  https://github.com/issp-center-dev/Komega}\BibitemShut {NoStop}%
\bibitem [{\citenamefont {Kato}(1949)}]{doi:10.1143/ptp/4.4.514}%
  \BibitemOpen
  \bibfield  {author} {\bibinfo {author} {\bibfnamefont {Tosio}\ \bibnamefont
  {Kato}},\ }\bibfield  {title} {\enquote {\bibinfo {title} {On the convergence
  of the perturbation method. i},}\ }\href@noop {} {\bibfield  {journal}
  {\bibinfo  {journal} {Progress of Theoretical Physics}\ }\textbf {\bibinfo
  {volume} {4}},\ \bibinfo {pages} {514} (\bibinfo {year} {1949})}\BibitemShut
  {NoStop}%
\bibitem [{\citenamefont {Sakurai}\ and\ \citenamefont
  {Sugiura}(2003)}]{Sakurai2003119}%
  \BibitemOpen
  \bibfield  {author} {\bibinfo {author} {\bibfnamefont {Tetsuya}\ \bibnamefont
  {Sakurai}}\ and\ \bibinfo {author} {\bibfnamefont {Hiroshi}\ \bibnamefont
  {Sugiura}},\ }\bibfield  {title} {\enquote {\bibinfo {title} {A projection
  method for generalized eigenvalue problems using numerical integration},}\
  }\href@noop {} {\bibfield  {journal} {\bibinfo  {journal} {Journal of
  Computational and Applied Mathematics}\ }\textbf {\bibinfo {volume} {159}},\
  \bibinfo {pages} {119--128} (\bibinfo {year} {2003})}\BibitemShut {NoStop}%
\bibitem [{\citenamefont {Ikegami}\ \emph {et~al.}(2010)\citenamefont
  {Ikegami}, \citenamefont {Sakurai},\ and\ \citenamefont
  {Nagashima}}]{ikegami2010filter}%
  \BibitemOpen
  \bibfield  {author} {\bibinfo {author} {\bibfnamefont {Tsutomu}\ \bibnamefont
  {Ikegami}}, \bibinfo {author} {\bibfnamefont {Tetsuya}\ \bibnamefont
  {Sakurai}}, \ and\ \bibinfo {author} {\bibfnamefont {Umpei}\ \bibnamefont
  {Nagashima}},\ }\bibfield  {title} {\enquote {\bibinfo {title} {A filter
  diagonalization for generalized eigenvalue problems based on the
  {S}akurai--{S}ugiura projection method},}\ }\href@noop {} {\bibfield
  {journal} {\bibinfo  {journal} {Journal of Computational and Applied
  Mathematics}\ }\textbf {\bibinfo {volume} {233}},\ \bibinfo {pages}
  {1927--1936} (\bibinfo {year} {2010})}\BibitemShut {NoStop}%
\bibitem [{\citenamefont {Shimizu}\ \emph {et~al.}(2016)\citenamefont
  {Shimizu}, \citenamefont {Utsuno}, \citenamefont {Futamura}, \citenamefont
  {Sakurai}, \citenamefont {Mizusaki},\ and\ \citenamefont
  {Otsuka}}]{Shimizu201613}%
  \BibitemOpen
  \bibfield  {author} {\bibinfo {author} {\bibfnamefont {Noritaka}\
  \bibnamefont {Shimizu}}, \bibinfo {author} {\bibfnamefont {Yutaka}\
  \bibnamefont {Utsuno}}, \bibinfo {author} {\bibfnamefont {Yasunori}\
  \bibnamefont {Futamura}}, \bibinfo {author} {\bibfnamefont {Tetsuya}\
  \bibnamefont {Sakurai}}, \bibinfo {author} {\bibfnamefont {Takahiro}\
  \bibnamefont {Mizusaki}}, \ and\ \bibinfo {author} {\bibfnamefont {Takaharu}\
  \bibnamefont {Otsuka}},\ }\bibfield  {title} {\enquote {\bibinfo {title}
  {Stochastic estimation of nuclear level density in the nuclear shell model:
  An application to parity-dependent level density in 58{N}i},}\ }\href@noop {}
  {\bibfield  {journal} {\bibinfo  {journal} {Physics Letters B}\ }\textbf
  {\bibinfo {volume} {753}},\ \bibinfo {pages} {13 -- 17} (\bibinfo {year}
  {2016})}\BibitemShut {NoStop}%
\bibitem [{\citenamefont {Falcioni}\ \emph {et~al.}(1982)\citenamefont
  {Falcioni}, \citenamefont {Marinari}, \citenamefont {Paciello}, \citenamefont
  {Parisi},\ and\ \citenamefont {Taglienti}}]{FALCIONI1982331}%
  \BibitemOpen
  \bibfield  {author} {\bibinfo {author} {\bibfnamefont {M.}~\bibnamefont
  {Falcioni}}, \bibinfo {author} {\bibfnamefont {E.}~\bibnamefont {Marinari}},
  \bibinfo {author} {\bibfnamefont {M.L.}\ \bibnamefont {Paciello}}, \bibinfo
  {author} {\bibfnamefont {G.}~\bibnamefont {Parisi}}, \ and\ \bibinfo {author}
  {\bibfnamefont {B.}~\bibnamefont {Taglienti}},\ }\bibfield  {title} {\enquote
  {\bibinfo {title} {Complex zeros in the partition function of the
  four-dimensional {SU}(2) lattice gauge model},}\ }\href@noop {} {\bibfield
  {journal} {\bibinfo  {journal} {Physics Letters B}\ }\textbf {\bibinfo
  {volume} {108}},\ \bibinfo {pages} {331 -- 332} (\bibinfo {year}
  {1982})}\BibitemShut {NoStop}%
\bibitem [{\citenamefont {Marinari}(1984)}]{MARINARI1984123}%
  \BibitemOpen
  \bibfield  {author} {\bibinfo {author} {\bibfnamefont {Enzo}\ \bibnamefont
  {Marinari}},\ }\bibfield  {title} {\enquote {\bibinfo {title} {Complex zeroes
  of the d = 3 {I}sing model: Finite-size scaling and critical amplitudes},}\
  }\href@noop {} {\bibfield  {journal} {\bibinfo  {journal} {Nuclear Physics
  B}\ }\textbf {\bibinfo {volume} {235}},\ \bibinfo {pages} {123 -- 134}
  (\bibinfo {year} {1984})}\BibitemShut {NoStop}%
\bibitem [{\citenamefont {Prelov{\v{s}}ek}\ and\ \citenamefont
  {Bon{\v{c}}a}(2013)}]{prelovvsek2013ground}%
  \BibitemOpen
  \bibfield  {author} {\bibinfo {author} {\bibfnamefont {P}~\bibnamefont
  {Prelov{\v{s}}ek}}\ and\ \bibinfo {author} {\bibfnamefont {J}~\bibnamefont
  {Bon{\v{c}}a}},\ }\bibfield  {title} {\enquote {\bibinfo {title} {Ground
  state and finite temperature lanczos methods},}\ }in\ \href@noop {} {\emph
  {\bibinfo {booktitle} {Strongly Correlated Systems}}}\ (\bibinfo  {publisher}
  {Springer},\ \bibinfo {year} {2013})\ pp.\ \bibinfo {pages}
  {1--30}\BibitemShut {NoStop}%
\bibitem [{\citenamefont {Steinigeweg}\ \emph {et~al.}(2016)\citenamefont
  {Steinigeweg}, \citenamefont {Herbrych}, \citenamefont {Zotos},\ and\
  \citenamefont {Brenig}}]{PhysRevLett.116.017202}%
  \BibitemOpen
  \bibfield  {author} {\bibinfo {author} {\bibfnamefont {Robin}\ \bibnamefont
  {Steinigeweg}}, \bibinfo {author} {\bibfnamefont {Jacek}\ \bibnamefont
  {Herbrych}}, \bibinfo {author} {\bibfnamefont {Xenophon}\ \bibnamefont
  {Zotos}}, \ and\ \bibinfo {author} {\bibfnamefont {Wolfram}\ \bibnamefont
  {Brenig}},\ }\bibfield  {title} {\enquote {\bibinfo {title} {Heat
  conductivity of the heisenberg spin-$1/2$ ladder: From weak to strong
  breaking of integrability},}\ }\href@noop {} {\bibfield  {journal} {\bibinfo
  {journal} {Phys. Rev. Lett.}\ }\textbf {\bibinfo {volume} {116}},\ \bibinfo
  {pages} {017202} (\bibinfo {year} {2016})}\BibitemShut {NoStop}%
\bibitem [{\citenamefont {Yoshitake}\ \emph {et~al.}(2017)\citenamefont
  {Yoshitake}, \citenamefont {Nasu}, \citenamefont {Kato},\ and\ \citenamefont
  {Motome}}]{PhysRevB.96.024438}%
  \BibitemOpen
  \bibfield  {author} {\bibinfo {author} {\bibfnamefont {Junki}\ \bibnamefont
  {Yoshitake}}, \bibinfo {author} {\bibfnamefont {Joji}\ \bibnamefont {Nasu}},
  \bibinfo {author} {\bibfnamefont {Yasuyuki}\ \bibnamefont {Kato}}, \ and\
  \bibinfo {author} {\bibfnamefont {Yukitoshi}\ \bibnamefont {Motome}},\
  }\bibfield  {title} {\enquote {\bibinfo {title} {Majorana dynamical
  mean-field study of spin dynamics at finite temperatures in the honeycomb
  kitaev model},}\ }\href@noop {} {\bibfield  {journal} {\bibinfo  {journal}
  {Phys. Rev. B}\ }\textbf {\bibinfo {volume} {96}},\ \bibinfo {pages} {024438}
  (\bibinfo {year} {2017})}\BibitemShut {NoStop}%
\bibitem [{\citenamefont {Mehlawat}\ \emph {et~al.}(2017)\citenamefont
  {Mehlawat}, \citenamefont {Thamizhavel},\ and\ \citenamefont
  {Singh}}]{PhysRevB.95.144406}%
  \BibitemOpen
  \bibfield  {author} {\bibinfo {author} {\bibfnamefont {Kavita}\ \bibnamefont
  {Mehlawat}}, \bibinfo {author} {\bibfnamefont {A.}~\bibnamefont
  {Thamizhavel}}, \ and\ \bibinfo {author} {\bibfnamefont {Yogesh}\
  \bibnamefont {Singh}},\ }\bibfield  {title} {\enquote {\bibinfo {title} {Heat
  capacity evidence for proximity to the {K}itaev quantum spin liquid in
  ${A}_{2}$ {I}r{O}$_{3}$ (${A}$={N}a, {L}i)},}\ }\href@noop {} {\bibfield
  {journal} {\bibinfo  {journal} {Phys. Rev. B}\ }\textbf {\bibinfo {volume}
  {95}},\ \bibinfo {pages} {144406} (\bibinfo {year} {2017})}\BibitemShut
  {NoStop}%
\bibitem [{\citenamefont {Ran}\ \emph {et~al.}(2017)\citenamefont {Ran},
  \citenamefont {Wang}, \citenamefont {Wang}, \citenamefont {Dong},
  \citenamefont {Ren}, \citenamefont {Bao}, \citenamefont {Li}, \citenamefont
  {Ma}, \citenamefont {Gan}, \citenamefont {Zhang}, \citenamefont {Park},
  \citenamefont {Deng}, \citenamefont {Danilkin}, \citenamefont {Yu},
  \citenamefont {Li},\ and\ \citenamefont {Wen}}]{PhysRevLett.118.107203}%
  \BibitemOpen
  \bibfield  {author} {\bibinfo {author} {\bibfnamefont {Kejing}\ \bibnamefont
  {Ran}}, \bibinfo {author} {\bibfnamefont {Jinghui}\ \bibnamefont {Wang}},
  \bibinfo {author} {\bibfnamefont {Wei}\ \bibnamefont {Wang}}, \bibinfo
  {author} {\bibfnamefont {Zhao-Yang}\ \bibnamefont {Dong}}, \bibinfo {author}
  {\bibfnamefont {Xiao}\ \bibnamefont {Ren}}, \bibinfo {author} {\bibfnamefont
  {Song}\ \bibnamefont {Bao}}, \bibinfo {author} {\bibfnamefont {Shichao}\
  \bibnamefont {Li}}, \bibinfo {author} {\bibfnamefont {Zhen}\ \bibnamefont
  {Ma}}, \bibinfo {author} {\bibfnamefont {Yuan}\ \bibnamefont {Gan}}, \bibinfo
  {author} {\bibfnamefont {Youtian}\ \bibnamefont {Zhang}}, \bibinfo {author}
  {\bibfnamefont {J.~T.}\ \bibnamefont {Park}}, \bibinfo {author}
  {\bibfnamefont {Guochu}\ \bibnamefont {Deng}}, \bibinfo {author}
  {\bibfnamefont {S.}~\bibnamefont {Danilkin}}, \bibinfo {author}
  {\bibfnamefont {Shun-Li}\ \bibnamefont {Yu}}, \bibinfo {author}
  {\bibfnamefont {Jian-Xin}\ \bibnamefont {Li}}, \ and\ \bibinfo {author}
  {\bibfnamefont {Jinsheng}\ \bibnamefont {Wen}},\ }\bibfield  {title}
  {\enquote {\bibinfo {title} {Spin-wave excitations evidencing the kitaev
  interaction in single crystalline
  $\ensuremath{\alpha}\text{\ensuremath{-}}${R}u{C}l$_{3}$},}\ }\href@noop {}
  {\bibfield  {journal} {\bibinfo  {journal} {Phys. Rev. Lett.}\ }\textbf
  {\bibinfo {volume} {118}},\ \bibinfo {pages} {107203} (\bibinfo {year}
  {2017})}\BibitemShut {NoStop}%
\bibitem [{Note3()}]{Note3}%
  \BibitemOpen
  \bibinfo {note} {We note that the energy unit {\protect \it A} in
  Ref.\protect \rev@citealpnum {PhysRevB.93.174425} is the half of the present
  energy unit. The label of the typical momenta is also different: The
  {\protect \it Y} point in Ref.\protect \rev@citealpnum {PhysRevB.93.174425}
  is denoted by {\protect \it M} in the present paper.}\BibitemShut {Stop}%
\bibitem [{\citenamefont {Coldea}\ \emph {et~al.}(2001)\citenamefont {Coldea},
  \citenamefont {Hayden}, \citenamefont {Aeppli}, \citenamefont {Perring},
  \citenamefont {Frost}, \citenamefont {Mason}, \citenamefont {Cheong},\ and\
  \citenamefont {Fisk}}]{PhysRevLett.86.5377}%
  \BibitemOpen
  \bibfield  {author} {\bibinfo {author} {\bibfnamefont {R.}~\bibnamefont
  {Coldea}}, \bibinfo {author} {\bibfnamefont {S.~M.}\ \bibnamefont {Hayden}},
  \bibinfo {author} {\bibfnamefont {G.}~\bibnamefont {Aeppli}}, \bibinfo
  {author} {\bibfnamefont {T.~G.}\ \bibnamefont {Perring}}, \bibinfo {author}
  {\bibfnamefont {C.~D.}\ \bibnamefont {Frost}}, \bibinfo {author}
  {\bibfnamefont {T.~E.}\ \bibnamefont {Mason}}, \bibinfo {author}
  {\bibfnamefont {S.-W.}\ \bibnamefont {Cheong}}, \ and\ \bibinfo {author}
  {\bibfnamefont {Z.}~\bibnamefont {Fisk}},\ }\bibfield  {title} {\enquote
  {\bibinfo {title} {Spin waves and electronic interactions in
  {L}a$_2${C}u{O}$_4$},}\ }\href@noop {} {\bibfield  {journal} {\bibinfo
  {journal} {Phys. Rev. Lett.}\ }\textbf {\bibinfo {volume} {86}},\ \bibinfo
  {pages} {5377--5380} (\bibinfo {year} {2001})}\BibitemShut {NoStop}%
\bibitem [{\citenamefont {Shao}\ \emph {et~al.}()\citenamefont {Shao},
  \citenamefont {Qin}, \citenamefont {Capponi}, \citenamefont {Chesi},
  \citenamefont {Meng},\ and\ \citenamefont {Sandvik}}]{Shao2017}%
  \BibitemOpen
  \bibfield  {author} {\bibinfo {author} {\bibfnamefont {H.}~\bibnamefont
  {Shao}}, \bibinfo {author} {\bibfnamefont {Y.~Q.}\ \bibnamefont {Qin}},
  \bibinfo {author} {\bibfnamefont {S.}~\bibnamefont {Capponi}}, \bibinfo
  {author} {\bibfnamefont {S.}~\bibnamefont {Chesi}}, \bibinfo {author}
  {\bibfnamefont {Z.~Y.}\ \bibnamefont {Meng}}, \ and\ \bibinfo {author}
  {\bibfnamefont {A.~W.}\ \bibnamefont {Sandvik}},\ }\href@noop {} {\enquote
  {\bibinfo {title} {Nearly deconfined spinon excitations in the square-lattice
  spin-1/2 heisenberg antiferromagnet},}\ }\Eprint
  {http://arxiv.org/abs/arXiv:1708.03232} {arXiv:1708.03232} \BibitemShut
  {NoStop}%
\bibitem [{\citenamefont {Price}\ and\ \citenamefont
  {Perkins}(2012)}]{PhysRevLett.109.187201}%
  \BibitemOpen
  \bibfield  {author} {\bibinfo {author} {\bibfnamefont {Craig~C.}\
  \bibnamefont {Price}}\ and\ \bibinfo {author} {\bibfnamefont {Natalia~B.}\
  \bibnamefont {Perkins}},\ }\bibfield  {title} {\enquote {\bibinfo {title}
  {Critical properties of the kitaev-heisenberg model},}\ }\href@noop {}
  {\bibfield  {journal} {\bibinfo  {journal} {Phys. Rev. Lett.}\ }\textbf
  {\bibinfo {volume} {109}},\ \bibinfo {pages} {187201} (\bibinfo {year}
  {2012})}\BibitemShut {NoStop}%
\bibitem [{\citenamefont {Price}\ and\ \citenamefont
  {Perkins}(2013)}]{PhysRevB.88.024410}%
  \BibitemOpen
  \bibfield  {author} {\bibinfo {author} {\bibfnamefont {Craig}\ \bibnamefont
  {Price}}\ and\ \bibinfo {author} {\bibfnamefont {Natalia~B.}\ \bibnamefont
  {Perkins}},\ }\bibfield  {title} {\enquote {\bibinfo {title}
  {Finite-temperature phase diagram of the classical kitaev-heisenberg
  model},}\ }\href@noop {} {\bibfield  {journal} {\bibinfo  {journal} {Phys.
  Rev. B}\ }\textbf {\bibinfo {volume} {88}},\ \bibinfo {pages} {024410}
  (\bibinfo {year} {2013})}\BibitemShut {NoStop}%
\bibitem [{\citenamefont {Samarakoon}\ \emph {et~al.}(2017)\citenamefont
  {Samarakoon}, \citenamefont {Banerjee}, \citenamefont {Zhang}, \citenamefont
  {Kamiya}, \citenamefont {Nagler}, \citenamefont {Tennant}, \citenamefont
  {Lee},\ and\ \citenamefont {Batista}}]{PhysRevB.96.134408}%
  \BibitemOpen
  \bibfield  {author} {\bibinfo {author} {\bibfnamefont {A.~M.}\ \bibnamefont
  {Samarakoon}}, \bibinfo {author} {\bibfnamefont {A.}~\bibnamefont
  {Banerjee}}, \bibinfo {author} {\bibfnamefont {S.-S.}\ \bibnamefont {Zhang}},
  \bibinfo {author} {\bibfnamefont {Y.}~\bibnamefont {Kamiya}}, \bibinfo
  {author} {\bibfnamefont {S.~E.}\ \bibnamefont {Nagler}}, \bibinfo {author}
  {\bibfnamefont {D.~A.}\ \bibnamefont {Tennant}}, \bibinfo {author}
  {\bibfnamefont {S.-H.}\ \bibnamefont {Lee}}, \ and\ \bibinfo {author}
  {\bibfnamefont {C.~D.}\ \bibnamefont {Batista}},\ }\bibfield  {title}
  {\enquote {\bibinfo {title} {Comprehensive study of the dynamics of a
  classical kitaev spin liquid},}\ }\href@noop {} {\bibfield  {journal}
  {\bibinfo  {journal} {Phys. Rev. B}\ }\textbf {\bibinfo {volume} {96}},\
  \bibinfo {pages} {134408} (\bibinfo {year} {2017})}\BibitemShut {NoStop}%
\bibitem [{\citenamefont {Takai}\ \emph {et~al.}(2016)\citenamefont {Takai},
  \citenamefont {Ido}, \citenamefont {Misawa}, \citenamefont {Yamaji},\ and\
  \citenamefont {Imada}}]{doi:10.7566/JPSJ.85.034601}%
  \BibitemOpen
  \bibfield  {author} {\bibinfo {author} {\bibfnamefont {Kensaku}\ \bibnamefont
  {Takai}}, \bibinfo {author} {\bibfnamefont {Kota}\ \bibnamefont {Ido}},
  \bibinfo {author} {\bibfnamefont {Takahiro}\ \bibnamefont {Misawa}}, \bibinfo
  {author} {\bibfnamefont {Youhei}\ \bibnamefont {Yamaji}}, \ and\ \bibinfo
  {author} {\bibfnamefont {Masatoshi}\ \bibnamefont {Imada}},\ }\bibfield
  {title} {\enquote {\bibinfo {title} {Finite-temperature variational monte
  carlo method for strongly correlated electron systems},}\ }\href@noop {}
  {\bibfield  {journal} {\bibinfo  {journal} {Journal of the Physical Society
  of Japan}\ }\textbf {\bibinfo {volume} {85}},\ \bibinfo {pages} {034601}
  (\bibinfo {year} {2016})}\BibitemShut {NoStop}%
\bibitem [{\citenamefont {Huang}\ \emph {et~al.}(2016)\citenamefont {Huang},
  \citenamefont {Liao}, \citenamefont {Liu}, \citenamefont {Xie}, \citenamefont
  {Xie}, \citenamefont {Zhao}, \citenamefont {Chen},\ and\ \citenamefont
  {Xiang}}]{huang2016generalized}%
  \BibitemOpen
  \bibfield  {author} {\bibinfo {author} {\bibfnamefont {Rui-Zhen}\
  \bibnamefont {Huang}}, \bibinfo {author} {\bibfnamefont {Hai-Jun}\
  \bibnamefont {Liao}}, \bibinfo {author} {\bibfnamefont {Zhi-Yuan}\
  \bibnamefont {Liu}}, \bibinfo {author} {\bibfnamefont {Hai-Dong}\
  \bibnamefont {Xie}}, \bibinfo {author} {\bibfnamefont {Zhi-Yuan}\
  \bibnamefont {Xie}}, \bibinfo {author} {\bibfnamefont {Hui-Hai}\ \bibnamefont
  {Zhao}}, \bibinfo {author} {\bibfnamefont {Jing}\ \bibnamefont {Chen}}, \
  and\ \bibinfo {author} {\bibfnamefont {Tao}\ \bibnamefont {Xiang}},\
  }\bibfield  {title} {\enquote {\bibinfo {title} {A generalized lanczos method
  for systematic optimization of tensor network states},}\ }\href@noop {}
  {\bibfield  {journal} {\bibinfo  {journal} {arXiv preprint arXiv:1611.09574}\
  } (\bibinfo {year} {2016})}\BibitemShut {NoStop}%
\bibitem [{\citenamefont {Nasu}\ \emph {et~al.}(2017)\citenamefont {Nasu},
  \citenamefont {Yoshitake},\ and\ \citenamefont {Motome}}]{nasu2017thermal}%
  \BibitemOpen
  \bibfield  {author} {\bibinfo {author} {\bibfnamefont {Joji}\ \bibnamefont
  {Nasu}}, \bibinfo {author} {\bibfnamefont {Junki}\ \bibnamefont {Yoshitake}},
  \ and\ \bibinfo {author} {\bibfnamefont {Yukitoshi}\ \bibnamefont {Motome}},\
  }\bibfield  {title} {\enquote {\bibinfo {title} {Thermal transport in the
  kitaev model},}\ }\href@noop {} {\bibfield  {journal} {\bibinfo  {journal}
  {Physical Review Letters}\ }\textbf {\bibinfo {volume} {119}},\ \bibinfo
  {pages} {127204} (\bibinfo {year} {2017})}\BibitemShut {NoStop}%
\bibitem [{\citenamefont {Rousochatzakis}\ and\ \citenamefont
  {Perkins}(2017)}]{PhysRevLett.118.147204}%
  \BibitemOpen
  \bibfield  {author} {\bibinfo {author} {\bibfnamefont {Ioannis}\ \bibnamefont
  {Rousochatzakis}}\ and\ \bibinfo {author} {\bibfnamefont {Natalia~B.}\
  \bibnamefont {Perkins}},\ }\bibfield  {title} {\enquote {\bibinfo {title}
  {Classical spin liquid instability driven by off-diagonal exchange in strong
  spin-orbit magnets},}\ }\href@noop {} {\bibfield  {journal} {\bibinfo
  {journal} {Phys. Rev. Lett.}\ }\textbf {\bibinfo {volume} {118}},\ \bibinfo
  {pages} {147204} (\bibinfo {year} {2017})}\BibitemShut {NoStop}%
\bibitem [{\citenamefont {Catuneanu}\ \emph {et~al.}()\citenamefont
  {Catuneanu}, \citenamefont {Yamaji}, \citenamefont {Wachtel}, \citenamefont
  {Kee},\ and\ \citenamefont {Kim}}]{arXiv:1701.07837}%
  \BibitemOpen
  \bibfield  {author} {\bibinfo {author} {\bibfnamefont {Andrei}\ \bibnamefont
  {Catuneanu}}, \bibinfo {author} {\bibfnamefont {Youhei}\ \bibnamefont
  {Yamaji}}, \bibinfo {author} {\bibfnamefont {Gideon}\ \bibnamefont
  {Wachtel}}, \bibinfo {author} {\bibfnamefont {Hae-Young}\ \bibnamefont
  {Kee}}, \ and\ \bibinfo {author} {\bibfnamefont {Yong~Baek}\ \bibnamefont
  {Kim}},\ }\href@noop {} {\enquote {\bibinfo {title} {Realizing quantum spin
  liquid phases in spin-orbit driven correlated materials},}\ }\Eprint
  {http://arxiv.org/abs/arXiv:1701.07837} {arXiv:1701.07837} \BibitemShut
  {NoStop}%
\bibitem [{\citenamefont {Gohlke}\ \emph {et~al.}()\citenamefont {Gohlke},
  \citenamefont {Wachtel}, \citenamefont {Yamaji}, \citenamefont {Pollmann},\
  and\ \citenamefont {Kim}}]{arXiv:1706.09908}%
  \BibitemOpen
  \bibfield  {author} {\bibinfo {author} {\bibfnamefont {Matthias}\
  \bibnamefont {Gohlke}}, \bibinfo {author} {\bibfnamefont {Gideon}\
  \bibnamefont {Wachtel}}, \bibinfo {author} {\bibfnamefont {Youhei}\
  \bibnamefont {Yamaji}}, \bibinfo {author} {\bibfnamefont {Frank}\
  \bibnamefont {Pollmann}}, \ and\ \bibinfo {author} {\bibfnamefont
  {Yong~Baek}\ \bibnamefont {Kim}},\ }\href@noop {} {\enquote {\bibinfo {title}
  {Signatures of quantum spin liquid in kitaev-like frustrated magnets},}\
  }\Eprint {http://arxiv.org/abs/arXiv:1706.09908} {arXiv:1706.09908}
  \BibitemShut {NoStop}%
\bibitem [{HPh()}]{HPhi}%
  \BibitemOpen
  \href@noop {} {}\bibinfo {note} {An ED program package $\mathcal{H}\Phi$ is
  available through https://github.com/QLMS/HPhi}\BibitemShut {NoStop}%
\bibitem [{\citenamefont {Kawamura}\ \emph {et~al.}(2017)\citenamefont
  {Kawamura}, \citenamefont {Yoshimi}, \citenamefont {Misawa}, \citenamefont
  {Yamaji}, \citenamefont {Todo},\ and\ \citenamefont
  {Kawashima}}]{Kawamura2017180}%
  \BibitemOpen
  \bibfield  {author} {\bibinfo {author} {\bibfnamefont {Mitsuaki}\
  \bibnamefont {Kawamura}}, \bibinfo {author} {\bibfnamefont {Kazuyoshi}\
  \bibnamefont {Yoshimi}}, \bibinfo {author} {\bibfnamefont {Takahiro}\
  \bibnamefont {Misawa}}, \bibinfo {author} {\bibfnamefont {Youhei}\
  \bibnamefont {Yamaji}}, \bibinfo {author} {\bibfnamefont {Synge}\
  \bibnamefont {Todo}}, \ and\ \bibinfo {author} {\bibfnamefont {Naoki}\
  \bibnamefont {Kawashima}},\ }\bibfield  {title} {\enquote {\bibinfo {title}
  {Quantum lattice model solver {$\mathcal{H}\Phi$}},}\ }\href@noop {}
  {\bibfield  {journal} {\bibinfo  {journal} {Computer Physics Communications}\
  }\textbf {\bibinfo {volume} {217}},\ \bibinfo {pages} {180 -- 192} (\bibinfo
  {year} {2017})}\BibitemShut {NoStop}%
\bibitem [{\citenamefont {Yamamoto}\ \emph {et~al.}(2008)\citenamefont
  {Yamamoto}, \citenamefont {Sogabe}, \citenamefont {Hoshi}, \citenamefont
  {Zhang},\ and\ \citenamefont {Fujiwara}}]{doi:10.1143/JPSJ.77.114713}%
  \BibitemOpen
  \bibfield  {author} {\bibinfo {author} {\bibfnamefont {Susumu}\ \bibnamefont
  {Yamamoto}}, \bibinfo {author} {\bibfnamefont {Tomohiro}\ \bibnamefont
  {Sogabe}}, \bibinfo {author} {\bibfnamefont {Takeo}\ \bibnamefont {Hoshi}},
  \bibinfo {author} {\bibfnamefont {Shao-Liang}\ \bibnamefont {Zhang}}, \ and\
  \bibinfo {author} {\bibfnamefont {Takeo}\ \bibnamefont {Fujiwara}},\
  }\bibfield  {title} {\enquote {\bibinfo {title} {Shifted
  conjugate-orthogonal–conjugate-gradient method and its application to
  double orbital extended {H}ubbard model},}\ }\href@noop {} {\bibfield
  {journal} {\bibinfo  {journal} {J. Phys. Soc. Jpn.}\ }\textbf {\bibinfo
  {volume} {77}},\ \bibinfo {pages} {114713} (\bibinfo {year}
  {2008})}\BibitemShut {NoStop}%
\bibitem [{\citenamefont {Saad}(2011)}]{saad2011numerical}%
  \BibitemOpen
  \bibfield  {author} {\bibinfo {author} {\bibfnamefont {Yousef}\ \bibnamefont
  {Saad}},\ }\href@noop {} {\emph {\bibinfo {title} {Numerical Methods for
  Large Eigenvalue Problems: Revised Edition}}}\ (\bibinfo  {publisher}
  {SIAM},\ \bibinfo {year} {2011})\BibitemShut {NoStop}%
\bibitem [{\citenamefont {Mead}\ and\ \citenamefont
  {Papanicolaou}(1984)}]{mead1984maximum}%
  \BibitemOpen
  \bibfield  {author} {\bibinfo {author} {\bibfnamefont {Lawrence~R}\
  \bibnamefont {Mead}}\ and\ \bibinfo {author} {\bibfnamefont {Nikos}\
  \bibnamefont {Papanicolaou}},\ }\bibfield  {title} {\enquote {\bibinfo
  {title} {Maximum entropy in the problem of moments},}\ }\href@noop {}
  {\bibfield  {journal} {\bibinfo  {journal} {Journal of Mathematical Physics}\
  }\textbf {\bibinfo {volume} {25}},\ \bibinfo {pages} {2404--2417} (\bibinfo
  {year} {1984})}\BibitemShut {NoStop}%
\bibitem [{\citenamefont {Williams}\ and\ \citenamefont
  {Maris}(1985)}]{PhysRevB.31.4508}%
  \BibitemOpen
  \bibfield  {author} {\bibinfo {author} {\bibfnamefont {Michael~L.}\
  \bibnamefont {Williams}}\ and\ \bibinfo {author} {\bibfnamefont
  {Humphrey~J.}\ \bibnamefont {Maris}},\ }\bibfield  {title} {\enquote
  {\bibinfo {title} {Numerical study of phonon localization in disordered
  systems},}\ }\href@noop {} {\bibfield  {journal} {\bibinfo  {journal} {Phys.
  Rev. B}\ }\textbf {\bibinfo {volume} {31}},\ \bibinfo {pages} {4508--4515}
  (\bibinfo {year} {1985})}\BibitemShut {NoStop}%
\bibitem [{\citenamefont {Yakubo}\ and\ \citenamefont
  {Nakayama}(1987)}]{PhysRevB.36.8933}%
  \BibitemOpen
  \bibfield  {author} {\bibinfo {author} {\bibfnamefont {K.}~\bibnamefont
  {Yakubo}}\ and\ \bibinfo {author} {\bibfnamefont {T.}~\bibnamefont
  {Nakayama}},\ }\bibfield  {title} {\enquote {\bibinfo {title} {Absence of the
  hump in the density of states of percolating clusters},}\ }\href@noop {}
  {\bibfield  {journal} {\bibinfo  {journal} {Phys. Rev. B}\ }\textbf {\bibinfo
  {volume} {36}},\ \bibinfo {pages} {8933--8936} (\bibinfo {year}
  {1987})}\BibitemShut {NoStop}%
\bibitem [{\citenamefont {Gegenwart}\ and\ \citenamefont
  {Trebst}(2015)}]{gegenwart2015spin}%
  \BibitemOpen
  \bibfield  {author} {\bibinfo {author} {\bibfnamefont {Philipp}\ \bibnamefont
  {Gegenwart}}\ and\ \bibinfo {author} {\bibfnamefont {Simon}\ \bibnamefont
  {Trebst}},\ }\bibfield  {title} {\enquote {\bibinfo {title} {Spin-orbit
  physics: Kitaev matter},}\ }\href@noop {} {\bibfield  {journal} {\bibinfo
  {journal} {Nature Physics}\ }\textbf {\bibinfo {volume} {11}},\ \bibinfo
  {pages} {444--445} (\bibinfo {year} {2015})}\BibitemShut {NoStop}%
\bibitem [{\citenamefont {Trebst}()}]{trebst2017kitaev}%
  \BibitemOpen
  \bibfield  {author} {\bibinfo {author} {\bibfnamefont {Simon}\ \bibnamefont
  {Trebst}},\ }\href@noop {} {\enquote {\bibinfo {title} {Kitaev materials},}\
  }\Eprint {http://arxiv.org/abs/arXiv:1701.07056} {arXiv:1701.07056}
  \BibitemShut {NoStop}%
\bibitem [{\citenamefont {Balents}(2010)}]{balents2010spin}%
  \BibitemOpen
  \bibfield  {author} {\bibinfo {author} {\bibfnamefont {Leon}\ \bibnamefont
  {Balents}},\ }\bibfield  {title} {\enquote {\bibinfo {title} {Spin liquids in
  frustrated magnets},}\ }\href@noop {} {\bibfield  {journal} {\bibinfo
  {journal} {Nature}\ }\textbf {\bibinfo {volume} {464}},\ \bibinfo {pages}
  {199--208} (\bibinfo {year} {2010})}\BibitemShut {NoStop}%
\bibitem [{\citenamefont {Sogabe}\ \emph {et~al.}(2007)\citenamefont {Sogabe},
  \citenamefont {Hoshi}, \citenamefont {Zhang},\ and\ \citenamefont
  {Fujiwara}}]{sogabe2007numerical}%
  \BibitemOpen
  \bibfield  {author} {\bibinfo {author} {\bibfnamefont {Tomohiro}\
  \bibnamefont {Sogabe}}, \bibinfo {author} {\bibfnamefont {Takeo}\
  \bibnamefont {Hoshi}}, \bibinfo {author} {\bibfnamefont {Shao-Liang}\
  \bibnamefont {Zhang}}, \ and\ \bibinfo {author} {\bibfnamefont {Takeo}\
  \bibnamefont {Fujiwara}},\ }\bibfield  {title} {\enquote {\bibinfo {title} {A
  numerical method for calculating the {G}reen's function arising from
  electronic structure theory},}\ }in\ \href@noop {} {\emph {\bibinfo
  {booktitle} {Frontiers of Computational Science}}}\ (\bibinfo  {publisher}
  {Springer},\ \bibinfo {year} {2007})\ pp.\ \bibinfo {pages}
  {189--195}\BibitemShut {NoStop}%
\end{thebibliography}%

\end{document}